# Virtual Windshields: Merging Reality and Digital Content to Improve the Driving Experience

**Michelle Krüger Silvéria Santos**

FINAL VERSION

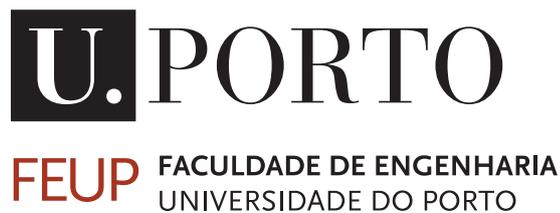

Mestrado em Multimédia

Supervisor: Michel Celestino Paiva Ferreira (PhD)

July, 2013

# Virtual Windshields: Merging Reality and Digital Content to Improve the Driving Experience

**Michelle Krüger Silvéria Santos**

Mestrado em Multimédia

Approved by:

President: Eurico Manuel Elias de Morais Carrapatoso (PhD)

External Referee: Mónica Sofia Santos Mendes (PhD)

Supervisor: Michel Celestino Paiva Ferreira (PhD)

July, 2013

# Abstract

During recent years, the use of the automobile as the primary mode of transportation has been increasing and the driving task performed daily by millions of people has become an important part of daily life. Driving is both a physical and a mental skill and is a multi-sensory experience. Drivers greatly rely on their senses to provide them with information about their surroundings to guide them to accurately and safely perform the driving task. Despite the effectiveness of human senses, in a vehicular context they are all too often limited and obstructed by various elements which can lead to misinterpretations of particular situations by drivers. Such a flaw may have negative outcomes and potentially result in accidents. Today, road accidents constitute the eighth leading cause of death, representing the first leading cause of death in young people between the ages of 15 and 29.

Vehicular safety has been actively explored in recent year and even before the appearance of motorized vehicles many devices had been developed and placed not only in vehicles, but also in the road environment as a means to regulate traffic, and to provide better awareness and increase road safety for all road users.

The rapid growth of technology has propelled novel ways in which driver's senses may be augmented. One such technology is that of Augmented Reality (AR) which offers the ability to provide drivers with a variety of information in an unobtrusive manner. Augmented Reality is starting to be widely explored in a vehicular context, mainly aimed at drivers, in providing them with different kinds of information which allows for more effective and safer driving. The enclosed aspect of a car, allied with the configuration of the controls and displays directed towards the driver, offer significant advantages for augmented reality systems when considering the amount of immersion it can provide to the user. The windshield itself may be seen as an augmented reality transparent canvas on which digital content may be super-imposed over the driving scenario. In addition, the inherent mobility and virtually unlimited power autonomy transform cars into perfect mobile computing platforms. However, automobiles currently present limited network connectivity and thus the created augmented objects are merely providing information captured by in-vehicle sensors, cameras, maps and other databases. A review of such systems currently available in vehicles has been conducted as part of this thesis, as well as a study of vehicular connectivity paradigms.

By combining the new paradigm of Vehicular Ad Hoc Networking (VANET) with augmented reality human machine interfaces, we aim to show that it is possible to design novel cooperative Advanced Driver Assistance Systems (ADAS), that base the creation of AR content on the information collected from neighbouring vehicles or roadside infrastructures. This digital content can then be super-imposed on a transparent windshield and create novel opportunities for the contextualized dissemination of digital content. Not only can the visual sense of the driver be explored through this AR paradigm, but also the auditory sense, creating acoustic digital content that can greatly improve the overall awareness of drivers.

We implement some prototypes of both visual and acoustic AR systems, using not only a simulation-based platform but also using real vehicles and highly immersive AR equipment such as surround sound systems and augmented reality glasses.

We believe our results contribute to the formulation of a vision where the vehicle is perceived as an extension of the body which permeates the human senses to the world outside the vessel, where the car is used as a better, multi-sensory immersive version of a mobile phone that integrates touch, vision and sound enhancements, leveraging unique properties of vehicular ad hoc networking.

# Resumo

Ao longo dos anos, o uso do automóvel como principal meio de transporte tem vindo a aumentar e a tarefa da condução realizada diariamente por milhões de pessoas tornou-se uma parte importante do dia-a-dia. Conduzir é uma tarefa tanto física como mental, e é também uma experiência multissensorial. Os condutores dependem essencialmente dos seus sentidos para obterem informação sobre o meio envolvente para que possam desempenhar a tarefa de condução de forma correcta e em segurança. Apesar da eficácia dos sentidos humanos, num contexto veicular estes são demasiadas vezes limitados e obstruídos por vários elementos que podem levar à má interpretação de uma situação por parte do condutor. Tal falha poderá ter um resultado negativo e poderá potencialmente resultar em acidentes. Actualmente, os acidentes de viação constituem a oitava causa de morte a nível mundial, representando a primeira causa de morte entre os jovens com idades compreendidas entre os 15 e os 29 anos.

A segurança veicular tem sido explorada de forma activa nos últimos anos. Contudo, mesmo antes do aparecimento dos veículos motorizados o desenvolvimento e colocação de dispositivos de segurança, tanto dentro dos veículos como no próprio ambiente rodoviário, (tem sido feito) como meio para regular o trânsito, melhorar a percepção global dos condutores e aumentar a segurança para todos os utilizadores das estradas.

O rápido crescimento da tecnologia tem impulsionado novos meios que permitem ampliar os sentidos dos condutores. Uma dessas tecnologias é a Realidade Aumentada (AR), que oferece aos condutores uma variedade de informação de forma discreta. A Realidade Aumentada tem vindo a ser amplamente explorada no contexto veicular, direccionada principalmente aos condutores, proporcionando-lhes diferentes tipos de informação que permite uma condução mais eficaz e segura. O aspecto confinado de um carro, aliado à configuração dos dispositivos de controlo e visores direccionados para o condutor, oferecem vantagens significativas para os sistemas de realidade aumentada, quando se considera a quantidade de imersão que este pode proporcionar ao utilizador. O pára-brisas em si pode ser visto como uma tela transparente de realidade aumentada sobre o qual conteúdo digital pode ser sobreposto sobre o cenário de condução. Além disso, a mobilidade inerente e autonomia energética virtualmente ilimitada transforma os carros em perfeitas plataformas de computação móvel. Contudo, os automóveis apresentam atualmente uma conectividade de rede limitada e, assim, os objetos criados limitam-se a fornecer informação obtida por sensores e câmaras contidas no veículo, mapas e outras bases de dados. Como parte desta tese foi feita uma revisão de tais sistemas actualmente disponíveis em veículos, bem como um estudo de paradigmas de conectividade veicular.

Ao combinar o novo paradigma das redes veiculares ad hoc (VANET) com interfaces homem-máquina de realidade aumentada, pretendemos mostrado que é possível desenhar novos sistemas avançados de assistência à condução (ADAS), que baseiam a criação de conteúdo de realidade aumentada na informação recolhida de veículos vizinhos ou infraestruturas na estrada. Este conteúdo digital pode então ser sobreposto sobre um pára-brisas transparente e criar novas oportunidades para a divulgação contextualizada de conteúdos digitais. Não apenas o sentido visual do mo-torista pode ser explorado através deste paradigma de realidade aumentada, mas também o sentido auditivo, através da criação de conteúdo digital acústico que pode melhorar significativamente a percepção global dos condutores.

Implementamos alguns protótipos de sistemas de realidade aumentada visuais e acústicos, utilizando não só uma plataforma de simulação, mas também utilizando veículos reais e equipamentos de realidade aumentada altamente imersivos, tais como sistemas de som surround e óculos de realidade aumentada.

Acreditamos que os nossos resultados contribuem para a formulação de uma visão na qual o veículo é entendido como uma extensão do corpo, permeando os sentidos humanos com o mundo exterior ao veículo. Uma visão onde o carro é utilizado como uma versão melhorada, multi-sensorial e mais imersiva do que um smartphone que integra toque, visão e som, alavancando nas propriedades únicas de redes veiculares ad hoc.

# Acknowledgements

Throughout the duration of my Masters program, there were many who supported me in their different ways. Now as I conclude two years of hard work I take this opportunity to express my appreciation to all who made this possible.

I praise and thank God for blessing me throughout my life and for strengthening me during this research project "I can do all this through him who gives me strength" (Philippians 4: 13).

I would like to express my deep and sincere gratitude to my supervisor Dr. Michel Ferreira, Assistant Professor at the Faculty of Sciences of the University of Porto, for encouraging me to under take the Masters in Multimedia program and for giving me the opportunity to develop this thesis in the scope of the DRIVE-IN and VTL projects, at Instituto de Telecomunicações do Porto. I thank him for his invaluable guidance and motivation and for always being available to assist me in any way possible.

I am extremely grateful to my husband for his love, understanding and support, for always letting me know that I could succeed in the Masters program and this research project. His constant motivation often provided me with the strength to push forward and not give up. To my parents for being role models in my life, for raising me to believe that with honesty and hard work I can achieve everything I set my mind to. Thank you for your love and support! To my sister and brother in-law, a special thank you for your words of encouragement. Despite the distance your love was felt. To all my family, near or far, your constant words of encouragement were heart felt.

I also thank my friends and family at Riverside Porto for remembering me in their prayers.

To my collogues of the DRIVE-IN and VTL projects I appreciate your help and advice. I also thank Dr. Cristina Olaverri-Monreal for encouraging me to apply for the research position and for encouraging me to undertake the Masters in Multimedia program.

I truly appreciate everyone's support and encouragement and without it I surely would not have been able to completes this work!

Thank you!

Michelle Krüger Silvéria Santos

# Contents

# List of Figures

# List of Tables

# Abbreviations and Symbols

| | |
|---|---|
| 2D | Two-Dimensional |
| 3D | Three-Dimensional |
| 1G | First-generation |
| 2G | Second-generation |
| 3G | Third-generation |
| 3GPP | 3rd Generation Partnership Project |
| 4G | Fourth-generation |
| ADAS | Advanced Driver Assistance Systems |
| AIDA | Affective, Intelligent Driving Agent |
| AR | Augmented Reality |
| AU | Application Unit |
| C2C | Car-to-Car |
| C2C-CC | Car2Car Communication Consortium |
| C2I | Car-to-Infrastructure |
| CAM | Cooperative Awareness Message |
| CE | Código da Estrada |
| CHMSL | Centre High Mounted Stop Lamps |
| co-ADAS | Cooperative Advanced Driver Assistance Systems |
| CSMA-CA | Carrier Sense Multiple Access with Collision Avoidance |
| CT | Computed Tomography |
| DIVERT | Development of Inter-VEhicular Reliable Telematics |
| DRIVE-IN | Distributed Routing and Infotainment through VEhicular Inter-Networking |
| DSRC | Dedicated Short-Range Communications |
| EDGE | Enhanced Data Rates for GSM Evolution |
| ESP | Electronic Stability Program |
| EP | Estradas de Portugal |
| ETC | Electronic Toll Collection |
| EU | European Union |
| GDP | Gross Domestic Product |
| GPS | Global Positioning System |
| HMD | Head-Mounted Display |
| HMI | Human Machine Interface |
| HMPD | Head-Mounted Projective Display |
| HUD | Head-Up Display |
| InIR | Instituto de Infra-Estrutura Rodoviária IP |
| IDM | Intelligent Driver Model |
| ITS | Intelligent Transportation Systems |
| LCD | Liquid-Crystal D isplay |
| LOD | Level-of-Detail |
| MAC | Medium Access Control |
| MEMS | Micro-Electro-Mechanical System |
| MR | Mixed Reality |
| MRI | Magnetic Resonance Imaging |
| OBU | On-board Unit |
| OpenGL | Open Graphics Library |
| OSG | OpenSceneGraph |
| PC | Personal Computer |
| PDA | Personal Digital Assistant |
| PHMD | Projective Head-Mounted Display |
| PHY | Physical |
| PND | Portable Navigation Device |
| QoS | Quality of Service |
| ROI | Region of Interest |
| RST | Regulamento de Sinalização de Trânsito |
| RSU | Road-side Unit |
| RTM | Road Trench Model |
| SAD | Spatial Auditory Display |
| STS | See-Through System |
| TMC | Traffic Message Channel |
| TOF | Time of Flight |
| UMTS | Universal Mobile Telecommunications System |
| VANET | Vehicular ad hoc Network |
| V2I | Vehicle-to-Infrastructure |
| V2V | Vehicle-to-Vehicle |
| VC | Virtual Continuum |
| VMS | Variable Message Signs |
| VNS | Vehicular Network Simulator |
| VR | Virtual Reality |
| VRD | Virtual Retinal Display |
| VSS | Virtual Surround Sound |
| VTL | Virtual Traffic Lights |
| WLAN | Wireless Local Area Network |

# Chapter 1

# Introduction

## 1.1 Context

During recent years, the use of the automobile as the primary mode of transportation has been increasing and the driving task performed daily by millions of people has become an important part of daily life. This scenario is more prominent in major urban centres where the increasing use of automobiles causes large traffic congestions. According to [29], these traffic jams constitute a significant problem and its cost in 2010 represented approximately 1% of the European Union's Gross Domestic Product (GDP).

Augmented reality applied to a vehicular context can open new opportunities for merging digital content with real world environment that is captured by a transparent windshield and can be explored in three distinct areas. Firstly, in the area of vehicular safety, this aims to reduce road accidents through driving assistance systems. Subsequently in the enhancement of traffic-flow through the development of more effective navigation systems which present the information to the driver more effectively. Furthermore the development of virtual traffic light systems may significantly improve road intersection management, with important consequences for improving traffic flow. The concept of virtual traffic lights based on augmented reality is also reflected in the previous aspect of reducing the road accident rate. Finally, in the field of entertainment, augmented reality is present through the virtualization of advertising. The exploration of augmented reality for advertising is a very strong trend nowadays. In the context of the automobile and in particular of the windshields, the creation of such augmented reality to display advertising content has an enormous potential.

## 1.2 Motivation and Goals

The main goal of the work developed within this thesis is to design novel driver infotainment systems based on the creation of augmented reality over the windshield, improving the overall driving experience, in terms of safety, navigational and traffic control efficiency, and entertainment/advertising exposure.

We are motivated in this thematic as we have been part of Instituto de Telecomunicações – Porto which is an important research group within the University of Porto and has been addressing the design of Intelligent Transportation System applications in the context of several FCT-funded research projects, such as DRIVE-IN – Distributed Routing and Infotainment through VEhicular Inter-Networking (CMU-PT/NGN/0052/2008) and VTL – Virtual Traffic Lights (PTDC/EIA-CCO/118114/2010).

The DRIVE-IN project was a research project approved within the 2008 Call for Proposals for the Information and Communication Technologies Institute (ICTI) and the Carnegie Mellon University - Portugal Program (Carnegie Mellon|PORTUGAL) by the Fundação para a Ciência e a Tecnologia (FCT). DRIVE-IN was included in the research program on New Generation Dependable Trusted Networks and Telecommunications Policy (NGN), addressed by the 2008 Call. This project's goal was to investigate how vehicle-to-vehicle communication could improve the user experience and the overall efficiency of vehicle and road utilization.

The VTL project is a natural follow-up of the DRIVE-IN project. Taking advantage of the new vehicle-to-vehicle (V2V) communication capability of modern cars, this project proposes to design and validate an alternative system of urban traffic control that envisions physical traffic lights as in-vehicle virtual signs. This would allow for the ubiquity of signalized intersections, as well as synchronizing light phases, cycle durations and green splits based on distributed and self-organizing techniques that govern a Vehicular ad-hoc Network (VANET).

As a research scholar within these projects, the availability of important research tools towards the experiments I intend to undertake within this thesis, further motivates my commitment into the design of novel systems based on advanced human-machine interfaces for the display of digital content. In particular, the availability of novel hardware equipments, such as transparent LCDs or vehicle-to-vehicle communication radios through the IEEE 802.11p norm [67], are particularly enticing in the context of the work proposed for this thesis.

Beyond the personal motivation that results from the context of my current position as a research scholar in the area of Intelligent Transportation Systems (ITS), this thesis is highly motivated by the pressing challenges faced by vehicular mobility today, where the rising costs of fuel or the increased urbanization of the world have questioned the sustainability of the current paradigm of privately-owned, car-based, transportation. Moreover, recent reports of the World Health Organization have highlighted car accidents as the ninth leading cause of death worldwide, with much amplified proportions if mortality is accounted in terms of Years of Life Lost [88].

If novel technology and state-of-the-art human-machine interfaces which display digital content can contribute to improve the efficiency of road transportation and reduce its current mortality toll, then this is the ultimate motivation that engages me into this work.

In a permanently connected society, where the quest for a share of Internet time is exploring all the minutes of the daily lives of people, from the small screens of a smartphone, to a tablet computer or a large smart TV, the 80" windshield screen results appeals to a number of Internet companies. More than the screen size, it is the overwhelming statistic of three daily hours of exposure of the average U.S. driver to this windshield screen that is motivating internet-based companies to look seriously into the technology of connected vehicles and autonomous driving [112]. This thesis aims to show that the windshield of connected vehicles, using different paradigms of wireless networking connectivity, exploring novel technologies that enable the creation of augmented reality through super-imposed digital content, can in fact capture a significant share of Internet time and reshape the car industry business in terms of its main players.

Of course, several challenges have to be met, as the displaying of informative content is especially critical during the driving task [75], during the transitory but long lasting phase of migration to autonomously driven vehicles.

## 1.3 Methodology

The proposed research methodology involved the study of relevant augmented reality libraries and techniques. Particular emphasis was given to the investigation of how to integrate digital objects representing roadside infrastructures in a realistic manner, such as traffic signs or advertising billboards, with the real-world vision, namely through the derivation of geographic coordinates for those objects through a process known as image registration.

Another important issue was the study of conventional traffic signs, in terms of rules for placement and visibility, types of traffic signs and the migration of these to in-vehicle display. Here we proposed new signs and develop 3D models of these signs as well as other road infra-structures.

A study of connectivity paradigms within vehicles was conducted, ranging from cellular based connectivity to the novel vehicle-vehicle and vehicle to infrastructure communication that is enabled by DSRC using the norm 802.11p.

Prototyping and evaluation was conducted through video see-through technology. Thus resorting to pre-recorded videos of driving environments to super-impose our digital content. As such, another task also involved the collecting a large database of video, collected through windshield-installed cameras. These videos supported the research in a laboratory environment prior to experimentation with laser holographic projection directly on the windshield, transparent LCD technology that would replace a conventional glass-based windshield and validation in real-driving scenarios. The recent acquisition of augmented reality goggles within the context of the projects where I am involved also provide an additional tool to design and evaluate the augmented content in the context of driving, in particular the augmented reality libraries that are provided with such equipment are also used within the work developed in this thesis.

## 1.4 Structure of the document

The remainder of this document is organized as follows. Chapter 2 presents a review of the literature and state-of-the-art of augmented reality, its techniques, technology and applications.

In chapter 3 the notion of driving as a multi-sensory experience is explored. We identify some traditional elements that stimulate the human senses and regulate traffic. Subsequently, augmented reality is introduced in the driving scenario, presenting various projects which enable an augmented driving experience. The projects are grouped according to the type of sense they explore: visual, auditory and tactile.

Chapter 4 introduces connected driving by exploring connectivity paradigms within vehicles, such as DSRC enabled through the 802.11p norm which allows vehicle-to-vehicle and vehicle-to-infrastructure communication. We will review wireless communication technologies, such as cellular and WiFi that are currently being explored in vehicular network. We conclude the chapter by identifying relevant applications for VANETs.

In chapter 5 we describe the implementation of some prototypes of AR-based ADAS and discuss some evaluation results. We start by describing the generic framework for the creation of acoustic and graphical content, embedded in the car and connected to its sensors and actuators. We divide our presentation between prototypes which have solely been implemented in simulated environments, and prototypes which have been implemented in real vehicles and experimented in real driving scenarios. Some evaluation results are also presented and discussed.

Finally in chapter 6 conclusions are drawn through the discussion of the obtained results. This chapter will also specify any steps needed for future work.

# Chapter 2

# Augmented Reality

This chapter introduces the topic of Augmented Reality technology and explores some of its characteristics. Firstly a definition and historical outline is presented and subsequently a review of the technologies that enable augmented reality, such as display and tracking, is detailed. Finally, some examples of projects which explore augmented reality in different fields are listed.

## 2.1  Definition and Overview

Augmented Reality or simply AR, is an innovative technology that has become widely popular over the past decade or so. Its use is currently being explored in numerous fields ranging from industrial contexts to medical research or even entertainment. Its uniqueness resides in its ability to place virtual objects into a real world environment complementing one's perception of reality. The exploration of artificially created virtual worlds is not new. Virtual Reality (VR), for example, is a well known and established technology having been explored for much longer. Although similar to Virtual Reality in the use of computer-generated elements, the two technologies differ in relation to the amount of virtual elements they use. While Virtual Reality provides users with the opportunity to be completely absorbed into a virtual scenario, losing all perception of reality, Augmented Reality on the other hand offers the possibility of superimposing digital elements over the real world, allowing the user to remain completely aware of his surroundings, as he still sees them [8, 19].

To better comprehend the similarities and differences between AR and VR and their relationship with the real world, [80] presented a taxonomy in which different environments were defined. In what is known as the Virtuality Continuum (VC), environments are located in conformity with the amount of virtual elements they present. As such, the real world is found at the one end of the continuum, containing no elements that alter the perception of it in any way. Virtual Reality on the other hand is found at the opposite end, consisting entirely of computer-generated virtual worlds. Augmented Reality finds itself in the area between real and virtual, in the space known as Mixed Reality (MR). Figure 2.1 illustrates this continuum and Augmented Reality is seen placed closer to the real world than to the virtual world as it contains a limited amount of virtual objects, which do not overpower reality. Also present in this MR space is Augmented Virtuality, which its turn is located closer to the virtual world and thus consisting mainly of virtual objects, only preserving some aspects of reality.

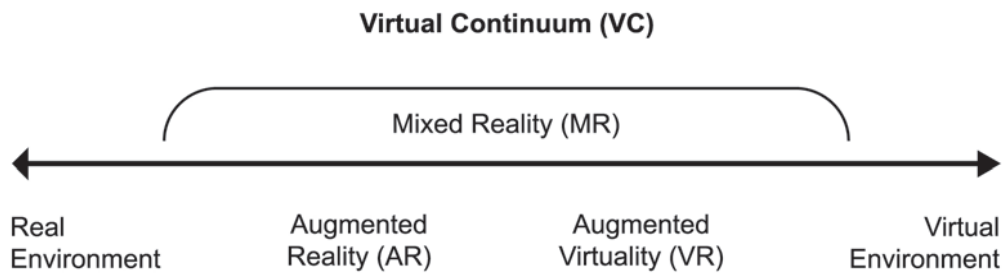

Figure 2.1: Representation of the "virtuality continuum" [80].

Although Augmented Reality has only become familiar among the general public in recent years, the idea of it seems to have been envisioned around a century ago by author L. Frank Baum in his novel of 1901 entitled "The Master Key" [13]. In this novel Baum introduces the idea of augmented reality spectacles known as "Character Marker" which overlay information onto real life. The next transcript taken from "The Master Key" presents this idea.

> "Therefore, that you may judge all your fellow-creatures truly, and know upon whom
> to depend, I give you the Character Marker. It consists of this pair of spectacles.
> While you wear them every one you meet will be marked upon the forehead with a
> letter indicating his or her character. (...) Thus you may determine by a single look the
> true natures of all those you encounter." (...) "All character sends out certain electrical
> vibrations, which these spectacles concentrate in their lenses and exhibit to the gaze
> of their wearer, as I have explained." [13]

However, it was only in the late 1960s that the first virtual and augmented reality head-mounted display (HMD) system was created by Ivan Sutherland, an American computer scientist. The system named "The Sword of Damocles" was so large and heavy that it was suspended from the ceiling to allow its usability. Figure 2.2 shows a close-up of the HMD in frame (a) and in frame (b) it is seen hanging from the ceiling.

The term "Augmented Reality" itself was only coined in the early 1990s by Tom Caudell, a senior researcher at Boeing, the American multinational aerospace and defence corporation while developing systems to assist in aircraft maintenance [105].

Given that head-mounted displays were the first type of AR displays to be developed, many suggested that this technology was solely applicable with HMD devices. However since the 1990s much progress has been made in the field of augmented reality, and since then a wide array of display devices have been developed. Following the technology's evolution [8] conducted a survey in 1997, in which he analysed the state-of-the-art of Augmented Reality at that time, and defined its implementation by three characteristics: 1) it combines the real world and virtual elements; 2) these elements are interactive in real time; 3) and they are registered in 3D. Therefore opening up the technology to much more then just Head Mounted Displays [8, 7].

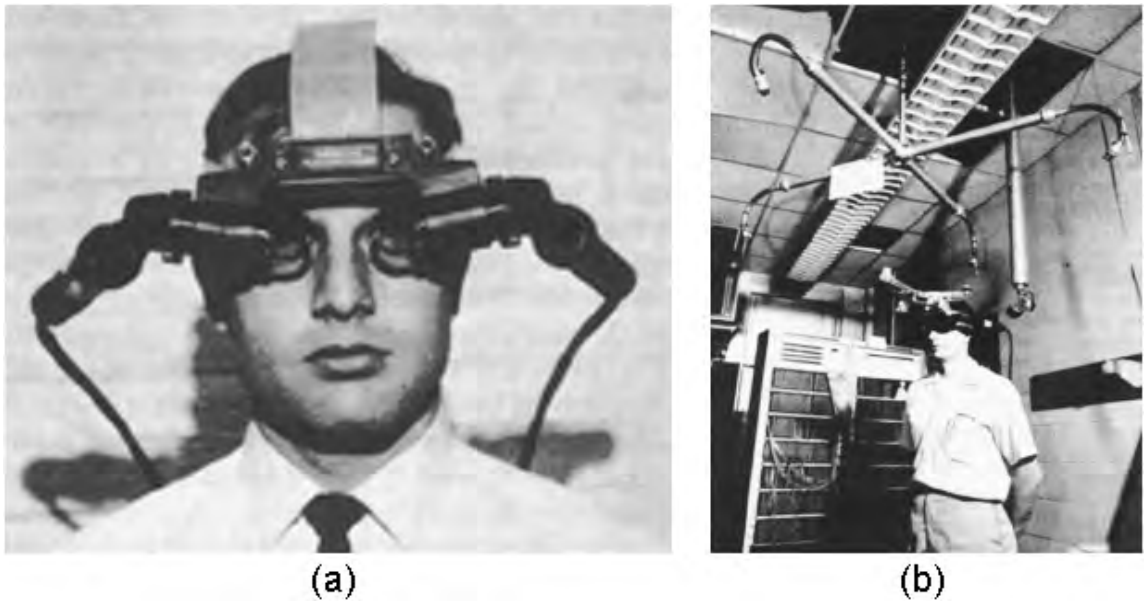

Figure 2.2: The Sword of Damocles [105]

## 2.2 Technology

The evolution of processors, displays and computing device sensors, as well as promising computer vision algorithms have all contributed to the spreading of Augmented Reality. Modern mobile devices such as Smartphones and tablet computers contain elements which include cameras and Micro-Electro-Mechanical Systems (MEMS) sensors such as accelerometer, Global Positioning System (GPS) and compass. These elements are all fitting for augmented reality platforms [5]. Next we will briefly describe the main technologies used in augmented reality systems.

### 2.2.1 Display Technology

> Augmented Reality displays are image-forming systems that use a set of optical, electronic and mechanical components to generate images somewhere on the optical path between the observer's eyes and the physical object to be augmented [19].

Currently, a great variety of technologies are used in Augmented Reality displays. Displays differ in their size, image resolution and even in the amount of immersion or reality they offer. Immersion may differ depending on the position of the displays between the observer and the real object. Figure 2.3 demonstrates different types of displays and their position relative to the user.

In [19], the authors discuss several types of augmented reality displays that we will summarize in the next section.

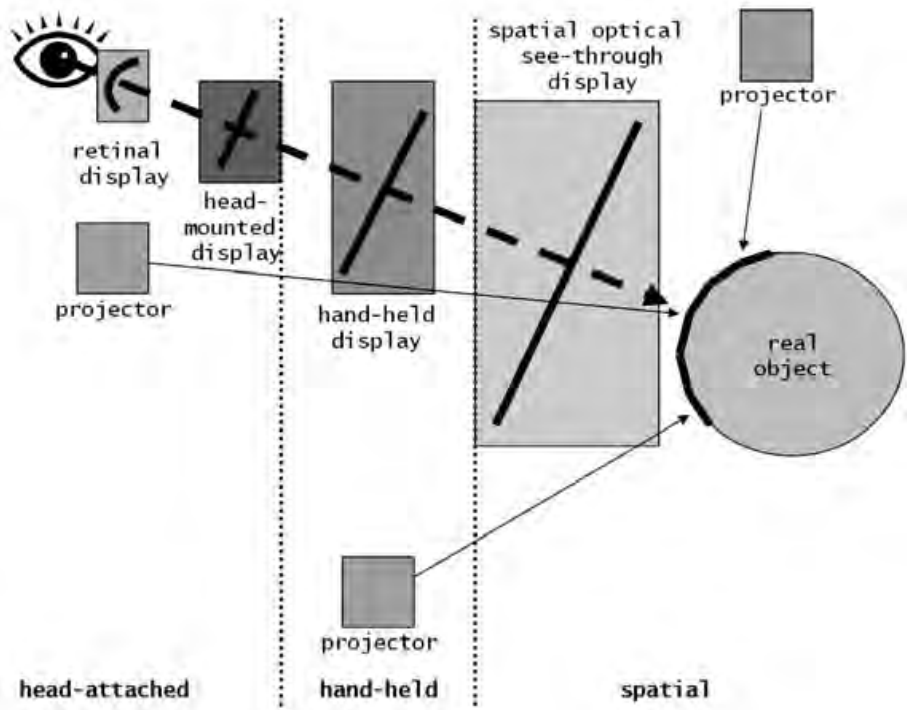

Figure 2.3: Image generation for augmented reality displays [19].

#### 2.2.1.1   Head-Attached Display

These types of displays commonly correspond to headsets such as helmets and glasses and require the user to wear them on their head. In this grouping three main types exist:

**Retinal Displays**
Retinal Displays use low-power lasers to scan light directly onto the viewer's retina. This technology produces a brighter and higher-resolution image and offers a wider field of view image than a screen-based display [19, 113]. In Fig. 2.4 frame a) shows a simplified diagram of how a retinal display works and frame b) presents a personal Virtual Retinal Display (VRD) that has been developed at the University of Washington's Human Interface Technology Laboratory (HITLab) in Seattle, Washington, USA [113].

**Head Mounted Displays (HMD)**
Head Mounted Displays or HMD's seem to be the most popular display devices used for augmented reality applications. According to [19] this technology appears in two forms: video see-through and optical see-through. Video see-through technology uses video-mixing to display images inside a closed-view head-mounted display such as a helmet, glass or screen. Optical see-through technology uses optical combiners such as half-reflective mirrors or transparent LCD displays, to impose the digital content over the viewer's field of view. Figure 2.5 shows the main differences between optical see-through shown in frame (a) and video see-through shown in frame (b).

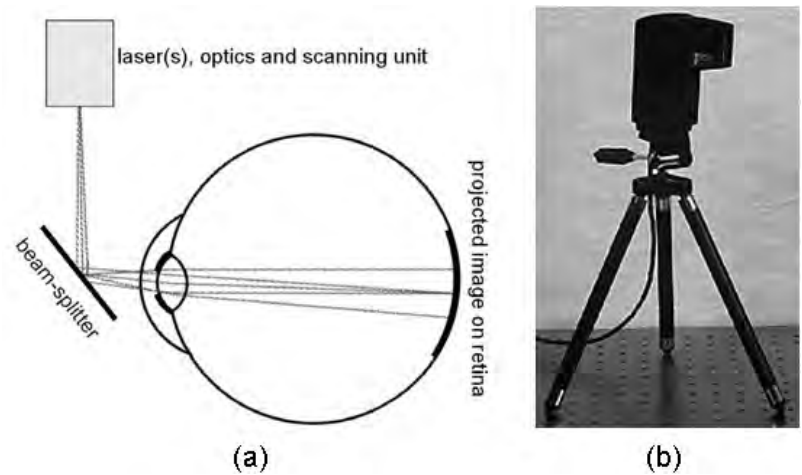

Figure 2.4: Simplified diagram of a retinal display (a) [19] and portable VRD unit (b) [113].

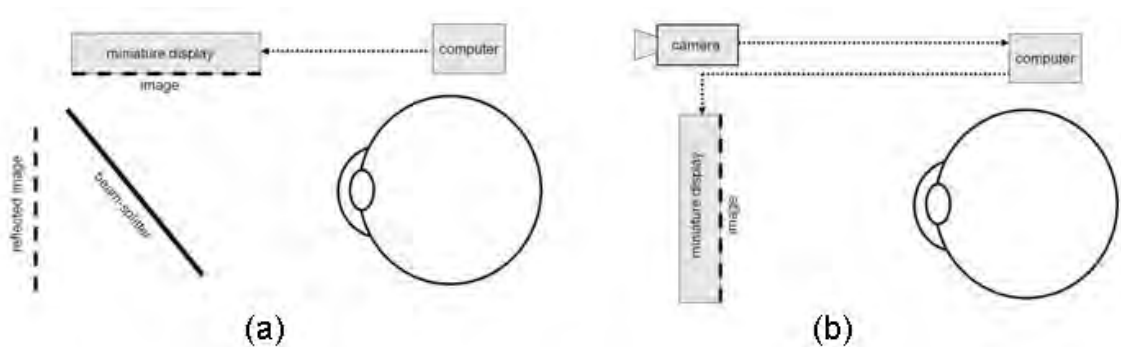

Figure 2.5: Simplified diagram of optical see-through (a) and video see-through (b) [19].

**Head Mounted Projectors**

As the name indicates, Head-mounted projectors use projectors that are attached to the head to display AR content. Overall, these displays present a larger field of view and they exist in two forms: Head-mounted projective displays (HMPDs) and projective head-mounted displays (PHMDs). Head-mounted projective displays beam images onto retro-reflective surfaces that are located in front of the viewer. According to [19] this is similar to the holographic films used for transparent projection screens. However, these films use back-projection whereas the retro-reflective surfaces used in HMPDs are front-projected. Projective head-mounted displays differ from HMPDs as they beam the images onto regular ceilings. The images are then integrated into the viewer's field of view through half-silvered mirrors. A simplified version of the head-mounted projector concept is shown in Fig. 2.6.

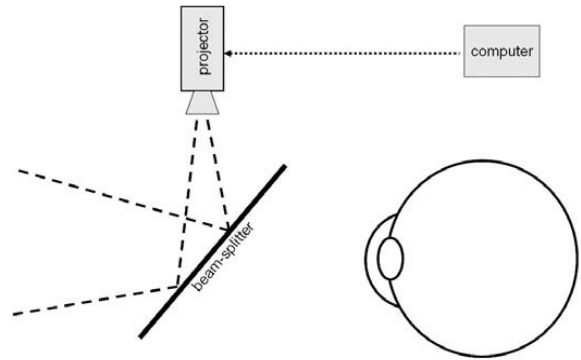

Figure 2.6: Simplified diagram of head-mounted projector [19].

### 2.2.1.2    Hand-Held Displays

Hand-held devices make use of small displays that fit in the user's hands. Examples of these are tablet PCs, personal digital assistants (PDA) and mobile phones. All of these devices combine processor, memory, display and integration technology into one and allow for unrestricted handling. Video see-through technology is mostly used with hand-held devices. Live video of the environment is captured with the integrated cameras of these devices and is then overlaid by graphic content creating an augmented scene. Some disadvantages of hand-held devices include physical constraints, such as the need for the user to hold the device in his hand the whole time and limited screen size which restrict the user's field of view [19]. Some hand held projectors are also sometimes used to display the AR content directly onto objects.

### 2.2.1.3    Spatial Display

Spatial Displays differ from body-attached displays (head-mounted or hand-held) in that they are not connected to the user, making them independent devices which integrate the AR content into the environment. This feature makes spatial AR suitable for group work, as the system can be used by several people at once. These types of displays augment the environment through three different approaches: Screen-Based Video See-Through, Spatial Optical See-Through and Projection-Based Spatial Displays. The main difference lies essentially in the way they augment the environment [19].

#### Screen-Based Video See-Through Displays

As mentioned previously, video See-Through makes use of video mixing for AR content display. Contrary to head-mounted or hand-held video See-Through, Spatial video See-Through is screen-based and as such the AR content is displayed on a regular monitor such as that of a standard desktop computer. The downside of this kind of display is that the user's field of view is restricted to the size of the monitor.

**Spatial Optical See-Through Displays**
Spatial Optical See-Through differs to head-attached or hand-held Optical See-Through as it aligns the generated images with the physical environment and does not need to be attached to the user. Again, mirror beam combiners, transparent screens or optical holograms are used in this type of display.

**Projection-Based Spatial Displays**
With Projection-Based Spatial devices the AR content is projected directly onto the surfaces of physical objects. This is achieved through different kinds of projectors (single static or steerable or even by the use of several projectors simultaneously) which amplify the display area.

### 2.2.2 Tracking Technology

Tracking is an important element in AR technology. It accurately integrates virtual and real objects, providing spatial uniformity between these elements. Tracking provides information about the position and orientation of objects and users as they move and interact within an environment. This technology combines different fields of engineering such as optics, electromagnetics, electronics, and mechanics and thus offers diverse systems which provide different degrees of precision and dimension. [101] provides an overview of tracking technologies, classifying them based on their physical principles of operation. These are: time of flight (TOF), spatial scan, inertial sensing, mechanical linkages, phase-difference sensing, and direct-field sensing.

**Time of flight (TOF)**
Time of flight or TOF systems measure the time that it takes for pulsed signals to travel from features placed on a referenced object to a moving target. The measurements are made with the assumption that the speed of propagation of the signals is constant. TOF systems may include ultrasonic trackers, infra-red laser beam trackers, GPS and optical gyroscopes [101].

**Spatial scan**
As specified in [101] Spatial scan trackers use optical devices to analyse 2D projections of image features so as to verify the position and orientation of a target. The optical sensors used for spatial scanning commonly consist of cameras, lateral-effect photodiodes or four-quadra detectors. It is possible to sub-classify these optical systems into two categories: inside-out and outside-in. The first presents the sensors attached to the target and the emitters are located on the reference. The second presents the sensors in form of cameras which are attached to a fixed reference and thus follow and record features of a moving target. Within outside-in systems, two methods are commonly employed: multiscopy and pattern recognition. Multiscopy use multiple imaging sensors (cameras) to calculate the spatial position of the target through triangulation. Pattern recognition on the other hand uses only one camera combined with a pattern. The pattern is made up of a 3D geometrical arrangement of features on a target which is captured by the camera. The recorded 2D pattern on the image combined with the 3D pattern make it possible to calculate the position and orientation of the target.

**Inertial sensing**

Inertial sensors use inertia to measure the rotation or position of the target. To calculate rotation, angular rate sensors such as mechanical gyroscopes are used. These function on the principles of angular momentum. Position, on the other hand, is measured by means of accelerometers. These devices measure the linear acceleration or translational motion of the target [101].

**Mechanical linkages**

These systems use the linkages between the reference and the target to measure the distance between them.

**Phase-difference sensing**

Phase-difference sensing measure the relative phase of signals coming from a target and compare them to signals located on the target. Although this method is similar to the TOF system, phase-difference trackers are able to generate higher data rates than the TOF ultrasonic trackers, as the phase can be measured continuously.

**Direct-field sensing**

Direct field sensing systems use a magnetic or gravity field to obtain the orientation or position of the target. These systems include the use of compasses or inclinometers.

**Hybrid systems**

Hybrid systems merely refer to the combination of two or more of the tracking technologies previously mentioned. In this way, different sensors may complement each other and overcome any limitations that they may present when used alone.

## 2.3   Applications

In 1997, [8] published a survey on augmented reality in which he proposed a definition for the field, identified registration and sensing issues and presented an overview of potential applications. The 1997 survey defines 6 classes for AR applications: medical, manufacturing, visualization, path planning, entertainment and military. Since then, many technological advances have been made and have allowed for applications to have potential in all fields. To accompany the technological advances, [7] presented a new survey in 2001 as a complement to the original, in which they indicate the rapid technological advancements. In this survey three new categories are added: mobile, collaborative and commercial applications. Additionally [58] provides an extensive overview of AR applications in education, as well as within fields such as movies and music, gaming, sports and others. A broad summary of current AR applications is also provided in [123].

As mentioned before, augmented reality has great potential for all fields where the rapid transfer of information is critical. In the next section we provide an overview of some of the fields where augmented reality has been explored and present some implementations.

### 2.3.1 Advertising and Marketing

The exploration of augmented reality for advertising is a very strong trend nowadays. Smartphone apps that overlay virtual ads over video captured through the camera of a smartphone abound in appstores. Many companies have implemented a variety of apps that integrate marketing material as they seek new ways to engage the interest of consumers.

Some applications are able to present the customer with a view of what is inside a product's packaging without the need for it to be opened. Lego has developed such augmented reality boxes with the idea of showing the consumer what the final Lego product (in the box they are holding) will look like when built [61]. An example of this is shown in Fig. 2.7.

Other advertising and marketing AR applications can be used to help users select products from a catalogue and even to allow them to visualize the products in a real environment in order to understand how they might look in a particular setting. The Swedish retail company IKEA has already launched applications where this is possible [120]. For their 2013 catalogue users can download an app which will allow them to peek inside the furniture to discover hidden products. Certain pages contain trigger images that, when scanned by an AR device activate the hidden content [9]. Examples of these two technologies by IKEA are shown in Fig. 2.8. Frame (a) presents a virtual chair placed in a real environment, while frame (b) shows how it is possible to see objects inside furniture presented in the IKEA catalogue.

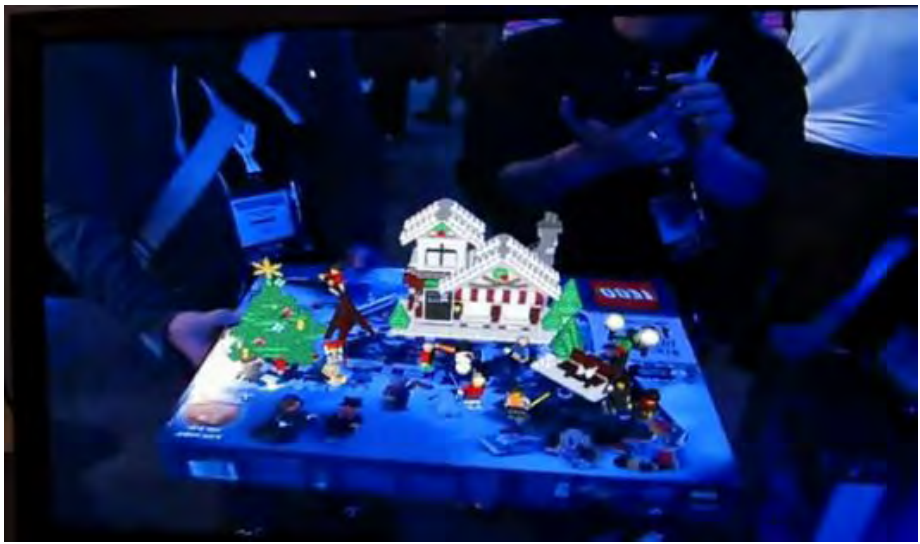

Figure 2.7: Lego augmented reality packaging [61].

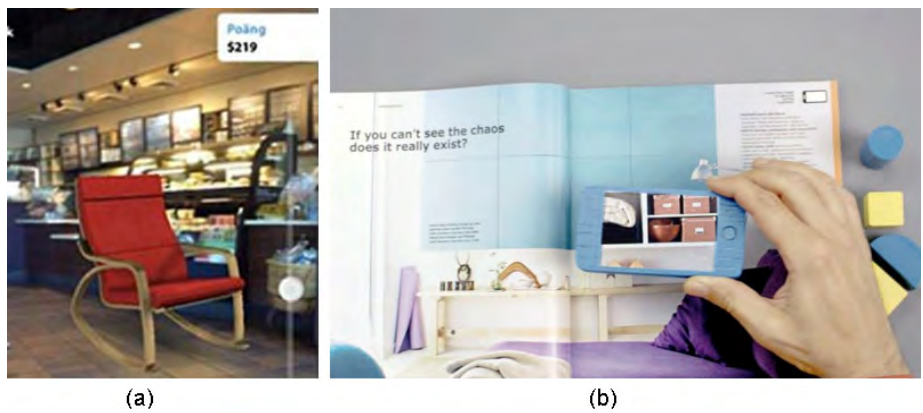

Figure 2.8: Augmented reality products by IKEA [120, 9].

### 2.3.2 Architecture and Construction

In the areas of architecture and construction, AR could become both a time and money-saving feature [123]. Assisting not only in the visualization of building projects, but also with the planning of construction jobs, as images of the buildings or even representations of utility pipes and electrical wires can be superimposed over the real life construction site. One such app, Urbasee, developed by Artefacto, functions not only as an in-the-field visualization tool, but also allows the visualization of 3D interactive models from their 2D plans. [68, 38]. In Fig. 2.9 it is possible to see a virtual building displayed next to a real building through means of a tablet pc.

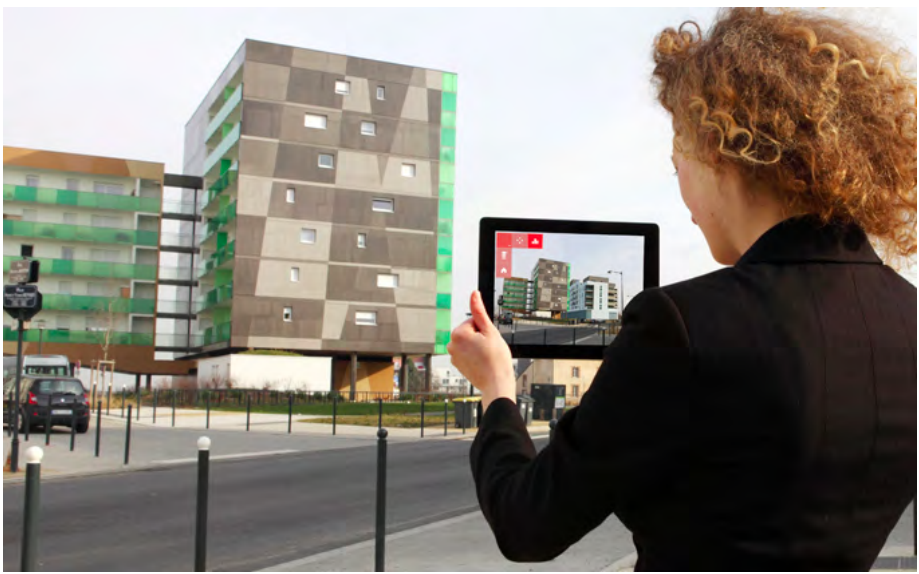

Figure 2.9: Virtual building in a real scenario seen through a tablet pc [68].

### 2.3.3 Education and Training

In [40] the author gives an insight into what future classrooms may look like and how teaching material might evolve to make learning more interactive. For example, regular textbooks may contain markers that when read by a computer or mobile phone present additional information about that specific subject. Total Immersion presents an AR book where children build a ladder while listening to voice information and pointing to the required object in the book [58]. Figure 2.10 shows this application.

AR can also be of use in skills training as it is able to provide powerful contextual learning experiences [123] in areas where real errors could be fatal. Examples of these may be the training of fire-fighters in different scenarios or airplane maintenance.

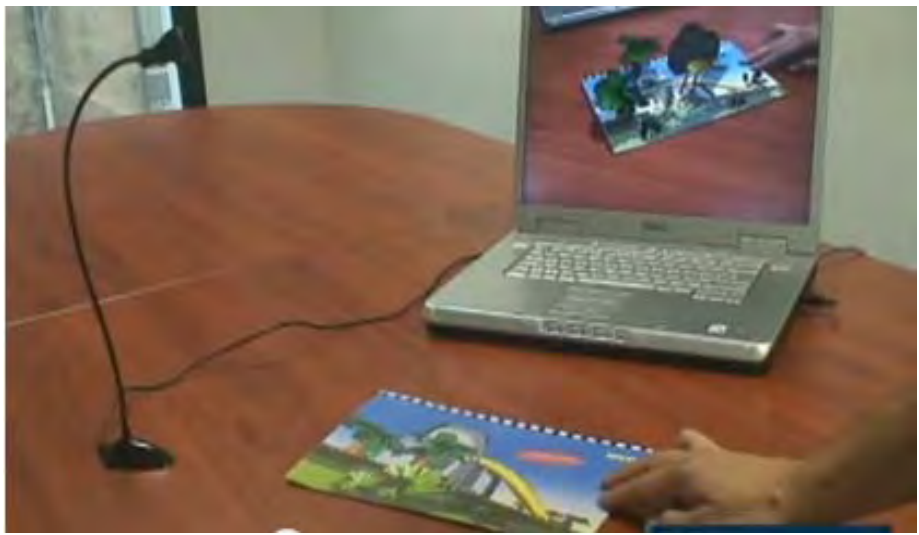

Figure 2.10: AR book presented by Total Immersion.

### 2.3.4 Entertainment

AR has become extremely common in the entertainment business. Areas such as sports and games have benefited from AR technology. In the case of televised sports, augmented reality is usually present in the form of coloured lines that indicate ball trajectory, in the case of football or snooker matches, or, in the case of swimming indicate the position of the standing world record holder. Additional information about car positions and statistics is also presented to viewers at race car events [7]. In 2009 at a Wimbledon tennis match, spectators were able to use the Wimbledon Seer app that allowed them to see augmented information about the match superimposed over the court with the aid of their Android Smartphone [58].

The gaming industry has also presented some interesting developments as AR games have been developed for outdoor and indoor environments. ARQuake is an Interactive Outdoor Augmented Reality game based on the original desktop Quake game. It was one of the first AR games in development and used a head mounted display, mobile computer, head tracker and GPS system.

With this AR game, players walk around the real world, playing against virtual monsters [71]. An indoor AR game was featured by Sony in 2010 for the PlayStation Move controller. The EyePet allowed users to interact with a virtual monkey through the TV screen. Figure 2.11 shows examples of these two games. Frame a) shows a player with the ARQuake game equipment. Frame b) shows the EyePet game for PlayStation.

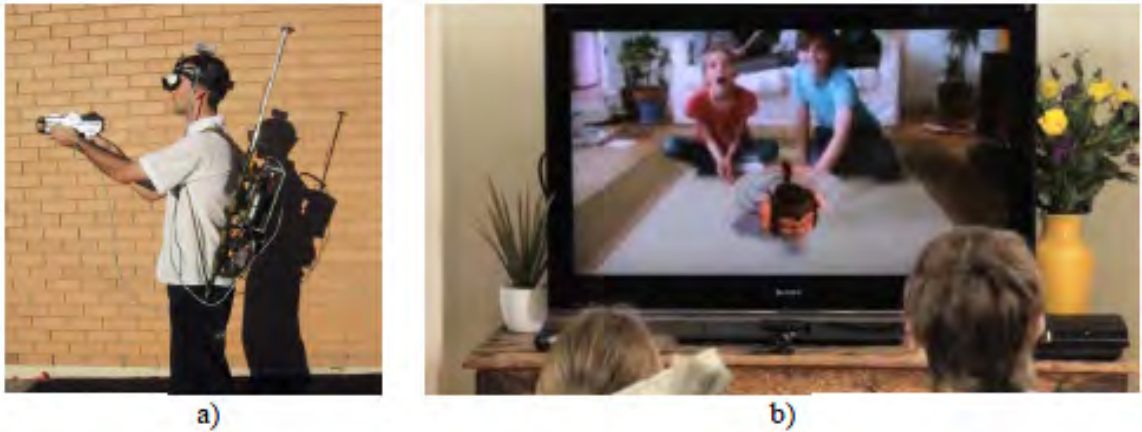

Figure 2.11: Augmented Reality games - ARQuake and EyePet.

### 2.3.5   Medical

Augmented reality for medical studies and applications has proved to be extremely effective and may even assist in the invention of new surgical procedures. [123] Surgeons could be presented with information in real time that otherwise might not be visible with the naked eye. In combining AR with traditional CT or MRI imaging, doctors will be presented with more advanced pre-operative images which could help them better prepare a surgical plan.

# Chapter 3

# Driving and Augmented Reality

In this chapter driving is identified as a multi-sensory experience. Some traditional elements found in the driving scenario, both in the vehicle and on the road, are identified in order explore the way in which they trigger the human senses. The use of augmented reality in a vehicular context is then introduced, identifying various projects that are already exploring this technology as a way to increase road safety and the overall driving experience.

## 3.1 Driving: a Multi-Sensory Experience

Human senses play an important role in daily life. People rely on their five senses: sight, hearing, touch, smell and taste to collect information about their surroundings. In the driving scenario some senses play a more predominant role than others and taste for example is not used at all. Drivers rely principally on sight to obtain information about the position and direction of their vehicle, the movement of pedestrians, or the location of traffic signs. Hearing promotes awareness of the position of other vehicles and detects warning signals produced, for example, by emergency vehicles. Touch may provide information about the mechanical state of the vehicle, such as a flat tire, or may sense changes in the road surface. Finally, the sense of smell may also be triggered during the driving task, alerting drivers to possible danger, such as burned out brakes.

In the next section some traditional elements found not only in automobiles, but also in the road environment are identified, which enhance the driver's natural senses, thereby providing better awareness of the road environment and creating a safer driving experience.

### 3.1.1 Visual Awareness: Rear-view Mirrors

Rear-view mirrors are an example of aids found in vehicles that increase the visual perception of drivers. They were first incorporated into automobiles in the early $20^{th}$ century and have since remained a permanent feature in motorised vehicles. Rear-view mirrors are fixed centrally onto the windshield and provide drivers with a view of the rear of the vehicle whilst remaining faced forward, thus minimising the time driver's eyes are off the road. They are complemented by side-view mirrors which offer a view of the lateral part of the vehicle. Together, these mirrors provide almost all-around visual perception for drivers, increasing road awareness and driving safety.

Other mirrors found in a vehicular context, though not within the vehicle itself, are convex roadside mirrors. These mirrors, found at the side of the road, provide visibility to drivers at obscured T-junctions, concealed driveways and exits and even tights bends. Figure 3.1 shows an example of such a mirror placed at a parking lot exit.

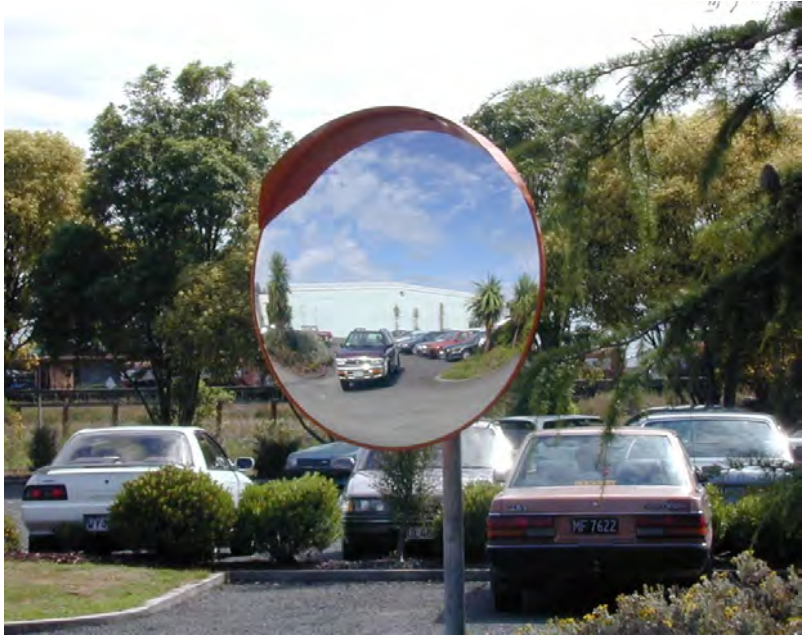

Figure 3.1: Convex mirror placed at a parking lot exit [104].

### 3.1.2 Conventional Traffic Signs

Traffic signs or road signs are an important part of the road environment as they provide visual messages not only for drivers, but for all road users. Their purpose is to insure the "safe, predictable, efficient and orderly movement of traffic" [97]. According to [27], traffic signs are used at specific locations on a road where traffic hazards may occur, where the road is subject to special restrictions and precautions, or whenever it is prudent to give any useful information to the road user.

Road signs have a long history and pre-date motor vehicles. The earliest form of road signs consisted of broken twigs, sticks or stones which marked routes along early paths. The Romans erected stone columns alongside their roads which served as milestones and showed the distance to Rome. Night time travelling presented different challenges and thus many towns lit torches or blew horns to guide travellers [74]. Modern road signs started appearing at the end of the 19$^{th}$ century when warning signs directed towards bicycle riders were put up by bicycle associations to warn them about potential dangers such as steep hills. With the development of automobiles, and in particular with its mass production due to the introduction of the moving assembly line by Henry Ford, urban traffic increased and traffic signs became more important and complex [78].

The increase of international travel by road led to attempts to standardize traffic signs, as many of these differed from country to country [60]. The system used in the United States, for example, relied rather on words [74], whereas in Europe signs with symbols were preferred [78]. A truly international system was proposed in 1968 at the Convention on Road Signs and Signals, held in Vienna. At this convention it was recognized that international uniformity of road signs, signals and symbols and of road markings is necessary in order to facilitate international road traffic and to increase road safety [47]. The convention acknowledged that both the European and the American systems were of equal value and in fact both these systems are in use today, though the European system with symbols is gradually becoming the world standard [74].

Since Portugal is a contracting party to the 1968 Vienna Convention on Road Signs and Signals [47], its national legislation is in accordance with the international rules and regulations specified within the Convention. Portugal's national legislation includes the following official documents: Código da Estrada (CE) [27] and Regulamento de Sinalização do Trânsito (RST) [98]. Other documents establishing norms and regulations for traffic signs can also be found on the website of Estradas de Portugal (EP) [42] and Instituto de Infra-Estruturas Rodoviárias IP (InIR) [65]. In Portugal all captions on traffic signs are written in Portuguese, except for those specified by international conventions [27].

### 3.1.2.1   Rules For Placement and Visibility

The placement and visibility of traffic signs are of great importance so as to ensure normal circulation and the safety of all road users. As stated in [2] traffic signs should meet five basic requirements to be effective: 1) Fulfil a need; 2) Command attention; 3) Convey a clear, simple meaning; 4) Command respect from road users; and 5) Give adequate time for proper response. To accomplish these requirements and provide adequate visibility, traffic signs should be placed within the road user's view [2]. Traffic signs should be appropriately located with regards to the location, object or situation to which they refer in order for their meaning to be easily recognized and to allow sufficient response time for road users [47].

### 3.1.2.2   Types of Traffic Signs

As specified in  [98] traffic signs in Portugal comprise vertical signs, road markings, traffic light signals, temporary signs, signals by traffic officers and signals by drivers.

#### Vertical Signs
Vertical signs are placed on the side of the road, on the right-hand side in the case of Portugal, or above it. They are positioned in relation to the location they identify and have a transversal placement, which is the spacing between the sign and the roadway or limit to the sidewalk; vertical placement, which corresponds to the height of the sign from the ground; and longitudinal placement, which is the distance between the sign and the place it identifies, varies according to the type of sign [100]. Portuguese regulation [98] stipulates that vertical signs should be placed no less than 50 centimetres, counting from the edge of the sign closest to the road and the edge of the road itself. Their height is measured from the signs' lowest edge to the ground and should be 150 centimetres when placed outside of urban areas, no less than 220 centimetres when in urban areas and 550 centimetres when placed over the road. [98] establishes five types of vertical signs used in Portugal, which vary in shape, colour, location and meaning.

- **Warning signs:** alert road users to the existence or possible occurrence of danger on the road which require special attention;

- **Regulatory signs:** express obligations, restrictions or prohibitions that road users must obey. These are sub-divided into priority signs, prohibitory signs, mandatory signs and special regulation signs;

- **Informative signs:** are intended to provide useful information to road users and are subdivided into information signs, pre-selection signs, direction signs, confirmatory signs, place identification signs, complementary signs and additional panels;

- **Variable message signs:** serve different road functions as they can display different types of information depending on need. These signs are able to transmit obligations, prohibitions or directions as well as alert drivers to dangerous road conditions;

- **Tourism and cultural signs:** inform road users about specific locations that have historical or cultural relevance.

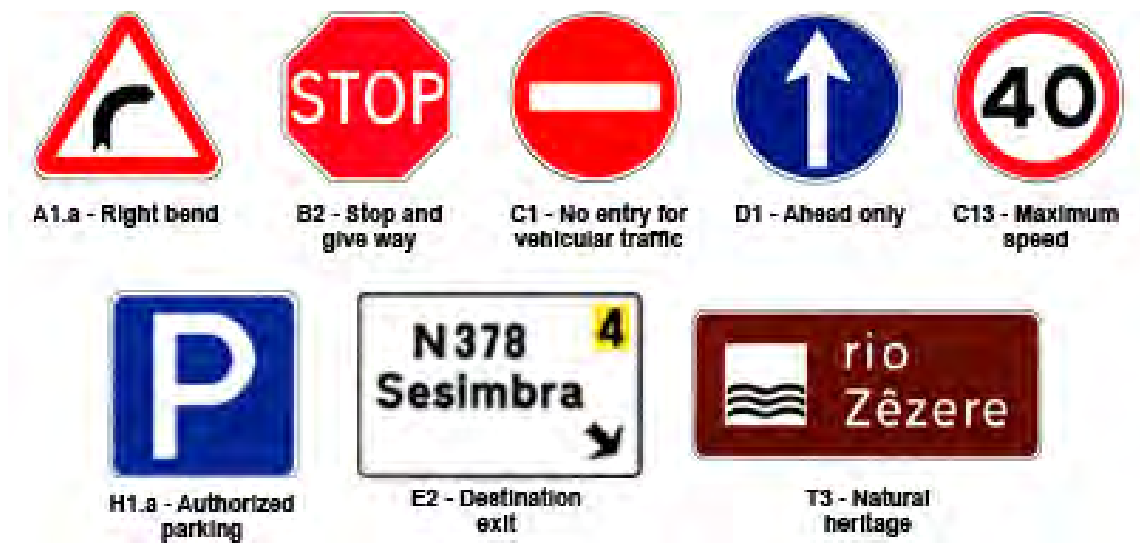

Figure 3.2: Examples of vertical traffic signs in Portugal. [98]

**Road Markings**

Road markings are used to regulate traffic flow or to warn or guide road users and may be complemented with other signs to reinforce their meaning. Road markings may be directly painted onto the road or applied in a manner that it is just as effective [47]. All road markings should be white, except for those used for temporary markings and standing regulations, which should be yellow [98].

The [98] identifies the following road markings in Portugal:

- **Longitudinal markings:** are placed throughout the length of a road and separate the two directions of traffic and also define traffic lanes;

- **Transverse markings:** are placed across the width of road lanes and may be accompanied by symbol or word markings on the road, as well as by vertical signs. They inform drivers of the location where they should stop, as well as inform pedestrians and cyclists where they may cross the road;

- **Standing and parking regulation markings:** regulate standing and parking restrictions and are marked on the roadway or on the sidewalk. They consist of continuous or broken lines either in yellow for standing regulations or white for parking regulations;

- **Arrow markings:** specify the direction drivers can and should take. Arrow markings are placed on the road lane they refer to and may be accompanied by word markings on the road, indicating destinations or speed limits;

- **Other markings and border lines:** consist of lines or symbols and are used to provide additional information to drivers or repeat information previously specified by other traffic control devices.

**Glow-in-the-dark Road Markings**

New approaches in the design of road markings are being explored in the Netherlands. Artist and designer Daan Roosegaarde along with Heijmans Infrastructure are building a "Smart Highway" [102] by leveraging on technology. Their idea is to create safer driving conditions on highways where there is a lack of street lighting to guide drivers at night, through the use of glow in the dark road markings, such as guide lines that delimit the road. Their approach consists of the creation of a photo-luminescent paint that contains glow-in-the-dark additives. These additives charge with sun light during the day and are able to glow for up to 10 hours at night.

Another approach is the creation of "Dynamic Paint" which will be used to paint snowflakes, that are invisible under normal weather conditions, on the road surface. When temperatures fall, the flakes become visible as the paint reacts to the cold weather, alerting drivers to dangerous conditions such as ice on the road. Figure 3.3 shows these two scenarios with the glowing guide lines in frame (a) and the ice marks in frame (b).

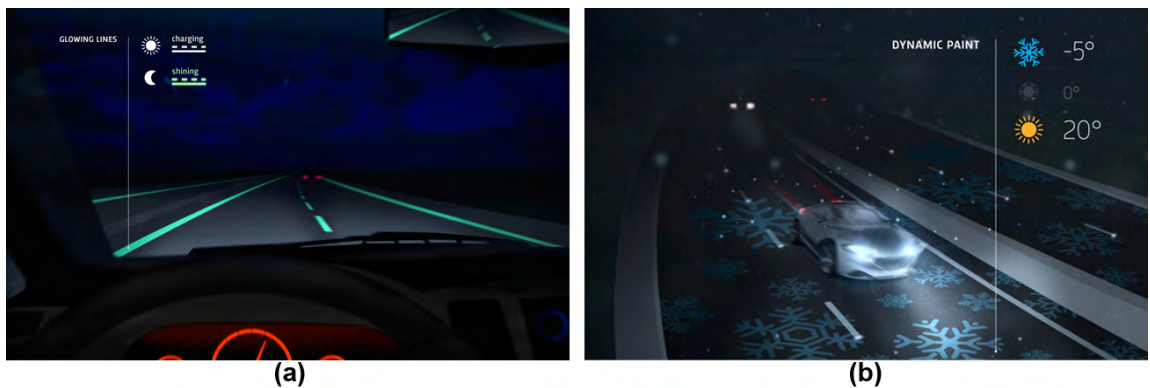

Figure 3.3: Smart Highway: Glow-in-the-dark Road Markings and Dynamic Paint [102].

**Traffic Light Signals**

Traffic light signals mainly regulate traffic flow in urban areas. Portuguese regulation [98] establishes a principle system consisting of three circular non-flashing lights, placed vertically with the following colour scheme, from top to bottom: red, amber and green. Whenever conditions do not allow a vertical positioning, the lights should be aligned horizontally from left to right: red, amber and green. Each colour has its own meaning and is as follows:

- **Red light:** indicates no transit. Drivers are forced to stop before they reach the location which the sign regulates;

- **Amber light:** signals the transition from the green to the red light. Drivers may not enter the area regulated by this sign, unless they are already very close to it when the sign appears and cannot safely stop their vehicle.

- **Green light:** indicates that passage is authorized. Drivers may enter the area regulated by the sign, except in situations of congested traffic when the driver might become immobilized in the area. A green light may not appear at the same time as an amber or red light.

This three-colour system may also be represented through arrows: black arrow over red circular background; black arrow over amber circular background; green arrow over black circular background. Signs with arrows maintain the meaning of the previously mentioned colour signs, although the prohibition or authorization only refers to the direction indicated by the arrow. Another possibility of traffic light signals containing icons are those found on bicycle lanes, for example. Instead of arrows, these lights have a bicycle icon. Figure 3.4 shows an example of the arrow variation in frame (a) and the bicycle variation in frame (b).

Similar to vertical signs, traffic lights should be carefully placed so that drivers and pedestrians may see them in time to react. In Portugal, traffic lights are located on the right side of the road and may be repeated on the left side of the road or above it. Traffic lights placed on the side of the road should be placed between 2 and 3.5 meters high, counting from the bottom edge of the sign to the ground. When placed over the roadway, they should be placed at a height of 5 meters.

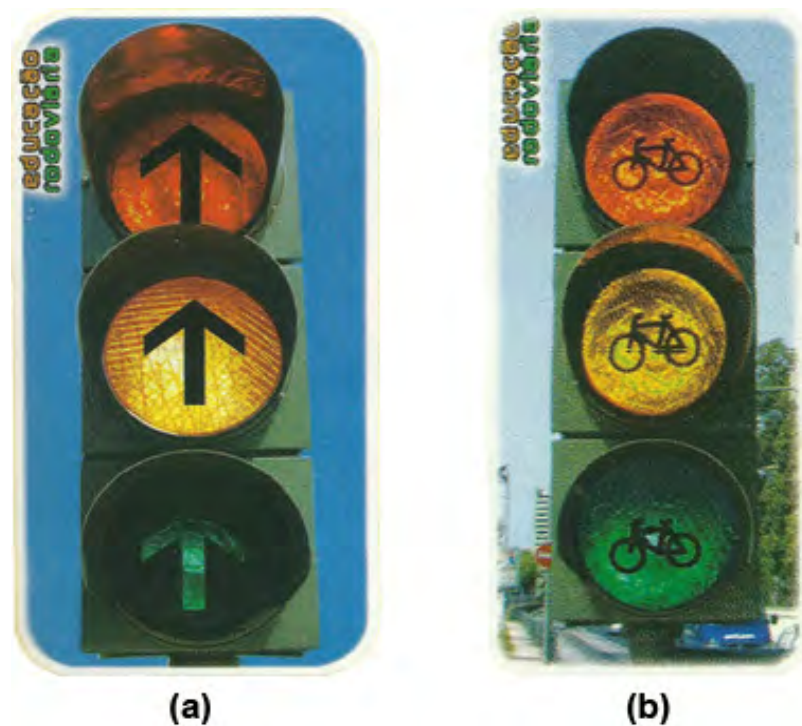

Figure 3.4: Traffic light signals with icons. [28]

Traffic lights aimed at regulating the flow of pedestrians consist of a two-colour light system: red and green. These lights are vertically aligned where the red light is located at the top and the green light at the bottom. A non-flashing red light prohibits pedestrians from crossing the road and is shown in the form of a standing man. Conversely, a non-flashing green light indicates that pedestrians may cross the roadway and is shown as a walking man. When the green light flashes pedestrians are warned that the time they have to cross the road is ending. An audible sound should accompany the green pedestrian light in order to facilitate the crossing of visually impaired pedestrians. Pedestrian control lights should be placed in a manner that does not confuse drivers into thinking that they are intended for them. These lights should be placed at a height of between 1.8 and 2.2 meters from the ground.

**Temporary Signs**

Temporary signs are used to warn drivers and other road users about the existence of occasional road works or obstacles in the road. They can appear in the form of vertical signs, road markings, light signals or additional devices. Even though temporary signs are represented by a different colour, amber, their meaning and value remains the same as that of the signals and road markings described earlier in this section. When temporary signs and markings are placed they supercede all permanent signs and markings.

Additional devices such as traffic cones or warning triangles may also be used to conveniently identify and delimit obstacles. The placement of the triangular warning sign is required whenever a vehicle is immobilized on the road, whether in the driving lane or at the road side, in order to warn drivers of other vehicles of its presence. The triangle warning sign should be placed behind the vehicle at a distance not inferior than 30 meters from the vehicle. Although its use may prevent other vehicles from colliding with the immobilized vehicle, it is only effective when seen in time.

**Signals by Drivers**

A vehicle's lighting system contains several lighting and signalling devices found at the front, sides, rear and in some cases on top of the vehicle. Besides aiding awareness of a vehicles location and allowing better visibility to drivers in low visibility scenarios such as night-time driving or bad weather conditions, a vehicle's lighting system also provides an effective communication line between others driver and road users.

Signals made by drivers used as a means of communication are very important in the driving scenario. Drivers may use signals to inform other road users of their intention to: slow down, stop, park, change lanes or driving direction, begin or conclude overtaking manoeuvres, or even alert to danger. It is important for a driver to leave no doubt as to his intentions so as to insure road safety and avoid accidents.

A driver may for example flash his headlights as an indication to another driver of his intention to allow him to enter the roadway in front of him or to alert them to a dangerous situation up ahead. Regular vehicles may also signal an emergency by flashing their high beam lights while in transit.

Turning signals on the other hand are used by drivers to indicate that they intend to change the direction of their vehicle, change driving lanes, or to signal the start or end of an overtaking manoeuvre. Turning signals are place at the front, side and rear of a vehicle, allowing other road users to see them from any location.

Lights placed at the rear of a vehicle communicate with road users behind it. Brake lights, for example, light up when the car's brakes are used and inform those behind the vehicle that it is slowing down or coming to a stop. It is important to note that signals made by drivers may not always be detected by other road users. For example, traditional brake lights may be effective for vehicles travelling immediately behind the braking car, though much less effective for cars travelling further back in the line of traffic. To change this scenario, centre high mounted stop lamps (CHMSL) were designed specifically to improve the detection of the brake signal, by positioning it centrally in the visual field of trailing drivers. The location of the signal allows it to be detached from other rear signals, as well as making it visible through other vehicles [81] as seen in Fig. 3.5. It has been shown that the CHMSL's effectiveness is not limited to night-time conditions and tests have shown a long-term effectiveness of 4.3% in the avoidance of collision [43].

Hazard lights are also effective for gaining drivers' awareness, as they blink in a pattern, and are mainly used when the vehicle may present itself as a danger to other road users. In situations where the vehicle is forced to stop due to an accident or breakdown, or when a sudden speed reduction occurs a driver should activate his vehicle's hazard lights to warn other road users of the danger.

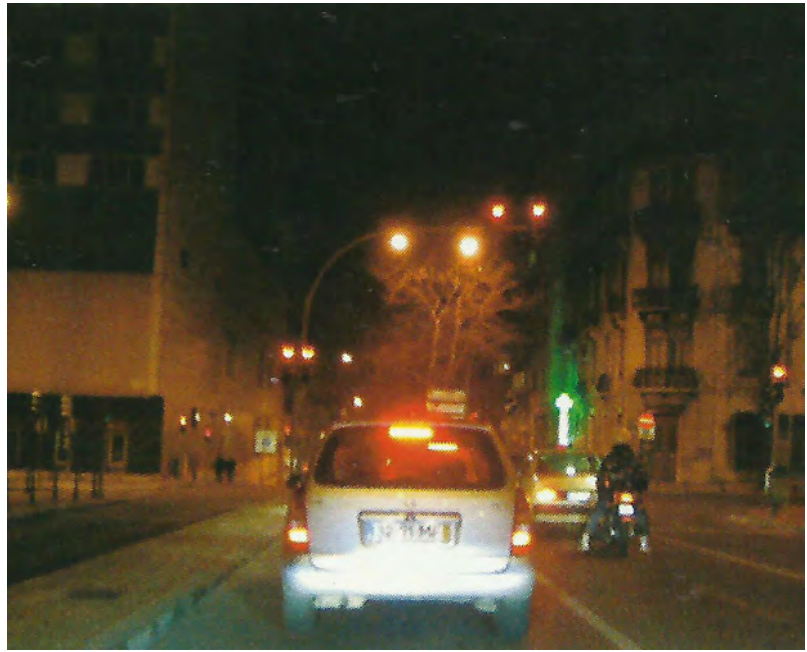

Figure 3.5: Centre high mounted stop lamp seen through another vehicle.

Emergency vehicles such as fire trucks, ambulances and police cars all contain other types of warning lights to distinguish them from regular vehicles, as well as to alert all road users of their approach. These lights consist of rotating or flashing blue light devices mounted on the top of the vehicles where they can easily be seen.

Other service vehicles such as road work vehicles or vehicles transporting unusual cargo are also equipped with warnings lights, although in this case of an amber colour. According to Portuguese regulation [27] it is prohibited for any vehicle other than emergency or service vehicles to be equipped with warning lights.

### 3.1.3  Auditory: Horns

Horns are a common feature in automobiles and are used to produce a sound capable of alerting other vehicles or pedestrians of its presence. It is particularly useful in situations where the danger is not in the road user's field-of-view, thus increasing their perception of the surrounding reality.

The use of horns pre-dates motor vehicles and was first used on horseless carriages. Early forms of these warning signals consisted of bells, whistles and bulb horns. However, they were not very effective as the sound they emitted did not surpass the noise made by vehicles themselves. In the 1920s powerful horns such as the Gabriel, an exhaust horn, and the Klaxon, with its vibrating metal diaphragm, replaced the softer devices. The Klaxon proved to be the most efficient and produced a sound which was hard to ignore. Its electrically vibrating metal diaphragm evolved into the horns used in standard automobiles today [51].

### 3.1.3.1   Sirens

Sirens date back to the late 1700s when they were first used as musical instruments. These early devices were pneumatic and their characteristic sound was produced by differences in air pressure, as air passed through small holes inside the device. Due to their ability to produce a powerful sound that could be heard from a great distance, sirens were adopted as emergency warning devices in the early $20^{th}$ century.

   In the vehicular context sirens are found in emergency service vehicles. Ambulances, police cars and fire trucks all use sirens when responding to an emergency call to alert other drivers and pedestrians to their presence. Modern day sirens used in emergency vehicles are mainly electrical and have replaced pneumatic and mechanical ones. With electrical sirens, features such as frequency, volume and even pattern can be controlled to produce better and more distinctive sounds.

   Although horns and sirens are a good way of gaining the attention of other road users, they are only truly effective when also heard from within a vehicle. Modern car manufacturers have strived at creating vehicles so isolated from outside noise that in some situations where the windows are closed and the radio is on, the sound from horns and sirens cannot be heard by drivers. This topic will be further discussed in section 3.2.2.

### 3.1.3.2   GPS Navigation Voices

In the previous section GPS navigation systems were presented as visual aids for driving. These systems provide drivers with information about their current position, as well as present them with visual driving instructions. However, visual instructions are also accompanied by auditory ones. Calm voices tell drivers when they need to turn for example. These navigation voices complement the visual information presented to the driver, and allow drivers to focus less on the image displayed on the device and more on the road, thereby reducing the amount of time the driver moves his eyes from the road.

### 3.1.4   Tactile: Rumble Strips

Rumble strips, also known as sleeper lines or audible lines, are road markings used to attract the attention of inattentive drivers. When a vehicle passes over these strips the frictions between the tire and the strip's specially designed surface will make the vehicle vibrate to a degree that it is felt by the driver, as well as producing a sound loud enough to be heard inside the vehicle. When applied longitudinally throughout the length of a road near its edges, these markings are directed towards drivers who, due to distraction or fatigue, depart from their lane and may possibly head off the road, or steer into oncoming traffic. The vibration of the car will thus alert drivers, providing them with an opportunity to correct the direction of the vehicle in time to avoid an accident. Rumble strips can also be applied transversely over the width of the road. In this configuration they intend to warn drivers of changes in the road that require their attention, such as an upcoming intersection, toll gate or pedestrian crossing [116].

## 3.2   Augmented Driving

Although the traditional elements mentioned in the previous section all contribute in controlling traffic flow and broadening a driver's overall awareness of the surrounding road scenario, and in many cases help avoid road accidents, they do not eradicate them. According to [88] through the "Global Burden of Disease" project of 2004, road accidents caused over 1.27 million deaths that year, and became the ninth leading cause of death worldwide. In 2013 [89] places road accidents as the eighth cause of death worldwide and identifies it as being the leading cause of death in young people aged 15 to 29. These numbers are alarming and of particular concern as estimations predict that if no action is taken road deaths will increase to 5th place as leading cause of death worldwide by 2030.

These figures demonstrate why the exploration of alternatives to traditional vehicular safety and traffic-flow methods have been widely explored in recent years. Researchers and car manufacturers have turned to technology to find answers and in the past few years many Intelligent Transportation Systems (ITS) have emerged. In the next section an overview of some of these up and coming systems is provided.

### 3.2.1   Visual: In-Vehicle Representation of Traffic Signs

Modern cars are already converging on the concept of a virtualized windshield. A basic approach is found in the replication of roadside traffic signs into in-vehicle virtual traffic signs, either projected on the windshield or displayed on LCD screens fitted on the dashboard. The fact that we are used to seeing traffic signs as roadside infrastructures is a matter of historical consequence. Today, new traffic technologies, known collectively as Intelligent Transportation Systems (ITS),through displays inside the vehicle, are capable of duplicating some or all of the information displayed through static traffic signs or Variable Message Signs (VMS) outside the vehicle [97]. By using such in-vehicle displays, the opportunity to capture the message is no longer limited to a single visual glance, opening opportunities for more complex traffic signs.

The earliest example of in-vehicle road signs only appeared in the 1990s with the introduction of GPS-based navigation systems. Digital road maps powered these navigation devices and included information about the speed limit of each road, which was displayed as a digital, in-vehicle traffic sign on the screen of the navigation device. The in-vehicle representation and awareness of the speed sign also allows the driver to check the current speed of the vehicle against the enforced speed limit, warning the driver about a violation. More sophisticated in-vehicle representations of roadside traffic signs resort to vehicular sensors other than GPS and the associated digital cartography. For example, in-vehicle radar-based systems are able to determine the distance to the preceding vehicle and warn the driver if the 2-second distance rule is being violated. Figure 3.6 shows the roadside-based sign for that rule and its in-vehicle version. The ubiquity and speed-awareness of the in-vehicle approach has obvious advantages compared to the traditional representation.

Windshield camera-based systems are another novel trend for the in-vehicle display of traffic signs. Such systems replace the visual sense of the driver by computer vision techniques that are able to recognize roadside traffic signs and automatically duplicate them on in-vehicle displays. Compared to map-based systems, the computer vision approach is able to detect transitory changes on the posted speed due to temporary road works, for instance. Computer vision in the context of vehicles has been a very active topic in ITS for the last two decades [16].

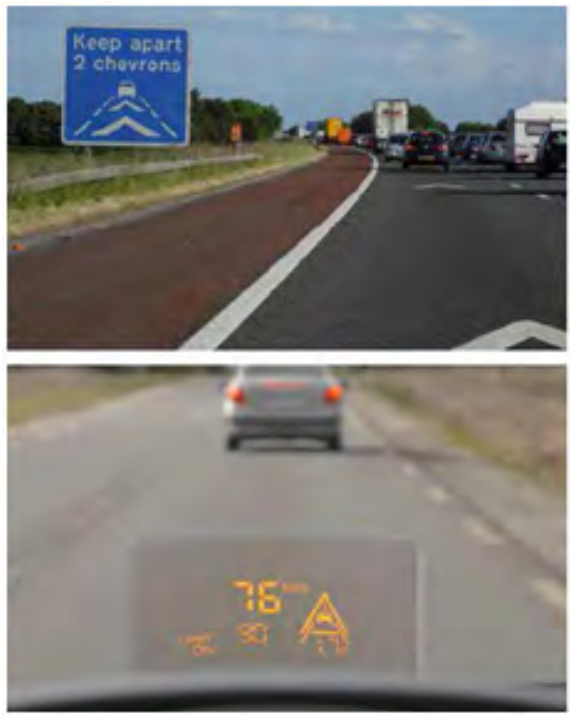

Figure 3.6: The 2-second distance rule through a roadside sign (top) and in-vehicle sign (bottom).

An example can be seen through the Traffic Sign Recognition technology developed by Siemens [56], which detects and reads speed limit signs and projects that information onto the car's windshield, creating an augmented environment. Figure 3.7 shows the detection of the road-based sign and its display over the windshield in frame (a). Car manufacturers such as Peugeot have also incorporated augmented reality technology into some of their vehicles. The Peugeot 3008 for example has a Head-Up Display located on the dashboard in front of the driver, seen in Fig. 3.7 in frame (b), where information related to the vehicle's speed is presented. The 3008 also features a radar-based distance alert system, informing the driver of the distance between his vehicle and the vehicle in front, also displaying this information on the HUD [94].

Most of the in-vehicle displaying systems for traffic signs described previously are merely duplicating traffic information found on existing road signs. A promising alternative source of traffic information will come from cooperative-based ITS technologies, namely V2V communication through the Dedicated Short Range Communications (DSRC) standard. [67] Vehicles and their sensing abilities will generate just-in-time and just-in-place traffic signs and transmit them through the wireless channel to the in-vehicle displays of their neighbours. This distributed creation of traffic signs and its broadcast to neighbour vehicles can dramatically increase the infor-mation that is conveyed to drivers. Excess information can result in a major problem in the driving environment and thus much research related to the in-vehicle display of traffic signs is concerned exactly with the human factors of the technology [24, 75]. The topic of cooperative-based ITS technologies will be explored in more detail in Chapter 4.

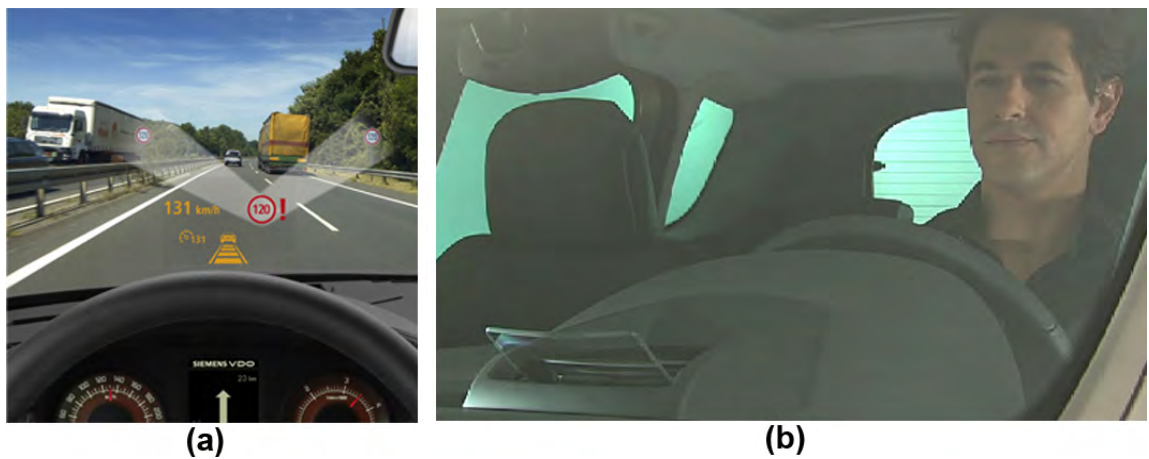

Figure 3.7: Siemens VDO Traffic Sign Recognition [106] and Peugeot 3008 Head-Up Display [94].

### 3.2.1.1 Navigation

Another area where the virtualization of the windshield is observed is that of navigation information. In Fig. 3.8 three examples of the evolution of GPS navigators are displayed. The traditional portable navigation devices (PND) shown in frame (a), were fixed onto the windshield and presented two-dimensional maps containing indications with regards to the route to follow. Since then the display of navigation information has evolved to become more embedded in the windshield, as shown in frame (c). This figure shows the navigational output of the BMW 5 Series, superimposing the information directly onto the windshield. An intermediate step is shown in frame (b), displaying the innovative Blaupunkt Travel Pilot, which merges a video stream of the actual road captured by a forward facing camera on the device with pictographic content created digitally, conveying navigational instruction in an augmented reality way.

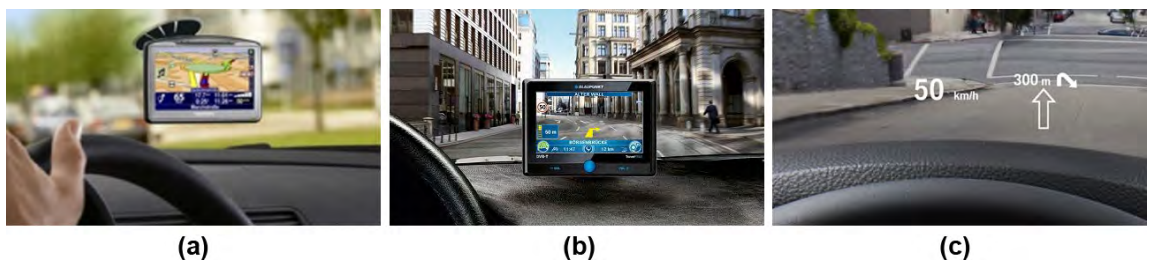

Figure 3.8: Evolution of navigation devices.

Novel proposals for the display of navigation information over the windshield through advanced projection techniques, such as laser holographic projection [23], have also appeared recently. One such proposal is the Virtual Cable [77], an application developed for the driving scenario that uses augmented reality to present navigation instructions, illustrated through a hanging virtual cable that is projected onto the windshield. The information is placed in the driver's line of sight in such a way that it gives the illusion that the digital content is part of the surrounding landscape. Therefore the driver keeps his eyes on the road, minimising distraction. A snapshot of this system in a field test is shown in Fig. 3.9 and Fig. 3.10 shows a comparison between the virtual cable and more traditional navigation devices.

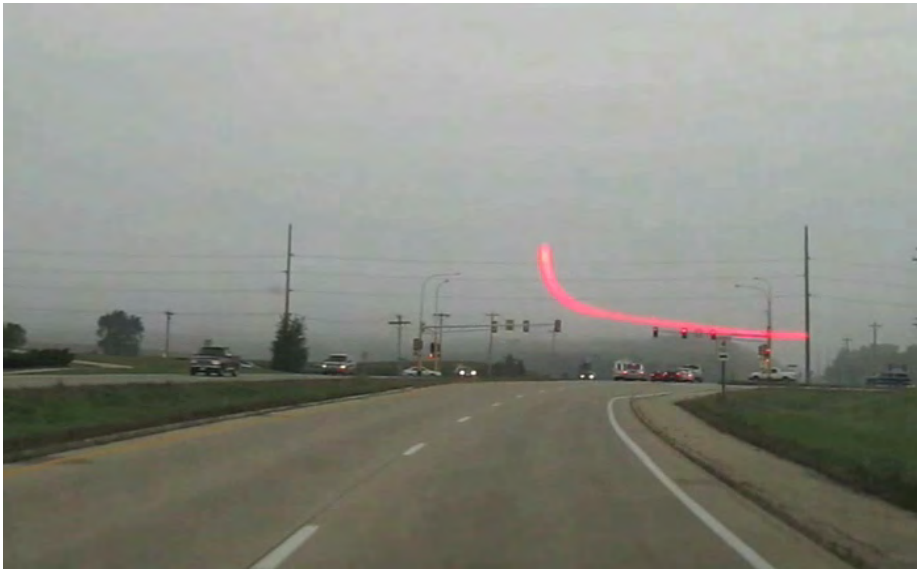

Figure 3.9: Snapshot of virtual cable field test [77].

Ways of presenting 3D virtual navigation objects, such as names of locations or address information, in a natural way in the same context as the driving scenario by proving natural-like shading and shadowing have been explored in [108]. They have created a system capable of generating a simplified model of the real road scene, by collecting information about its geometry, lighting and other attributes. Their Road Trench Model (RTM) as it is known, then uses these attributes to accurately apply shading and shadowing to the virtual content according to the vehicle's position and provides the reference coordinates to present the virtual information. Thus a sign comprising of 3D wording, indicating the distance to a destination, can be overlaid onto the windshield providing a sense that it is actually placed on the road surface and present shadows and shadings equal to those of the elements contained in the real scenario such as trees or physical road signs. An example of this can be seen in Fig. 3.11.

Going much further than simply displaying navigation information over the windshield, new forms of vehicle intelligence are being explored to provide drivers with innovative driving experiences by changing the way they interact with their vehicle. The AIDA project (Affective, Intelligent Driving Agent) has developed an interactive system capable of filtering information about the city in real-time, providing destination suggestions based on the driver's mobility patterns. The project is developed by MIT SENSEable City Lab, in collaboration with the Personal Robots Group - Media Lab and the Volkswagen Group of America, Electronics Research Lab. It features two prototypes: the first, AIDA [72], seen in frame (a) of Fig. 3.12 was a friendly little robot embedded in the vehicle's dashboard, developed in order to transform traditional navigators into travel companions capable of learning the drivers habits, understanding the vehicle and the city and providing useful information to the driver. The robot was replaced by AIDA 2.0 [73], an interactive 3D map displayed over the whole of the dashboard, seen in frame (b) of fig 3.12, presenting not only routing information to the driver, but also keeping him updated on traffic conditions, events in the city or even Twitter status updates. Similar to AIDA, AIDA 2.0 is also able to filter information according to the driver's preferences and mobility patterns. The interactive 3D interface is controlled by hand gesture recognition, which facilitates its operation while driving. The AIDA 2.0 is opening the way for the dashboards of the future, bringing digital maps closer to where you see the city through the windshield.

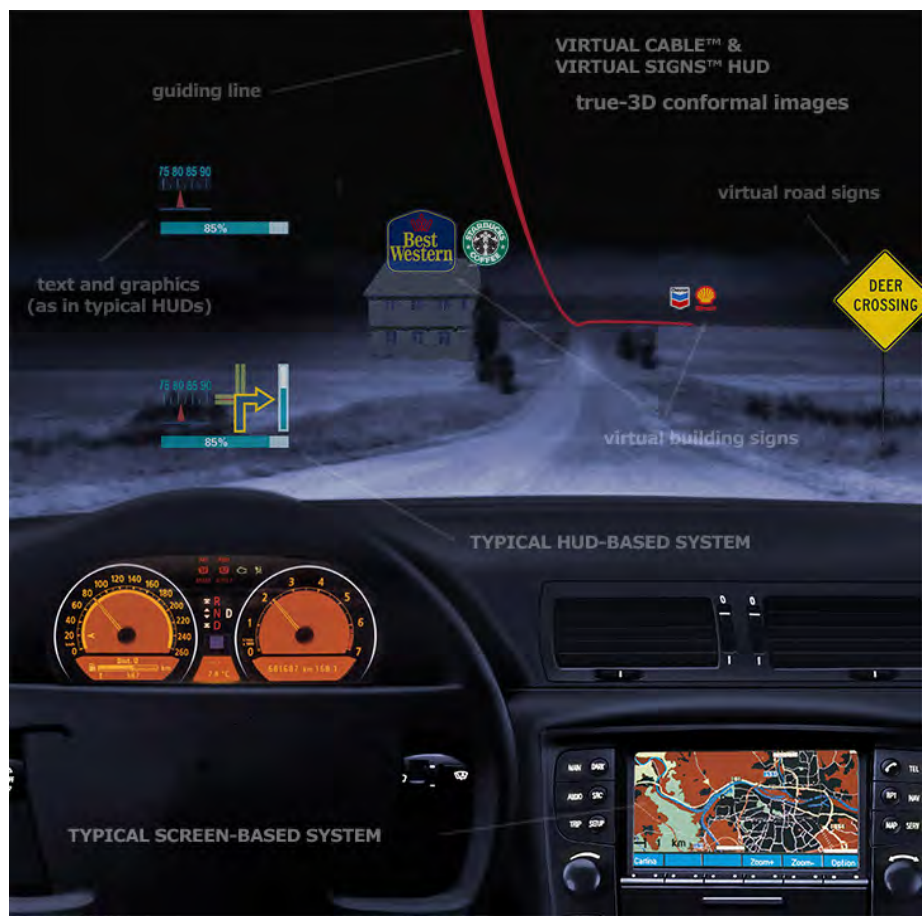

Figure 3.10: A comparison of the Virtual Cable with existing car navigation systems [77].

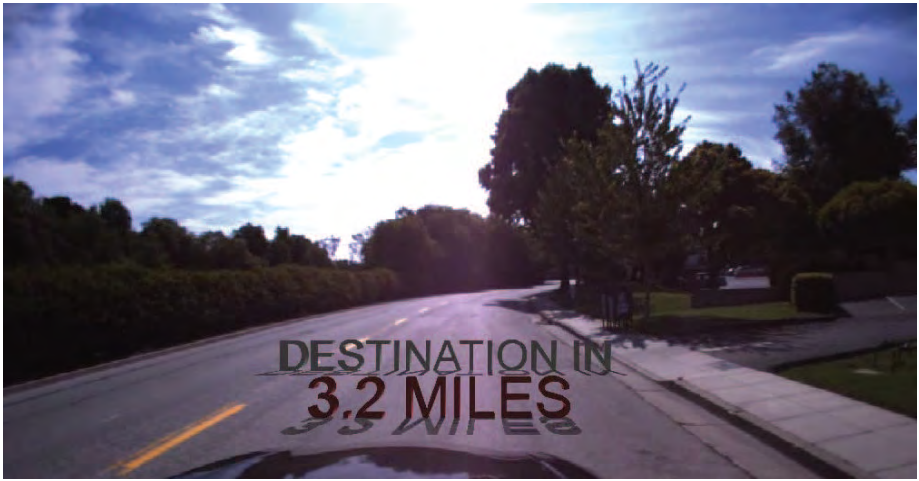

Figure 3.11: Correct shadowing of distance information [108].

#### 3.2.1.2   Visual Advanced Driver Assistance Systems

Advanced Driver Assistance Systems (ADAS) have been developed to aid drivers in the driving process. These systems use sensors or cameras found in and outside the vehicle that collect information about the road environment and the vehicle itself and present an augmented awareness to drivers. ADAS are currently found in a wide variety of vehicles and assist drivers in many ways. Traffic sign recognition and in-vehicle navigation systems have already been mentioned previously, and are only some of the possible uses for ADAS.

#### Parking assistance systems

Parking assistance systems aid drivers while parking their vehicles. Lexus [92] for example has introduced cameras into some of their models to provide drivers with a surround view of their vehicle through monitors, thus eliminating blind spots as they park. Real-time images are obtained through cameras installed above the licence-plate and beneath the side-mirrors, showing the areas directly behind the vehicle as seen in Fig. 3.13 and to the side of the vehicle. The images are then displayed either on the multi-display screen located on the centre console of the vehicle or on a small monitor incorporated in the rear view mirror. Beyond merely providing a video image of the space behind the vehicle, this system also presents superimposed guidelines that show the predicted path of the vehicle.

#### Enhanced Vision System

General Motors has presented a concept built with safety in mind which also uses several sensors and cameras to provide information about the vehicle and its environment to the driver. The Enhanced Vision System [39] as it is known, uses head up display systems to overlay this content over the windshield. This system can successfully detect objects that are in front of the car and can even create marks over the windshield that represent lane lines in foggy driving conditions as shown in Fig. 3.14.

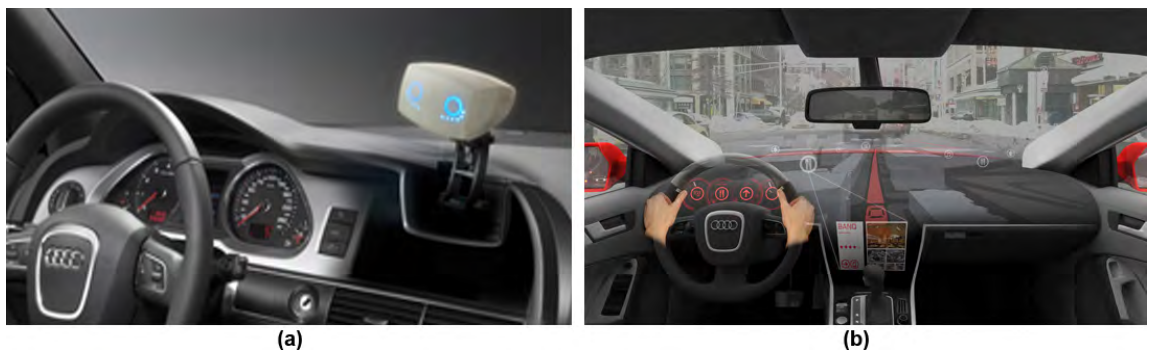

Figure 3.12: AIDA - Affective, Intelligent Driving Agent [72, 73].

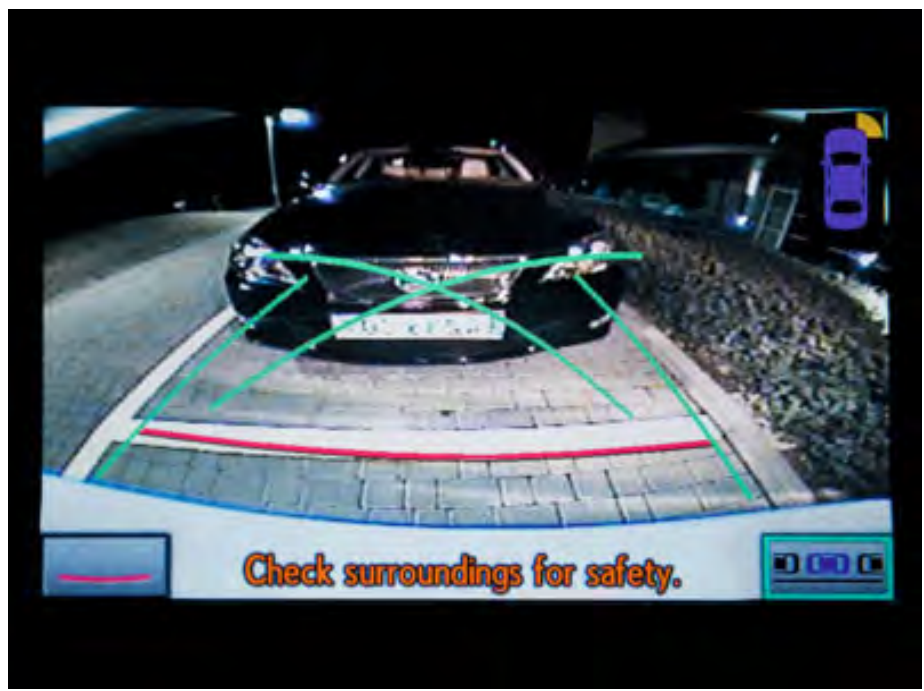

Figure 3.13: Lexus RX 450h Rear Parking Assist Monitor [92].

**Night Vision Driving**

Night vision technology is also currently appearing in several high end cars. The Audi A7, launched in 2012, features a Night Vision Assistant system [15]. The system uses infrared cameras to capture the road environment and the night vision view appears on a screen located behind the steering wheel, as seen in Fig. 3.15 in frame (a). The system presents a clear image with a great amount of detail and is completed with people and animal detection as seen in frame (b).

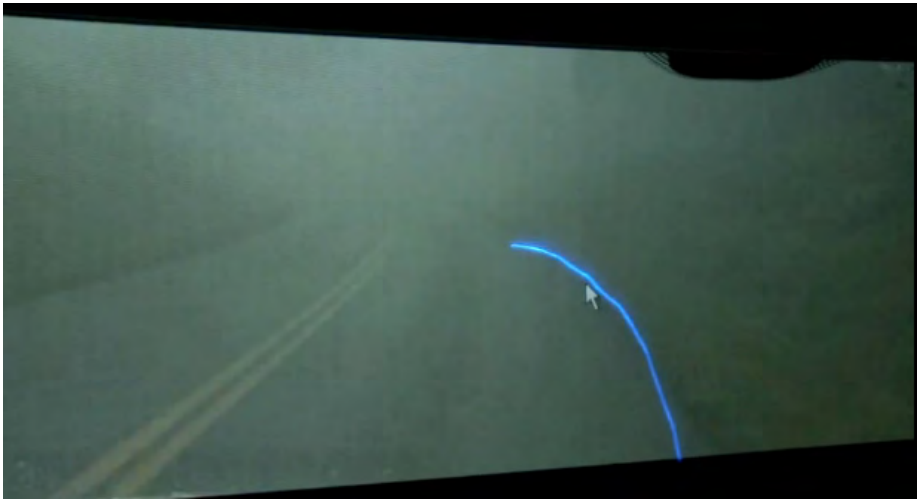

Figure 3.14: General Motors' augmented reality whindshiel [39].

### iOnRoad

Smartphone apps that offer several driver assistance functions resorting to augmented reality have also been developed and offer a more affordable solution to drivers without high-end cars. iOn-Road [66] is one such application, and offers functions such as forward collision warning, lane departure warning and speeding alert. The app runs in real time and for the forward collision warning it uses the smartphone's camera and sensors to detect objects and vehicles located in front of it. The smartphone's GPS calculates the vehicle's speed and the app then measures the distance between the vehicle and the one directly in front of it. This information is displayed on the smartphone's screen, showing not only the distance to the other vehicle, but also highlighting the lane the driver is in. If the app considers the distance to be safe, then the lane is presented with a green colour as shown in Fig. 3.16. However when the distance is considered too small and unsafe the Smartphone emits an audio-visual alert. A piercing sound is emitted and the lane colour on the screen chances to red.

### Virtual Transparency

The concept of virtual transparency has also been explored recently. In a vehicular context, possessing the ability to see through occluded surfaces could greatly reduce accidents, as this technology would offer clear visibility at intersections with poor or no visibility. One approach has been explored in [12]. The basic idea consists of collecting information about an occluded area through a perspective in which it is visible and then transferring that information to the perspective in which it is occluded. This is achieved through the use of video captured by several cameras placed in strategic positions. One camera views the occluded area, while another views the surface that is occluding the area. The combination of the information captured through both cameras is sufficient to provide a sense of transparency as the video captured by the camera viewing the occluded area is overlaid over the occluding surface. Similarly, [122] present a wall see-through system directed at drivers to help eliminate blind corners at intersections, and potential collisions with vehicles which are occluded by walls, for example. The main focus of these authors is to determine the least amount of information sufficient to present to a driver for him to predict a crossing collision. To determine this, several levels of visualization are compared, each one presenting more information than the previous one, as seen in Fig. 3.17. The level with the least information (excluding the level that presents no extra information to the driver) merely presents the direction of the occluded vehicle as a white circle, while the level that presents the most information shows everything behind the occluding wall. Their results show that it is possible for a driver to estimate a collision when only the direction of the occluded vehicle is displayed as a small white circle. This type of information would surely be of great value to drivers, although the authors make no mention of how it would be presented in a real driving scenario, as their experiments were carried out with the aid of LCD monitors.

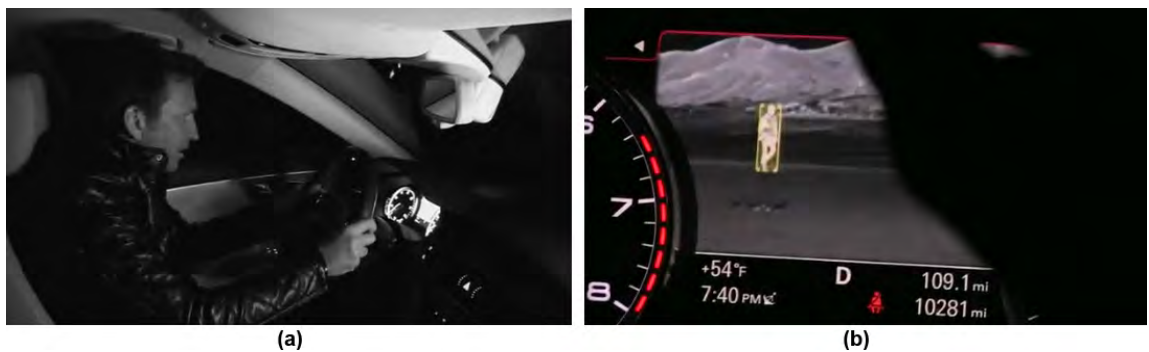

Figure 3.15: Night Vision Technology – Audi A7 [15].

**Visual Assistance for Bicycles**

Devices aimed at bettering road safety should not be aimed solely at car drivers but to all road users. [89] states that 50% of all deaths on the road comprise of pedestrians, cyclists and motorcyclists. Thus, it is equally important to provide tools that increase awareness of the road environment for these individuals, as well as help car drivers better locate them on the road, especially at night time. Some interesting projects aimed at cyclists have recently surfaced.

Lumigrids [96] is a project developed at the Sichuan University, Department of Industrial Design and is essentially a LED projector placed on bicycles, which projects a grid composed of equal sized squares onto the ground. This grid provides cyclist with a better awareness of any irregularities on the surface. When the ground is flat the squares that make up the grid will be perfectly aligned. However, when the ground surface is uneven the grid will deform, helping cyclists identify bumps and holes in the road. Figure 3.18 shows the Lumigrids projector in frame (a), the projection of the grid onto a flat surface in frame (b) and the deformed grid over an irregular surface in frame (c). In addition to being useful for cyclists, in identifying the type of surface being ridden on, the Lumigrids also provides awareness for pedestrians and other vehicles of the bicycles location.

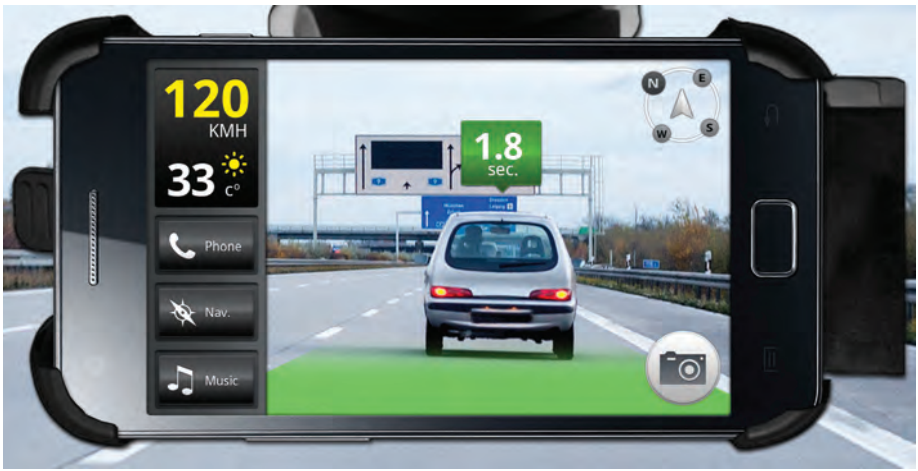

Figure 3.16: iOnRoad – Display of safe distance from proceeding vehicle [66].

In the same scope, Czech artists Vladimir Turner and Ondřej Mladý have presented a performance art piece entitled Safety First [17] to bring awareness to the limited amount of bicycle lanes in the city of Prague. The artists mounted a handheld projector onto a bicycle that projects a looping clip of a bicycle lane in front of the rider, thus guiding him as well as providing additional lighting for better visibility and awareness of the bicyle's location to nearby vehicles and pedestrians. A similar concept has also been presented in [50]. Lanelight projects a bicycle lane from the back of the bicycle, informing drivers coming from behind as to the location of the bicycle lane and thus the cyclist. Figure 3.19 presents a snapshot of Safety First in frame (a) [17] and Lanelight in frame (b) [50]. These examples show the positive exploration of multimedia content in a vehicular scenario, presenting alternatives to traditional static signs, by creating dynamic signs where they are really needed.

### 3.2.2   Auditory

As seen in the previous section, in-vehicle visual augmented reality systems can greatly improve the driving experience and road safety. However, driving is not a task that solely relies on sight, nor do pedestrians exclusively rely on this sense when walking in a road environment, for example. All road users are confronted with sounds and noises ranging from the roar produced by the engine of an oncoming vehicle, to the siren of an emergency vehicle in transit or simply by the honk of a horn.

In the driving scenario, auditory cues are able to provide further information to a driver without overloading his cognitive function. Moreover, it has been shown that dividing tasks through different sensory modalities has a better outcome than when two tasks are processed through the same sense [119]. For example, a driver may be faced with two visual tasks such as driving a vehicle and reading a map to find a destination. The reading of the map will surely interfere with the visual attention needed to safely drive the vehicle, whereas if the information provided by the map were to be presented to the driver via audible means such as the audible sound directions provided by portable navigation systems, he would not need to divide his visual attention.

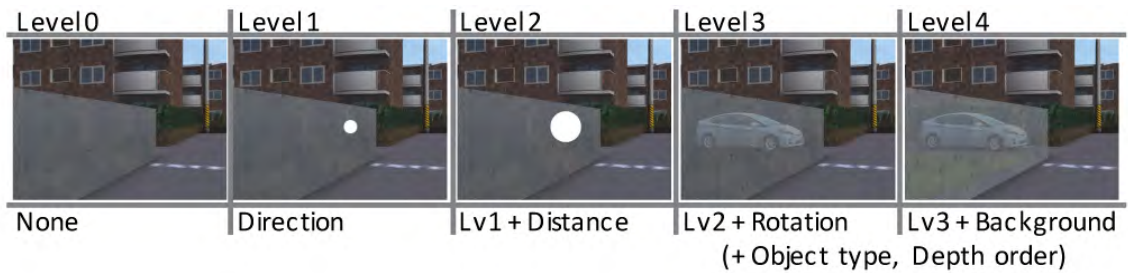

Figure 3.17: Comparison of visualization levels [122].

### 3.2.3 Parking Audio Cues

As an alternative to the visual parking assist systems presented previously, audio parking cues are a simpler warning system which can be fitted into vehicles. Parking audio cues use ultrasonic technology to determine the distance of a vehicle to an object. The system bounces high-frequency sound waves off objects, which echo back and are captured by sensors. The information is then analysed and the distance to the object is calculated. Warning sounds are emitted inside the vehicle and these change pitch or frequency depending on the distance of the car to the object.

### 3.2.4 FM-Radio with RDS-TMC Receiver

Traffic Message Channel (TMC) is a technology that provides drivers with information related to traffic and road conditions through a digital radio channel. TMC information can be integrated with GPS navigation devices to receive dynamic and contextualized information of the route to the destination. This information can be heard over the vehicle's standard FM-Radio.

### 3.2.5 Spatial Auditory Displays

Auditory augmented reality explored in the vehicular context may significantly improve road safety by reducing cognitive load and reducing driver reaction time, as audible stimuli is perceived much faster than visual stimuli [115]. In-vehicle spatial auditory displays are able to provide drivers with an increased awareness of where a sound is coming from, thus allowing for a more accurate response. Contrary to traditional vehicle sound systems which mainly present vehicular sound from one location, whether navigation information or a warning signal, spatial auditory displays (SAD) present the driver with the notion that the sound is coming from different locations within the vehicle or right outside it. For example, if an audible warning sign indicating engine problems sounds like it is coming from the engine rather than from the steering wheel, a drive is able to associate the sound to the engine much faster. Also, in a situation where an emergency vehicle may be approaching it is sometimes difficult to identify from which direction it is coming from. The directionality of the sound produced by the emergency vehicle may be wrongly interpreted when for example its origin is located to the left of the driver, but the driver perceives it as coming from the right, as his right-hand window is down. In this scenario, a SAD is able to display the sound of the emergency vehicle in the same direction as its origin. Similar to visual augmented reality, the type of display used in auditory AR can influence the amount of immersion a driver may feel inside a vehicle. For example a surround sound system may offer a greater immersion sense than a single loudspeaker. Considering that most current vehicles are equipped with an array of speakers, vehicles become ideal containers for the exploration of auditory augmented reality in a vehicular scenario [93].

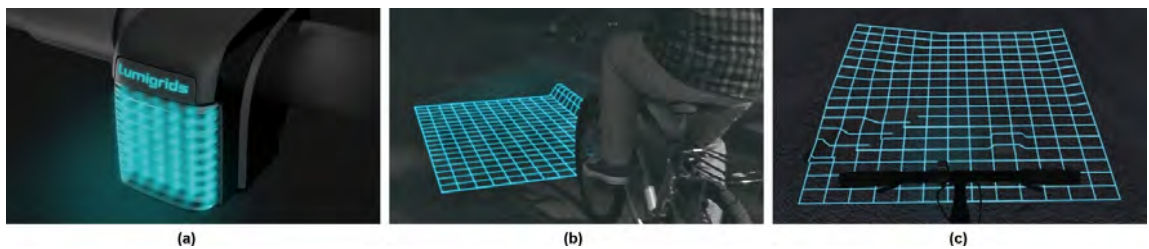

Figure 3.18: Lumigrids - LED projection system for bicycles [96].

### 3.2.6 Silent Vehicles

Augmented reality may also present a solution for increasing road safety for pedestrians and cyclists as current electrical and hybrid vehicles have become so silent that blind and even fully sighted pedestrians, as well as cyclists do not hear them approaching, thus placing them at risk of being hit by these vehicles [103]. The Pedestrian Safety Enhancement Act of 2010 [117] has instructed that a motor vehicle safety standard be created to ensure the minimal noise a vehicle must make to increase the safety of blind and other pedestrians, by allowing them to hear the approach of the vehicle. Audi has been exploring a solution to this problem through a synthetic approach: Audi e-sound [79]. The e-sound consists of the sound of a combustion engine found in a series-production Audi that is played through a loudspeaker located under on the under-side of vehicle. The Audi e-sound thus produces a loud enough engine rumbling sound that alerts pedestrians and other road users of its approach. Following a more promotional and entertaining approach, but none the less a safety aware approach, Domino's Pizza in Amsterdam has equipped its electrical delivery scooters with a recording of a fake engine sound to alert other motorists and pedestrians of its whereabouts [49]. The noise of the fake engine sounds as if it were made by a person and the words "Domino's" and "Pizza" can also be heard.

Modern cars have also become incredibly isolated to outside noise. The 2012 BMW M5 has such good isolation capabilities that the sound of the engine is hardly heard from the inside. BMW is exploring augmented reality to provide a stronger and clearer virtual sound of the roar produced by the twin-turbo V8 engine. The Active Sound Design [63] consists of a pre-recorded

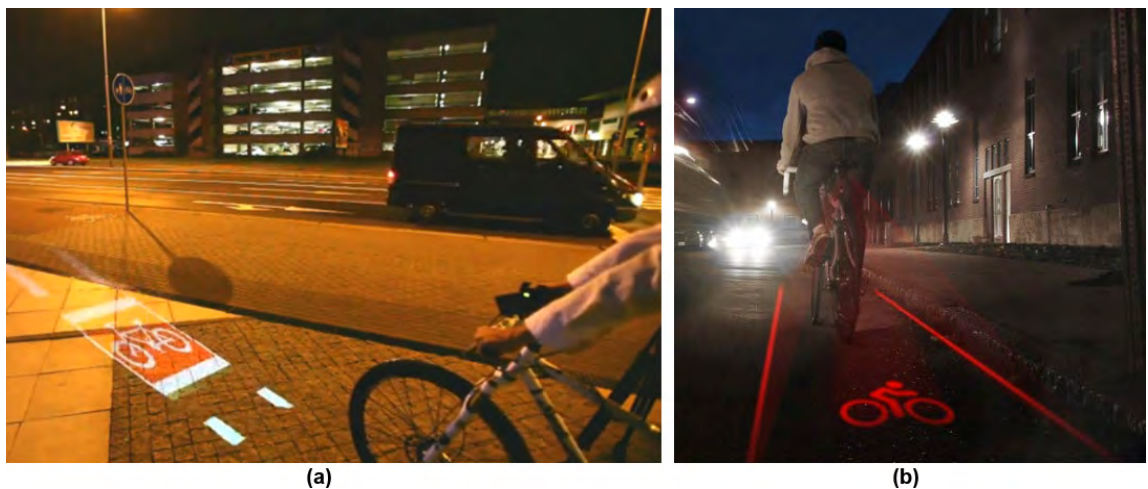

**(a)** **(b)**

Figure 3.19: Virtual lanes for bicycles.

sound, based on the current rotations-per-minute of the engine, that is heard through the vehicle's surround sound system.

### 3.2.7  Tactile

Although not nearly as widely explored as visual AR systems or even auditory AR systems, tactile cues for in-vehicle information systems may present a promising alternative to in-vehicle auditory warning signals. [25] have compared auditory and tactile cues for in-vehicle information systems to determine which better attracts the driver's attention. They concluded that there is no clear winner and that each one is best for specific tasks. On the one hand tactile cues through vibration proved best as they interfered less with the visual aspect of driving and were more accurately identified than the sound-based cues. On the other, sound cues were identified faster and different sound types were easier to distinguish than different vibrations. Thus the authors recommend that tactile cues such as vibration should be used in scenarios where the radio is on, for example, or when loud noise is found in or around the vehicle. Vibration is also best when the message being transmitted is not very urgent. Contrarily, sound should be used to identify urgent messages or whenever the vibration is not felt due to an irregular driving surface for example.

Some forms of tactile feedback systems have been explored in the automotive industry, mainly as lane departure warning systems. BMW presents drivers with a feature that is able to detect when the vehicle strays from the driving lane [21]. An integrated camera reads the road lanes, constantly tracking their position. When the vehicle is close to passing over the lane, a vibration is produced through the steering wheel alerting drivers to correct the vehicle's trajectory.

Audi offers another system in which tactile feedback is provided to the driver. Audi's dynamic steering adapts its steering ratio according to the vehicle's speed. For example, in a scenario where the vehicle is on a highway driving a great speed, the system helps keep the car under control with the electronic stability program (ESP) [6].

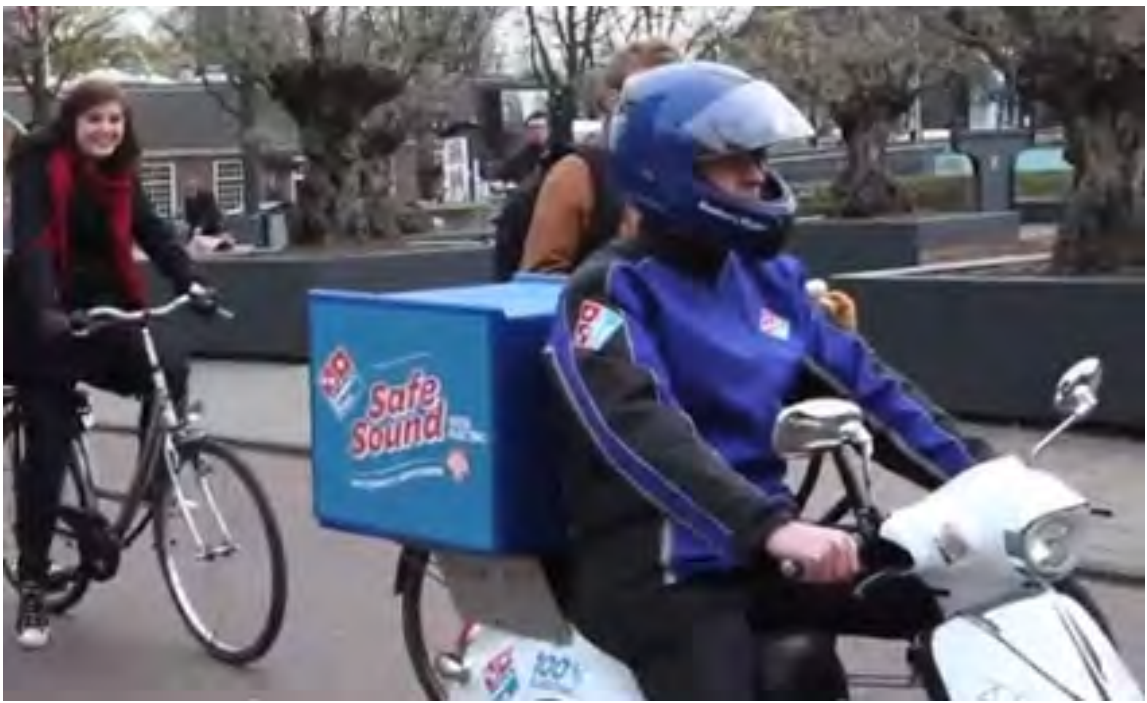

Figure 3.20: Domino's Pizza Safe Sound [49].

# Chapter 4

# Connected Driving

This chapter presents connectivity paradigms within vehicles, ranging from cellular based connectivity to the novel vehicle-to-vehicle and vehicle-to-infrastructure communication that is enabled by DSRC through the 802.11p norm. An overview of Vehicular Ad Hoc Networking and possible applications for VANET are also presented.

## 4.1 Connectivity paradigms within Vehicles

As seen in the previous chapter, many ITS systems have already made their way into vehicles, augmenting drivers' senses and providing an overall safer driving environment. However, the information that these systems offer is merely captured by advanced in-vehicle sensors, cameras, long- and short-range radar, maps and other databases. These sensor-based systems, despite being effective in many aspects, present some limitations. For example, even with highly sophisticated artificial intelligence capabilities, these systems are not able to perceive the road environment in the same manner as humans and thus cannot anticipate or predict certain scenarios. Also, equipping vehicles with a variety of cameras and sensors is costly and consumers may not be willing to pay for them.

As such, road transportation requires safer and more affordable solutions. A promising alternative source of traffic information is being explored through vehicular networking and the creation of cooperative-based ITS technologies. These systems are able to collect information from neighbouring vehicles or roadside infrastructures and are creating novel opportunities for the contextualized dissemination of digital content. Both Vehicle-to-Vehicle(V2V) and Vehicle to Infrastructure (V2I) may use wireless communication technology to exchange data either with other vehicles or road infrastructures. The combination of sensor-based systems with cooperative and connected vehicles can be more effective in reducing road accidents and provide better traffic flow. The combination of both may also facilitate the mimicking of human senses as well as reduce the need to equip vehicles with an assortment of expensive sensors [69].

Cooperative systems focus mainly on:

- **Comfort:** by providing contextual traffic information, weather forecasts or general access to internet services.

- **Safety:** passenger safety can be improved by transferring relevant information directly to the driver, such as road conditions or alerts when drivers breach rules.

- **Efficiency:** traffic may be re-routed in road congestion scenarios.

- **Capacity:** cooperative systems can promote a better road usage optimizing the capability of the road network.

### 4.1.1   Vehicular Network Architecture

In recent years interest in vehicular networks has increasingly grown, mainly due to the demand for safer and more efficient roads. A number of deployment architectures have been made possible each being more adequate in specific scenarios such as urban, rural or highway environments. [33] proposes a reference architecture and identifies three domain alternatives for vehicular networks: in-vehicle, ad hoc and infrastructure. Figure 4.1 illustrates this reference architecture.

- **In-vehicle domain:** Vehicles contain an on-board unit (OBU) and application units (AUs) which allow vehicles to communicate with one another. The OBU compasses the communication capabilities, through wireless or wired communications, while the AU devices, such as portable computers or PDAs, execute the applications by using the OBUs communication capabilities.

- **Ad hoc domain:** Vehicles are also equipped with OBUs, although not with AUs. Instead static road-side units (RSU) are located along the road way where they can be attached to an infrastructure network, which in return is connected to the internet. In an ad hoc domain both OBUs and AUs are seen as mobile and static nodes of an ad hoc network.

- **Infrastructure domain:** OBUs can communicate with the Internet through two infrastructure domain access exits: RSU and Wi-Fi hot spot. When no RSUs or hot spots are available, OBUs can also communicate through cellular radio networks such as GSM, GPRS, UMTS, WiMax and 4G.

As seen, connected driving can be made possible through the exploitation of wireless communication technologies and although these were not originally created for a vehicular scenario they have been adapted to them and are showing great promise in this area. For example, vehicular networks can be integrated into existing cellular systems or be connected to Wi-Fi hotspots positioned along a road. Considering the C2C-CC architecture reference along with these communication technologies it is clear that vehicular networks can be explored in two scenario types: car-to-car (C2C) and car-to-infrastructure (C2I). In the following sections a brief description of these wireless technologies is presented, exploring some of their characteristics and applications.

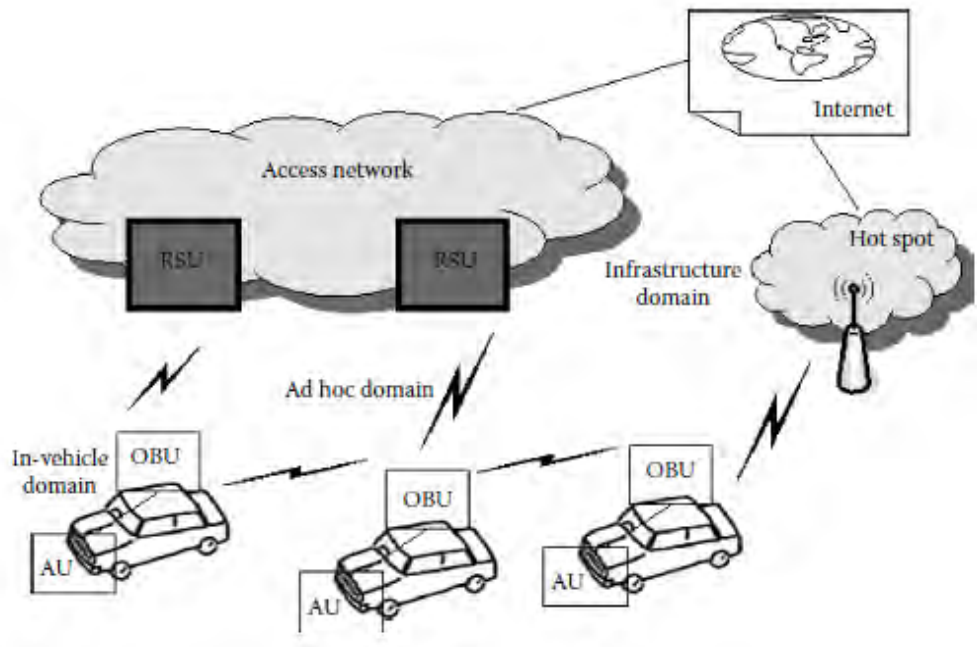

Figure 4.1: C2C-CC reference architecture [82].

### 4.1.2 Cellular-Based and It's Applications

Cellular communication technology has developed greatly in the past few decades and has had great success in changing the way people communicate. The mobility it offers allows people to make and receive phone calls from practically anywhere and more recently this technology has also allowed wireless internet access on mobile and smartphones.

A cellular network is a radio network that is able to provide network coverage over an ample land area. An area covered by this type of network is usually divided into smaller areas known as cells. Each cell has a fixed-location radio transceiver, commonly known as a base station. This base station is able to communicate with a great number of mobile devices, such as mobile phones, located in its area of influence. Communication between the base station and mobile devices is achieved via frequency channels. Each cell has a certain number of channels assigned to it. These can be forward channels (or downlink), used to transfer information from the base station to the mobile devices, or reverse channels (or uplink) to transmit back to the base station. Groups of cells are known as clusters and within each cluster no frequency may be repeated. A key aspect of cellular networks is introduced with the concept of frequency reuse. This allows for better coverage and capacity, particularly as radio channel frequencies are limited in wireless communications. Although frequencies may be reused, neighbouring cells may not be assigned the same channel, thus only cells that are far apart may be assigned the same channel. In rural areas cells may occupy larger areas and to better the quality of service in urban areas cells tend to be smaller. In cellular networks mobile devices cannot communicate with each other directly. As such, a base station is needed to establish the connection between individual mobile devices [124, 125]. Figure 4.2 shows a typical cellular network in which the cells take on a hexagonal shape and are arranged in clusters. Cells with the same number are assigned the same frequencies.

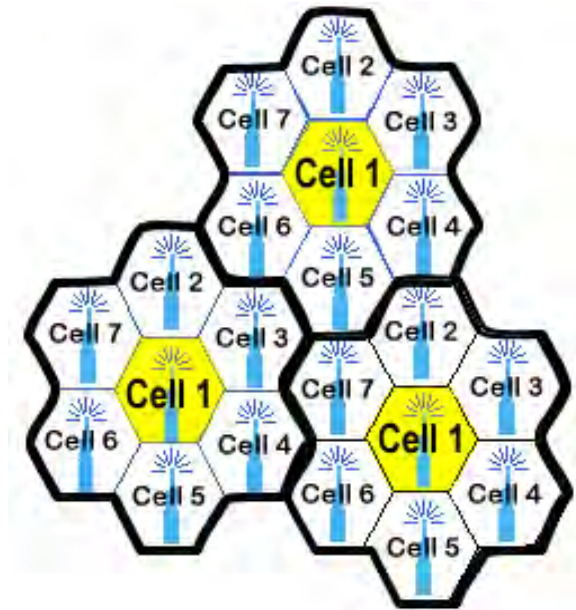

Figure 4.2: Example of cellular network [57].

### 4.1.2.1   Evolution of cellular networks

Over the years cellular technology has evolved rapidly and has experienced enormous growth. The first cellular systems also known as First-generation (1G) used analog transmission and offered voice services. This service was fairly limited and presented disadvantages such as the inability for cellular networks to communicate between different countries.

Technological advances brought forth the Second-generation (2G) of cellular technology which used digital multiple access technology, breaking from the analog nature of 1G. The Global System for Mobile Communications (GSM) standard, developed in Europe, describes the protocols for 2G and has become the global standard for cellular phones. GSM is characterized by a circuit-switched data channel and when compared to 1G presents higher spectrum efficiency, more evolved data services and enables communication across Europe through international roaming. Further developments lead to General Packet Radio Services (GPRS), also knowns as 2.5G and opened the gateway for wireless internet access. GPRS is a radio technology for GSM networks and is packet switched rather than connection oriented as GSM. It supports packet data transport and improves wireless access to the network. GPRS is also able to transmit position and speed information to the network for monitoring, traffic or weather applications. Enhanced Data rates in GSM (EDGE) is another evolution in 2G technology. Also commonly known as 2.75G, the transmission capabilities of EDGE introduced video and multimedia applications to cellular networks as it is able to move high volumes of data.

Another evolution in cellular technologies presented Third-generation or 3G. Universal Mobile Telecommunication Systems (UMTS) is one of the 3G technologies standardized by the third generation partner project (3GPP). UMTS provided users with a vast range of services and brought even better multimedia capabilities such as music, images and real-time video. Some characteristic of UMTS are higher bit rates, increased flexibility in resource usage, low delays, mobility for packet data applications, Quality of Service (QoS), simultaneous voice and data capabilities and interworking with existing networks such as GSM, GPRS, or EDGE [70, 82].

The most recent evolution in cellular technology consist of the Fourth generation (4G) of mobile communication technology standards. Once more, user demands for more freedom, flexibility, better user experience, as well as affordable prices have propelled this advancement.

### 4.1.2.2 Applications

Currently cellular networks present some limitations when applied to vehicular networks. For example it is not able to provide good quality service to a large number of users for long periods of time, and is also infrastructure dependent (base station) to establish transmissions between mobile nodes. On the other hand, this technology does show promise in the creation of vehicular networks as it does cover large areas. This is particularly useful when vehicles travel on highways or are far from large cities. Cellular communication is currently a well established and mature technology and as such vehicular networks are able to exploit the existent cellular infrastructure successfully, avoiding the installation of new and expensive infrastructures. Furthermore, according to [14] 4G will not only be able to support vehicular applications, but turn vehicles into wireless gateways that manage and collect data from a number of devices or sensors deployed in the vehicles and highway infrastructure, utilizing 4G cellular as the last-mile wide-area connection to the cloud.

An example of a simple cellular system in a vehicular context is that of the radio system used in taxis several years ago. Throughout a city taxi companies would have several antennas (base systems) installed and as taxi drivers drove around the city they would manually change the frequency channels on their taxi radio until the correct frequency was found. This system does present limitations as taxi drivers could only speak one at a time when connection with the base system had been established.

### 4.1.2.3 eCall System

The eCall System is another example of an application using cellular-based systems. To help quicken the response time of emergency services in road accident scenarios on European roadways, the European Commission has initiated the eCall project which aims at equipping vehicles with a wireless emergency call system. The eCall System will allow vehicles to automatically dial the European emergency number 112 in the event of a serious accident. The system relies on cellular technology to establish a free telephone connection with the emergency call centre, allowing the vehicle's occupants to communicate with the centre through the vehicle's loudspeakers and microphone. Sensors within the vehicle, such as airbag sensors, are able to detect when the vehicle suffers a crash and will activate the system automatically and send detailed information about the vehicle's condition and location through GPS to the emergency centre. The transmission of critical information will be much faster and may be more accurate, as in many cases, victims may be unconscious or not have the ability to call emergency services. The eCall may also be activated manually, allowing drivers to alert the emergency centre about road accidents they have witnessed. Such a system may prove extremely effective, and estimations predict quicker response times of 40 to 50%. The European Commission intends for all new models of vehicles to be equipped with the eCall system and be fully functional in all countries of the European Union by 2015 [30]. Figure 4.3 details the functioning of the eCall system.

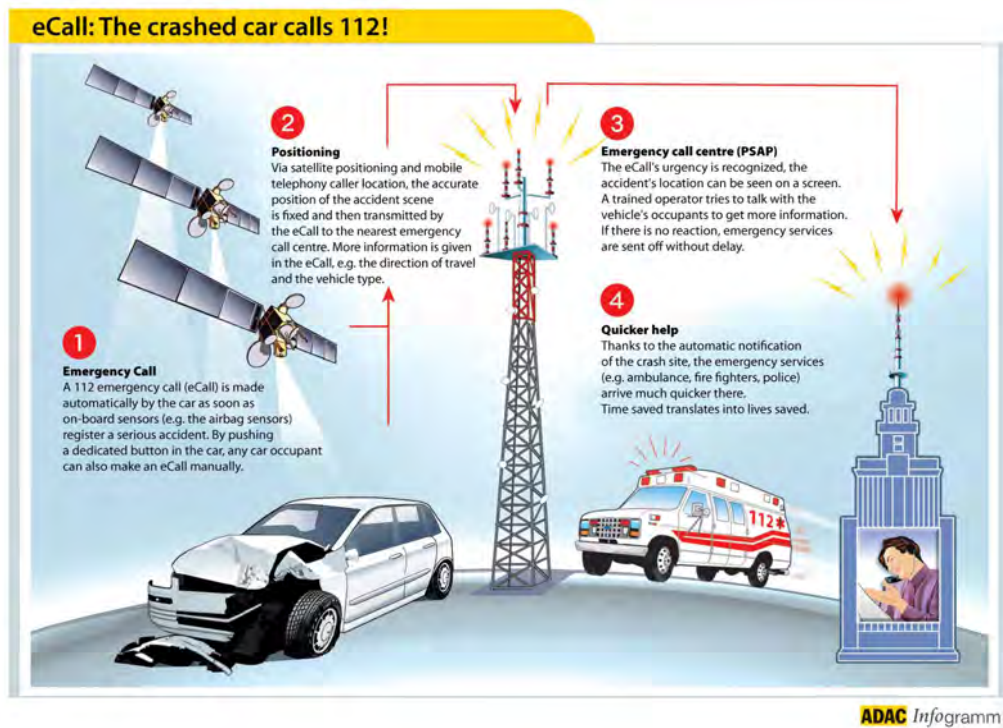

Figure 4.3: Functioning of the eCall System [31].

### 4.1.3 WiFi-Based

A Wireless Local Area Network (WLAN), allows electronic devices to connect to the internet or connect to other devices without resorting to wires. Connection is established through access points (or hotspot) which offer users the possibility to move around within a coverage area. Devices that are able to communicate via WiFi are referred to as stations and are equipped with wireless network interface controllers. Stations may be of two types: access points and clients. An access point is the base station for a wireless network and refers to the equipment (routers) that provides access to the network. Clients on the other hand refer to the devices which connect to the network. These may be laptops, PDAs, smarphones, printers and any other devices equipped with wireless network interface controllers. WLANs have become very popular mainly due to the flexibility and mobility they offer as well as easy installation.

WLANs are defined by a set of standards that guide their implementation and assure compatibility between different protocols. The IEEE 802.11 [62] was first established in 1997, but is currently the standard of reference for wireless local area networks. It defines a medium access control (MAC) and a physical (PHY) layer. At the MAC level a decentralized approach was defined through a Carrier Sensing Multiple Access with Collision Avoidance (CSMA-CA) which allows for ad hoc communication. As such, the IEEE 802.11 standard supports two network modes: infrastructure mode and ad hoc mode. In infrastructure mode, stations communicate through access points while in ad hoc mode they communicate directly with one another, without resorting to any access point.

Several amendments have been made to the IEEE 802.11 standard to accommodate new services. Namely, the IEEE 802.11p [67] standard has been specifically designed to function in vehicular networks. The main motivation behind it was the development of safety applications to better traffic flow and create safer driving environments. It is able to cope with the dynamic conditions, as well as high velocities present in a vehicular environment and thus is favourable for the creation of Intelligent Transportation Systems. Some of the main features the IEEE 802.11p standard supports are faster network recognition and setup, longer distance coverage (up to 1000 meters) and tolerance of high speeds. [82, 18]

### 4.1.3.1 Applications

Vehicle-to-Infrastructure communications enabled through DSRC can already be found in commercial applications such as Electronic Toll Collection (ETC) on highways, congestion charges in major cities.

**Electronic Toll Collection**   Electronic Toll Collection is a road traffic management system which allows the electronic payments of toll fees on highways without vehicles having to stop, thus minimizing delays and cues at toll gates. Currently this technology is quite mature as it was one of the first cooperative ITS systems to be developed. ETC systems rely on roadside infrastructure to scan each vehicle and an on-board unit containing information such as the type and model of the vehicle. When a vehicle passes through an electronic toll gate it is scanned by the roadside equipment to check if it contains the on-board unit placed on the windshield. ETC systems require on-board units and rely on four main components [48]:

- **Automated Vehicle Identification:** a vehicle's identity needs to be determined in order to bill the correct driver.

- **Automated vehicle classification:** identifies different classes of vehicles as these are normally charged different rates.

- **Transaction proceedings:** ensures the transaction in made through the right accounts.

- **Violation enforcement:** identifies vehicles that do not have on-board units.

Some of the main benefits of these systems are: reduction in waiting times, reduction of fuel consumption and pollutant emission, and reduction in costs.

In Portugal, such a system is for example the Via Verde technology. This system also allows automatic wireless payment at gas stations and parking lots. Figure 4.4 shows an ETC infrastructure in Portugal and in Fig. 4.5 the on-board unit is seen.

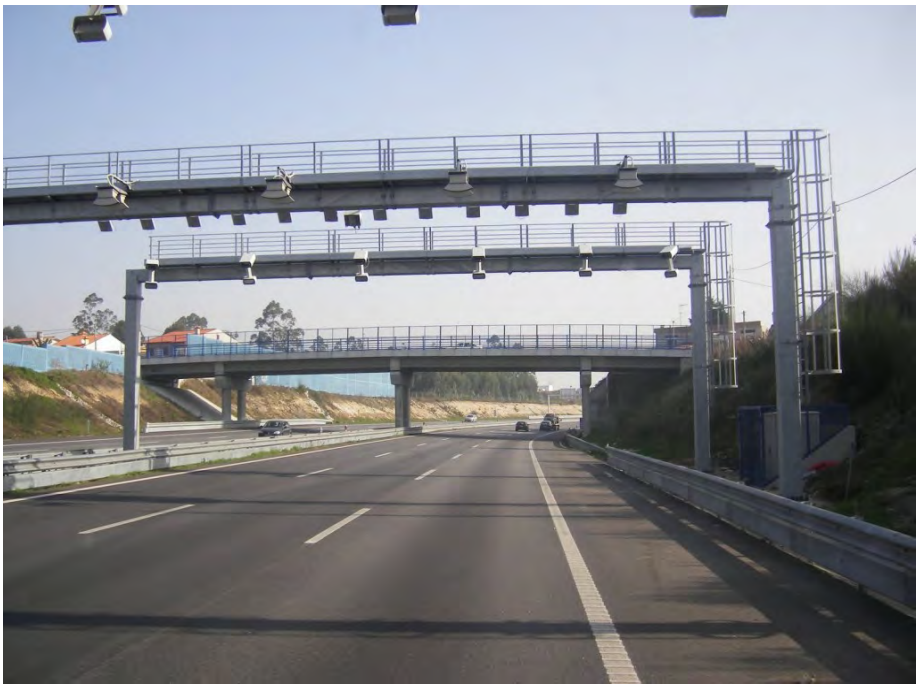

Figure 4.4: Electronic Toll Collection infrastructure [26].

## 4.2   Vehicular Ad Hoc Networking

Vehicular Ad Hoc Networking (VANET) is a new type of wireless communication that has emerged in recent years. In a VANET network each node is a moving vehicle equipped with a wireless interface. These networks are considered to be promising for a multitude of applications, enabling communication with neighbouring vehicles as well as with nearby fixed equipment (Roadside Units). Vehicular ad hoc networking uses the 801.11p standard that was specifically designed to operate in the vehicular environment. It operates on a specific spectrum in the 5.9 GHz band, and the allocated spectrum varies from 30 MHz in the European Union (EU) to 75 MHz in the United

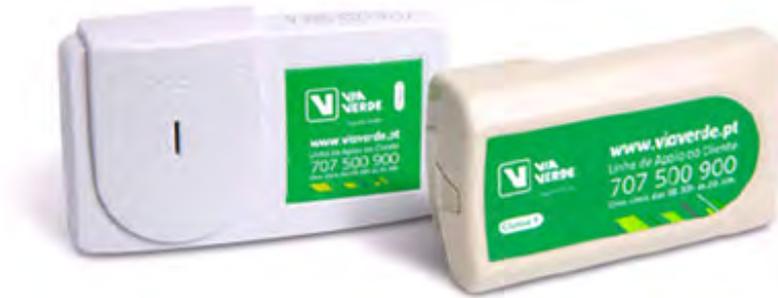

Figure 4.5: Via Verde on-board equipment for Electronic Toll Collection [118].

States (USA). VANETs present unique characteristics which set them apart from other wireless communication networks. These are:

- **Unlimited transmission power:** A node can power it self since in vehicular networks nodes are vehicles, which have virtually unlimited power autonomy.

- **Higher computational capability:** Operating vehicles can afford significant computing, communication, and sensing capabilities.

- **Predictable mobility:** The mobility of nodes (vehicles) can generally be predicted, as their movement is mainly limited to roadways.

- **Potentially large scale:** Vehicular networks are extendible over the entire road network.

- **High mobility:** Vehicular networks operate in dynamic and extreme configurations. Node speed and density can vary greatly and to extremes, for example due to rush-hour traffic.

- **Partitioned networks:** Gaps between vehicles may occur in a less dense cluster of nodes.

- **Network typology and connectivity:** Vehicular network typology changes constantly as the link between nodes connect and disconnect.

One of the building blocks for the design of VANET-enabled safety applications is the enhanced awareness of surrounding vehicles provided by the Cooperative Awareness Messages (CAM) [64] that is defined for DSRC communications. This provides presence, position and status information to neighbouring vehicles within a single hop distance. While the standard defines scenarios such as emergency vehicle approaching with the flags LightBarInUse and SireneInUse, it is beyond the scope of the standard to specify how this information should be presented to the user. The standard [64] does define timing requirements for different types of CAM messages, which must all have a maximum latency of 100ms and frequency between 1 and 10Hz depending on the message type priority. From the richness and immediateness of information provided by CAM messages, it is possible to virtually recreate the vehicle's surrounding environment.

### 4.2.1 Applications for VANET

Due to its infrastructureless nature and having very low association times, VANETs enable ubiquitous V2V and V2I communications. V2V communications are mostly driven by safety related applications. Examples of some proposed safety applications using V2V communications are forward collision warning [41, 121] and the intersection collision warning [111].

#### 4.2.1.1 See-Through System

The See-through System, presented in [85] is a system that assists the driver with an overtaking manoeuvre, by means of augmented reality. Initiating an overtaking manoeuvre is typically a challenge due to the lack of visibility, especially if the vehicle which one aims to overtake does not have a window at the same level as the driver's viewpoint allowing him to see traffic through it. This is particularly true in the case of large trucks and buses. The length of these vehicles present a further challenge. As these vehicles usually drive at a slower speed, they become the main reason for other vehicles to initiate an overtaking manoeuvre. The See-through System thus aims to transform these large and opaque vehicles into transparent objects, allowing the driver to see "through" them and better evaluate if it is safe to perform an overtaking manoeuvre.

To enable this, the See-through System uses radio vehicle-to-vehicle communication technology based on Dedicated Short Range Communication (DSRC). These are based on short to medium range uni or bi-directional wireless communication channels, specifically designed for automotive applications. The vehicle in front sends a video stream captured by a camera placed on its windshield to the vehicle that wants to initiate the overtaking manoeuvre. The receiving vehicle then uses sensors that measure the distance through computer vision to determine the geometry of a translucent image that is projected over its windshield by means of a holographic projector. This image overlaps the vehicle that is occluding the driver's vision, allowing him to "see through" the vehicle which he aims to overtake.

## 4.3 Possible Traffic Signs Enabled by VANET

As seen previously in Chapter 3, systems that are able to detect road traffic signs and duplicate them on devices inside the vehicle have already been developed. However, these systems are not effective when road signs do not exist. A new approach of traffic sign creates signs just in time and just in place when they are needed.

### 4.3.1 Virtual Traffic Lights

The Virtual Traffic Lights (VTL) project aims to reduce traffic congestion and improve traffic flow through a more efficient exploration of road network intersections. To this end, this project proposes the migration of existing physical traffic lights to virtual signs located inside the vehicle, solely supported through vehicle-to-vehicle communications. By placing traffic lights in a virtual way within vehicles through means of augmented reality will allow the entire road network to be presented with traffic lights and not only a reduced number of intersections, as is the case today. According to [45], the city of Porto, which is the second largest city in Portugal, has a total of 2,000 intersections, of which only 328 are regulated by traffic lights. This corresponds to a mere 16%. As such, [45] present the VTLs as a much more feasible response when pondering the possibility of implementing a traffic light at every intersection since it is not dependent on any physical infrastructure. With the VTL system elected vehicles act as temporary traffic lights and broadcast messages to other vehicles.

# Chapter 5

# Prototype Implementation and Evaluation

In this chapter we describe the implementation of some prototypes of AR-based ADAS and discuss some evaluation results. We start by describing the generic framework for the creation of acoustic and graphical content, embedded in the car and connected to its sensors and actuators. We divide our presentation between prototypes which have solely been implemented in simulated environments, and prototypes which have been implemented in real vehicles and experimented in real driving scenarios. Some evaluation results are also presented and discussed.

## 5.1 The OpenSceneGraph Framework

The creation of augmented reality based on digital objects requires the ability to create computer graphics that can be placed in a three-dimensional space. The OpenSceneGraph framework [87] is an open-source 3D graphics application programming interface that is widely used by a community of application developers in areas such as visual simulation, computer games, scientific visualisation and modelling, and virtual reality.

The toolkit is written in standard C++ using OpenGL [83], running on several platforms and operating systems, such as Microsoft Windows, Linux and iOS X. The latest version of OpenSceneGraph also supports application development for mobile platforms, namely iOS and Android.

### 5.1.1 OpenSceneGraph Architecture

Figure 5.1 shows an overview of the generic architecture of the OpenSceneGraph framework. OpenSceneGraph implements a feature-rich scene graph [10] data structure, which arranges the logical and spatial representation of a graphical scene into a collection of nodes in a graph or tree structure. A node can have multiple children, but often only a single parent node. Operations performed on a parent node propagate to all its child nodes. This is very useful for geometric operations, where a transformation matrix provides an efficient and natural way to process a group of nodes organised in the scene graph structure.

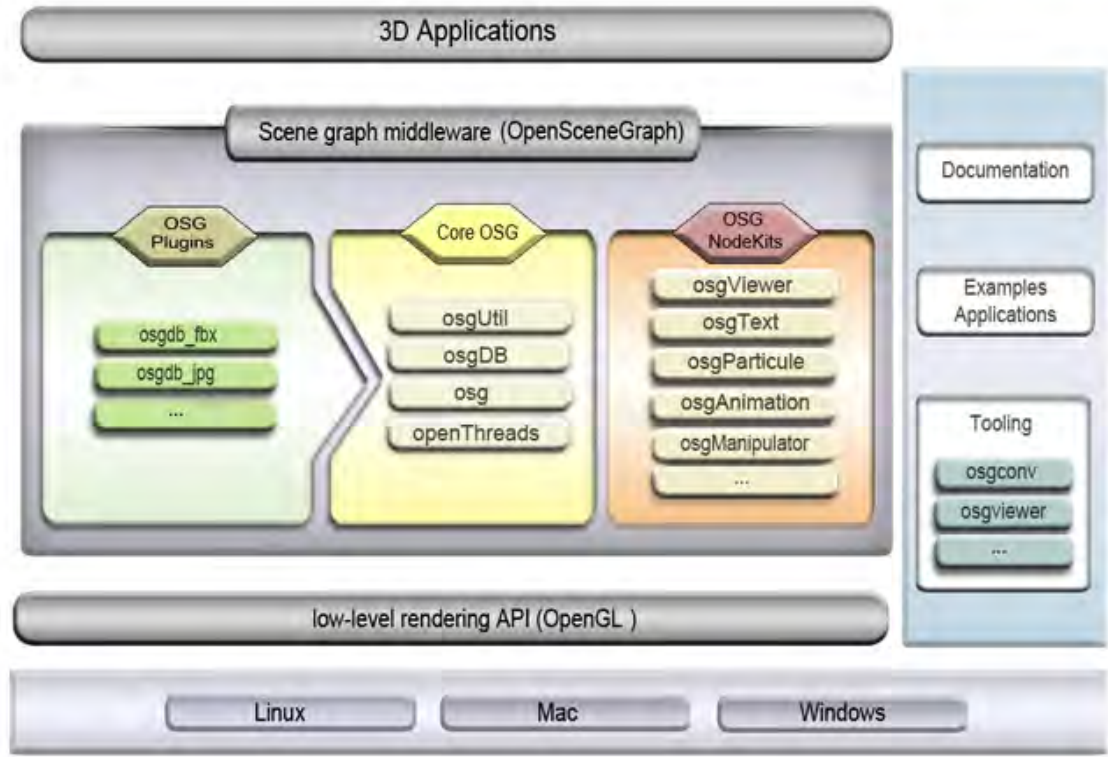

Figure 5.1: Architecture of the OpenSceneGraph framework [3].

In large scene graphs, the use of linear operations can become noticeably slow, as they are able to deal with a very large number of nodes. In such cases, a common data structure is the tree, which is organised in a hierarchical representation of group nodes and leaf nodes. Group nodes can have any number of children attached to it and include transformation and switch nodes. Leaf nodes are nodes that are actually rendered or see the effect of an operation. Leaf nodes can also include objects such as sound, allowing the creation of acoustic AR content, as will be presented later in this chapter.

OpenSceneGraph includes several plugins that simplify the creation of realistic AR. It supports a wide range of 2D image and 3D database formats, with loaders available for well-known standards such as OBJ, 3DS, JPEG or PNG.

OpenSceneGraph node kits also include native capabilities such as particle effects (supporting special effects such as fire or smoke), support for anti-aliased TrueType text and animations/simulations (skybox, lights, time-of-day, etc).

In a layer-based organisation, 3D applications interact with OpenSceneGraph, which communicates with the OpenGL layer, which in turn communicates with the graphics hardware.

### 5.1.2 OSG Core Classes

The scene graph tree structure in OSG has several classes of objects. A subset of some of the most important of these classes is described as follows:

- The *Node* class is the base class;
    - *Group*, holds a set of child nodes;
        * *Transform*, transforms all children by a 4x4 matrix;
        * *Switch*, switches between children, (e.g. to implement a traffic light);
        * *LOD*, level-of-detail, switch based on distance to viewer;
        * *LightSource*, leaf node defining a light in the scene;
    - *Geode*, lead node for grouping drawables;
        * *Billboard*, orients drawables to always face the viewer;
    - *Drawable*, abstract base class for drawable graphics;
        * *Geometry*, holds vertices, normals, faces, texture coords, etc;
        * *Text*, for drawing text;
        * *DrawingPixels*, encapsulates drawing images using the OpenGL primitive Draw-Pixels;
    - *StateSet*, encapsulates OpenGL states and attributes;
    - *Texture*, encapsulates OpenGL texture functionality;

### 5.1.3 A Simple Example: Showing a Traffic Sign

The following example shows the main parts of the required code to create a scenario with a traffic sign. The *Scenario* represents the root node of the scene graph, and thus, all objects like roads, trees, signs or vehicles are added to the *Scenario* as child nodes. Then, a traffic sign with different levels of detail (10%, 50% and 100%) is created. This way, the fully detailed version of the traffic sign only needs to be rendered when it is near the camera node. Once the traffic sign becomes distant, simplified versions of the object should be rendered, leading to an optimization of the rendering system. We also add a *CameraNode* to a specific position on the scene. This node can be controlled by the user and is responsible for our view within the scenario. Finally, we setup a viewer with our scenario.

```
int main(int argc, char** argv)
{

    osg::Vec positionVec = ...;
    osg::Vec myPosition = ...;
    ...
```

```
Scenario* scenario = new Scenario(); // world

osg::Node* sign1 = osgDB::readNodeFile("models/sign_detailed.osg");
osg::Node* sign2 = osgDB::readNodeFile("models/sign_50%detail.osg");
osg::Node* sign3 = osgDB::readNodeFile("models/sign_10%detail.osg");

osg::LOD* lodNode = new osg::LOD();
lodNode->addChild( sign3.get(), 200.0f, FLT_MAX );
lodNode->addChild( sign2.get(), 50.0f, 200.0f );
lodNode->addChild( sign1.get(), 0.0f, 50.0f );
scenario->addNode(lodNode,positionVec,dirAngle);

CameraNode* cameraNode = new CameraNode();
cameraNode->setPosition(myPosition);
scenario->addChild( cameraNode );

Viewer viewer;
viewer.setCameraManipulator(cameraNode->getCameraManipulator());
viewer.setSceneData( scenario );
return viewer.run();
}
```

## 5.2   A Driving Simulator in OSG

OpenSceneGraph is used in a variety of applications, such as computer games[35], virtual terrain projects [95] or flight simulators [91]. A number of driving simulators are also based on the OSG framework. In Fig. 5.2 we show a screenshot of the scenario created for the OSG powered driving simulator used by Vires Simulationstechnologie GmbH.

Driving simulators are very important tools to assist in learning to drive and also to test novel driving assistance systems in safe environments. Driving can be a life-threatening activity and it is impossible to test prototype systems in real and public driving environments without creating dangerous situations. Driving simulators are also widely used by car manufacturers [99, 34].

As part of this thesis, and in order to test and implement AR driver assistance systems we have been involved in the development of a driving simulator that is partially supported by OSG.

### 5.2.1   The Driving Cockpit

The simulation-based evaluation of novel driver assistance systems requires the driving environment to be replicated with as much realism as possible. A critical component is the driving cockpit, which should resemble a real vehicle cockpit. To this effect, we have used a Smart ForTwo space frame, where the dashboard, steering wheel, pedals and additional driving controls have been replaced by specific peripherals which are connected to computers and provide inputs to our driving simulator. Two very important components are the steering wheel and the braking and acceleration pedals. We have modified the Smart ForTwo space frame to seamlessly install a Logitech G27 Racing Wheel [76], which includes pedals and gear shifting controls. This steering wheel provides force feedback and is supported by various operating systems, being easily integrated in our OSG powered driving simulator. In Fig. 5.3 we show the general aspect of the modified Smart ForTwo used to implement the driving cockpit. Optical systems such as the rearview mirror installed on the windshield have been replaced by LCD displays, providing the appropriate view of an OSG viewer parametrized with the reflected view of the driver.

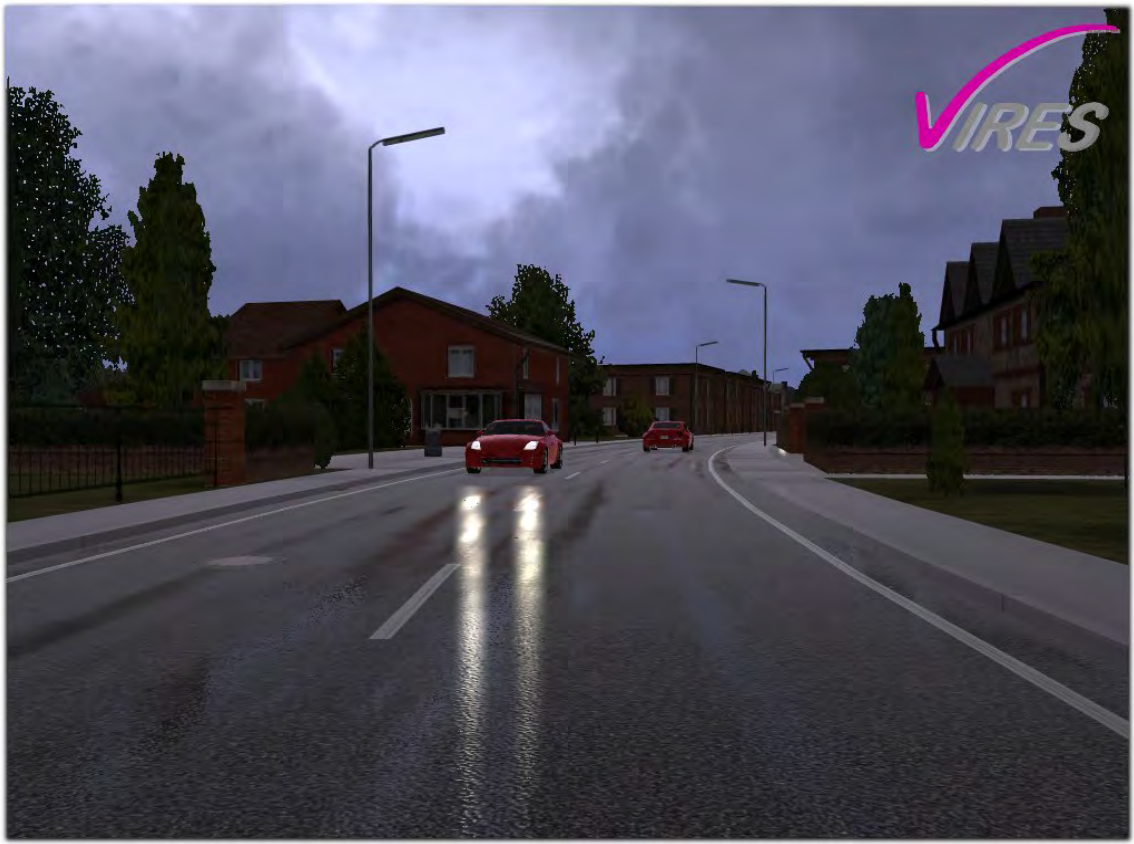

Figure 5.2: Vires driving simulator [52].

A Pioneer S-HS100 surround sound set of speakers were also installed in the Smart ForTwo space frame, to provide the acoustic augmented reality content.

Note that in terms of AR, the confined space of an automobile provides an immersion level that is far superior to the AR experiences that can be provided by smartphones. In terms of AR embodiment [20], an automobile as a wearable computer [11] offers a seat for the user, usually restrained by a seat belt, a comprehensive set of controls, operated by hand and foot (steering wheel, pedals), surround sound systems, and a variety of optical and electronic displays, where the transparent windshield is also included. Enabling wireless connectivity, namely with neighbouring vehicles, through VANET, is a key step in presenting automobiles as wearable computers, opening up a major application area for AR. In [46] we recently presented this perspective of vehicular embodiment and AR.

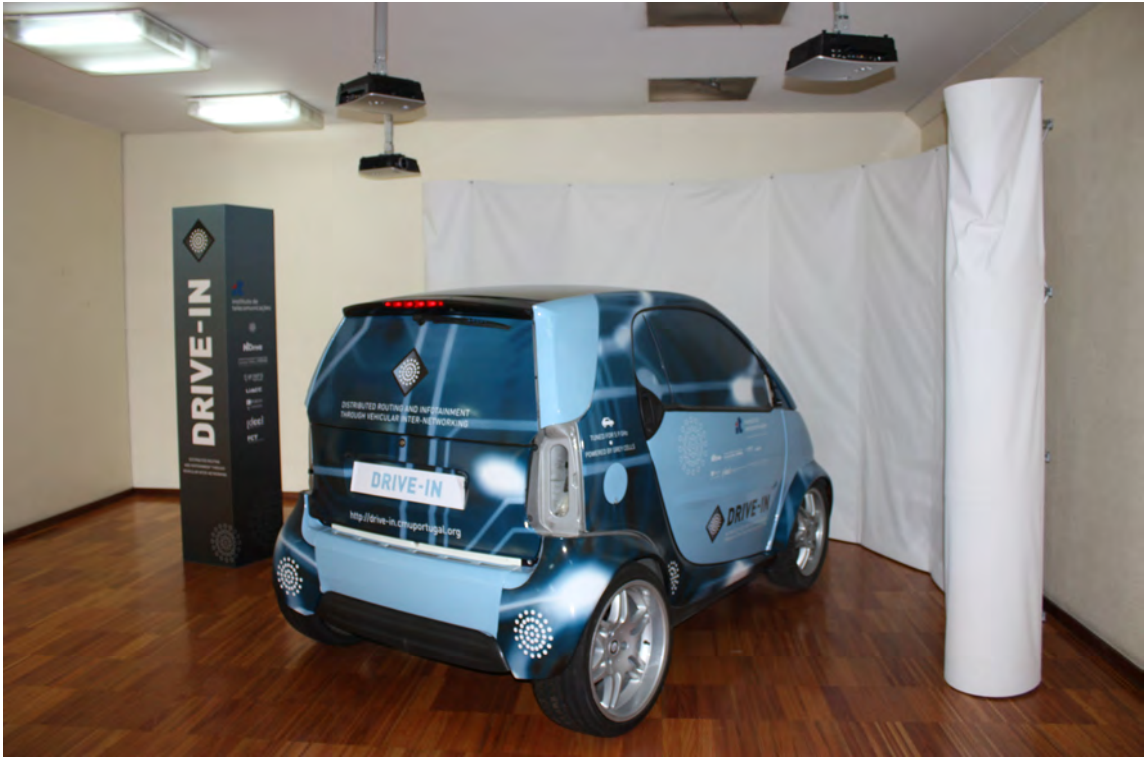

Figure 5.3: General aspect of the modified Smart ForTwo driving simulator.

### 5.2.2   Projection Environment

To project the virtual scenario in which the Smart ForTwo space frame is immersed, we opted to design a large curved canvas that partially surrounds the driving cockpit. We combined three LCD projectors, which are connected to a Matrox Triple Head graphics card, producing a combined image with a resolution of 4080x768 pixels. We computed the placement of the projectors, which hang over the driving cockpit, to cover the driver's field-of-view as much as possible, as illustrated in Fig. 5.4. We covered part of the side windows in the doors of the Smart ForTwo, to completely immerse the driver in the projected scenario.

In Fig. 5.5 we show the projection setup, together with the Smart ForTwo space frame, displaying a driving scenario in the semi-oval canvas.

### 5.2.3   Realistic 3D Scenarios

To construct realistic scenarios we have resorted to databases of three-dimensional objects, such as traffic signs, traffic lights, buildings, vehicles, trees and vegetation. We have also resorted to databases of common textures, using images to represent road pavements, road markings, building walls, billboards, etc. To create the complete scenario, we used Google Sketchup, where the different objects where arranged and placed geographically. In Fig. 5.6 we show an example of an urban scenario used in the experiments evaluating advanced driver assistance systems. To maximise the driving duration in our experiments that used small scenarios, we configured the road geometry and topology to permit cyclic circuits, such as the "eight" layout that is highlighted in Fig. 5.7.

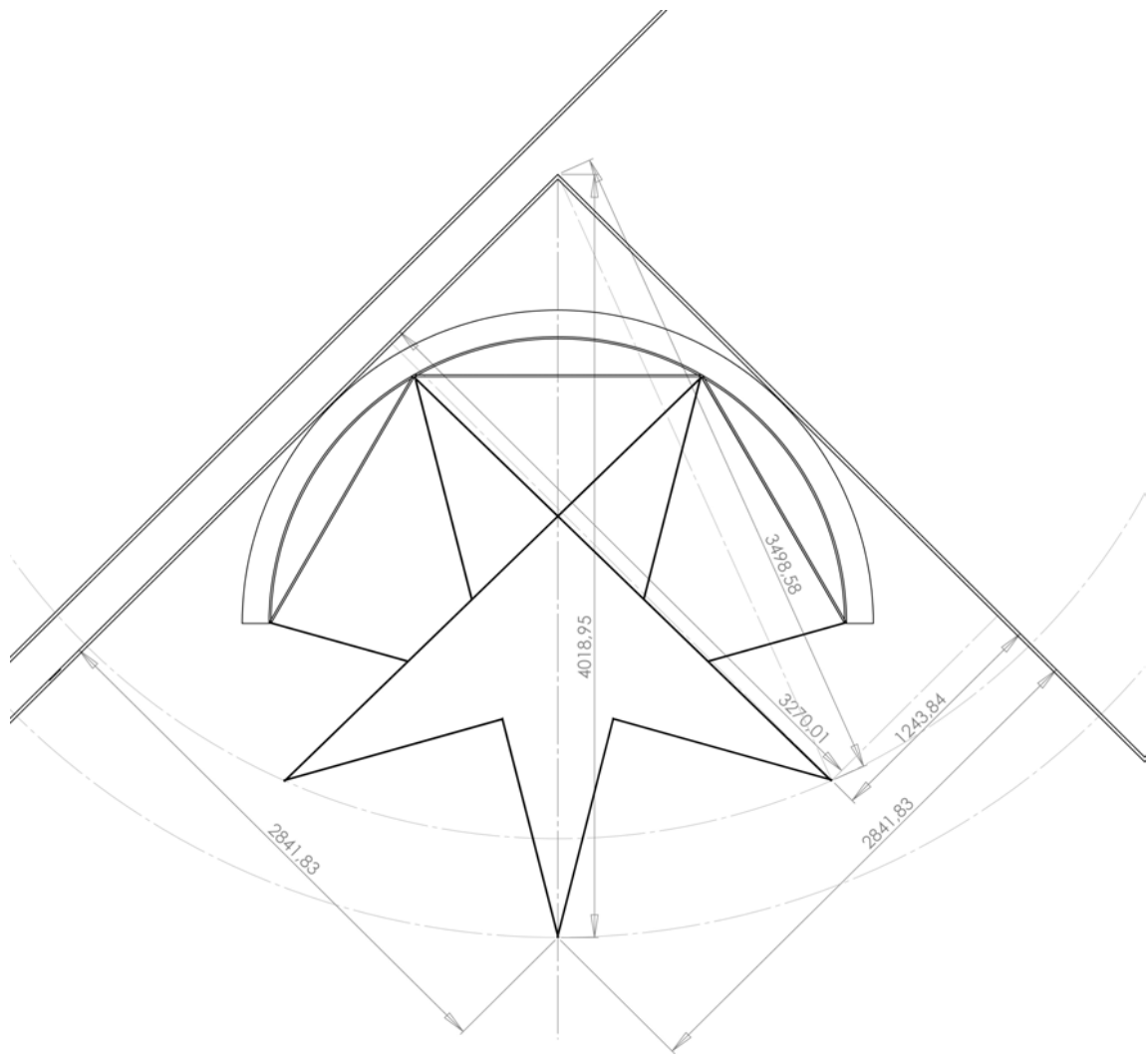

Figure 5.4: Projection calculation.

### 5.2.3.1 Creating the Scene Graph from 3D Objects

The scenario model exported by Google Sketchup is not prepared to be rendered efficiently by the rendering engine. Basically, the entire scenario is represented in a single and static model, instead of a collection of well organized models in a scene graph. Thus, we built a simple program that traverses all the objects of the original model, building a paginated scene graph with different levels of detail and culling features. Thus, we are allowed to render large realistic city scenarios.

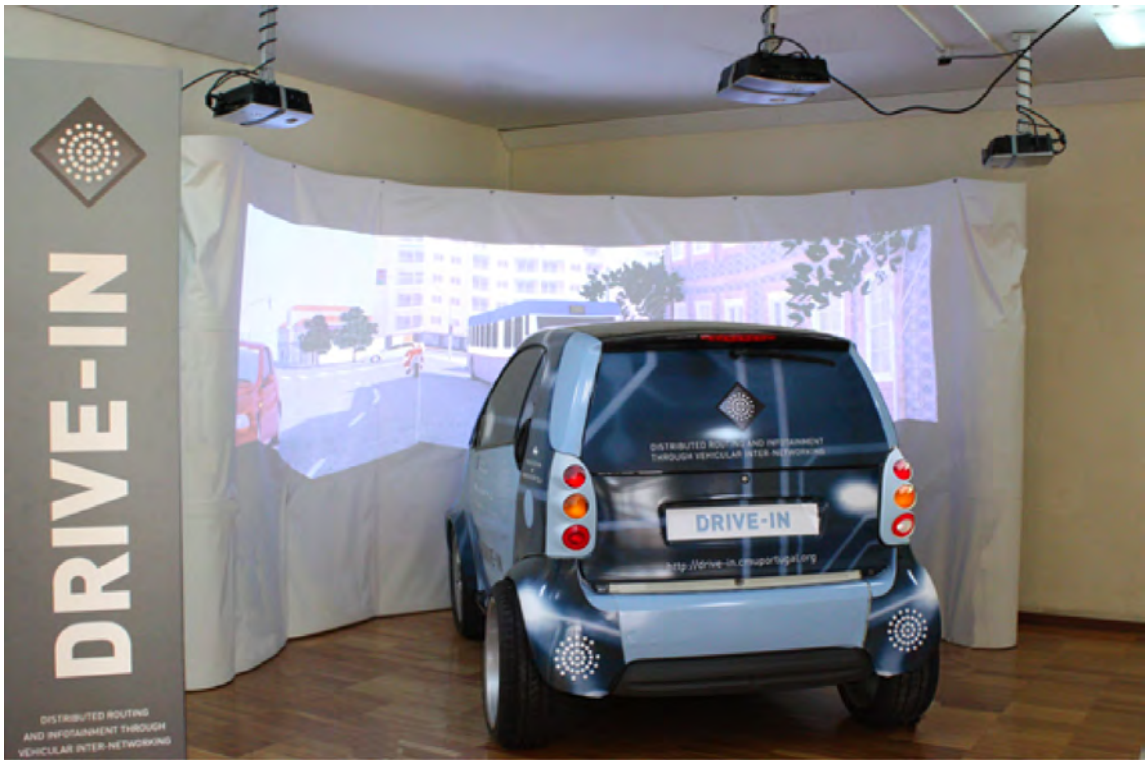

Figure 5.5: Projected driving scenario.

### 5.2.4   Coupling with a Vehicular Network Simulator

In addition to the driving simulator and realistic driving scenarios, a fundamental component of the simulation framework designed to evaluate novel driver assistance systems is the generation and realistic animation of other vehicles that cause interactions with the vehicles being driven through the Smart ForTwo driving cockpit. Furthermore, and given that the driver assistance systems we design leverage importantly on inter-vehicle communication, it is also fundamental that these additional vehicles have their V2V communications realistically simulated as well. To this extent, the DRIVE-IN (Distributed Routing and Infotainment through VEhicular Inter-Networking) project [36], involved researching and developing very large-scale VANET simulation frameworks, such as the DIVERT and VNS (Vehicular Networks Simulator) [32, 44] simulators.

#### 5.2.4.1   Mobility Simulation

The DIVERT simulator is responsible for the realistic simulation of vehicles, implementing a microscopic mobility model based on the Intelligent Driver Model (IDM) [114]. The driver-centric VANET simulation framework described in [54] has extended this model to refine mobility into a sub-microscopic level of detail, being able to provide realistic feedback to a human driver that is interacting with the simulated vehicles through a driving simulator, as describe above.

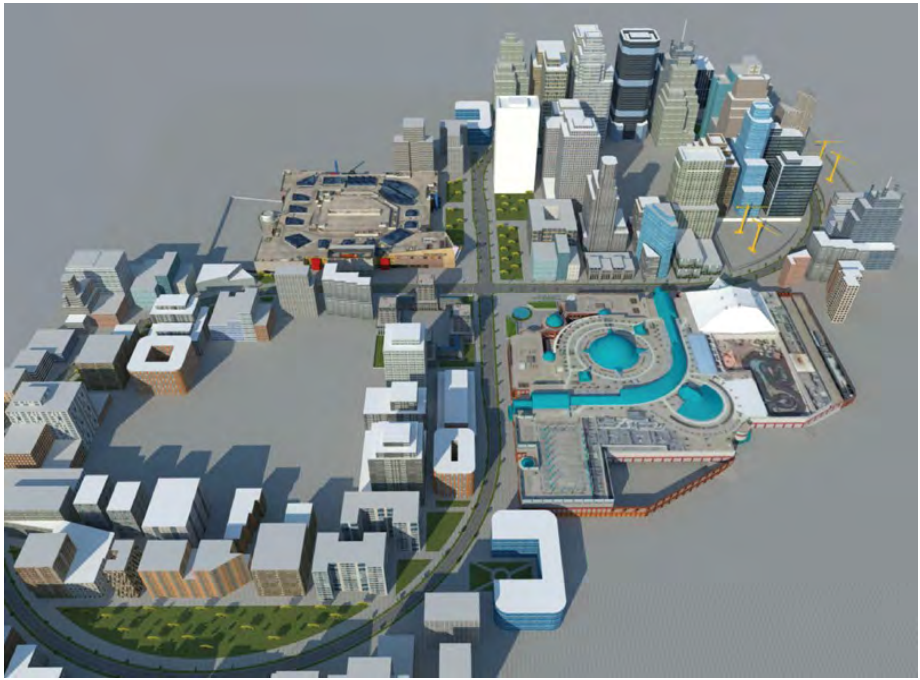

Figure 5.6: Urban scenario of driving simulator.

The coupling architecture between the DIVERT simulator and the driving simulator is a typical TCP client-server architecture, where DIVERT acts as a virtual GPS server, providing rich data sets that include the vehicle's position, heading and speed information of each vehicle being simulated. Since DIVERT provides a TCP server thread for external applications and the driving simulator is implemented as an autonomous external application, both can easily be connected. In this context the driving simulator acts as a client which connects to DIVERT to get information related to each simulated surrounding car. Any action performed by the driver in the driving simulator affects all his neighbours in DIVERT. In Fig. 5.8 we show how a snapshot of a DIVERT simulation state (right frame), generates the vehicles that are represented as detailed 3D objects in our OSG scenario.

### 5.2.4.2 Network Simulation

The VNS (Vehicular Networks Simulator) [44] framework couples DIVERT with a widely used network simulator, NS-3 [84], providing a complete and transparent integration between these two systems. Network simulation is critical, as it provides the information that enables the functioning of the driver assistance systems. Ultimately, network simulation will affect mobility simulation, as many vehicular control systems rely on network information to manage acceleration or braking systems (e.g. platooning systems [110], virtual traffic lights [45]). A tight and bi-directional coupling between the mobility and the network simulators is thus very important to achieve realistic results [107]. Note that in VANET, where the communication range is limited and the impact of the distribution of vehicles plays an important role [22], the mobility pattern clearly affects the performance of the network layer, and thus the bi-directionality in the coupling between these two components becomes fundamental.

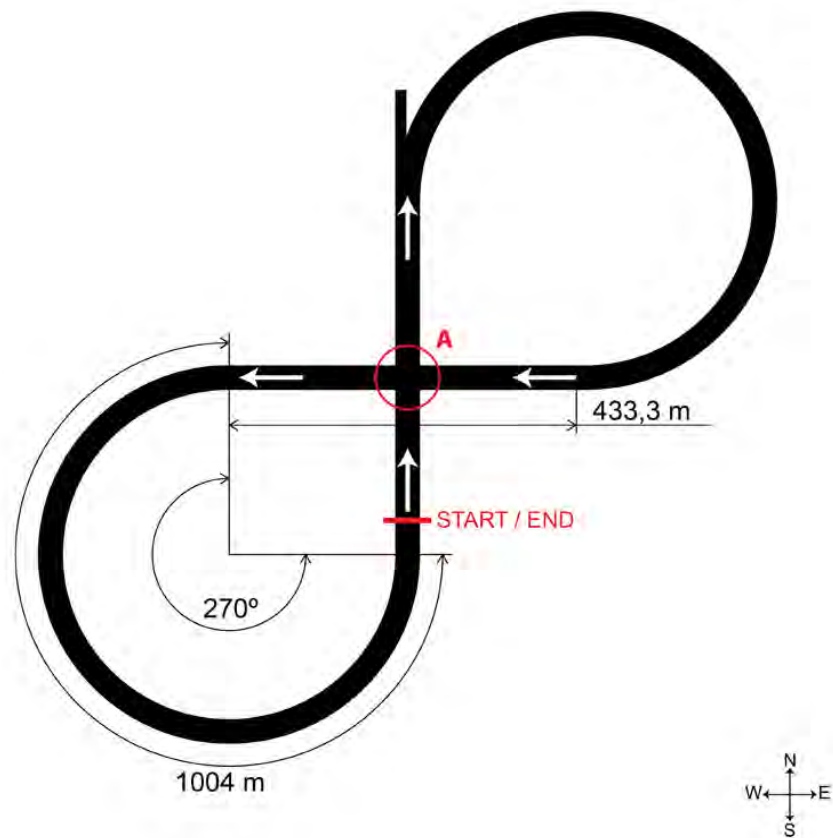

Figure 5.7: Driving circuit shaped as an "eight".

NS-3 is a discrete-event network simulator widely used by the research community. It provides various protocol modules and several radio access technologies such as 802.11p, UMTS or WiMAX. It is implemented as a C++ library, which allows its use in other software. VNS has focused on improving the performance of NS-3 in the simulation of large-scale VANETs, featuring thousands of nodes, by defining specialised indexing structures that provide efficient methods to determine which nodes are reachable by other nodes and to predict when a node will be in the communication range of another node. VNS uses Quadtrees to store the vehicles in the network, greatly improving the list-based access natively provided by NS-3.

## 5.3   Simulator-Based Prototyping

Leveraging on the simulation framework described in the previous section, we now describe the design and evaluation of AR driver assistance systems, which use VANET-based communication, namely in the V2V form. We start with a previously presented system, by Gomes et al. [53], describing its implementation and evaluation in the context of the simulation framework described previously. We next present our proposal of Virtual Surround Sound, describing two driver assistance systems, virtual siren and virtual emergency braking warning, that leverage on this acoustic form of AR. We then present the Virtual Traffic Lights concept and describe a possible in-vehicle virtual representation, discussing results related to a driver-centric evaluation.

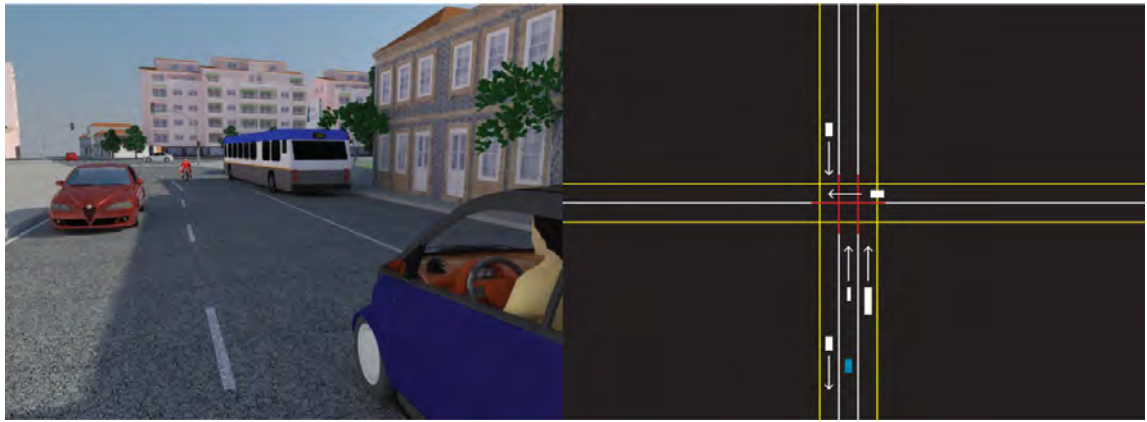

Figure 5.8: Simulated urban environment from the city of Porto populated through the DIVERT road scenario [54].

### 5.3.1 The See-Through-System (STS)

The See-Through System (STS) is a video-based co-ADAS designed to enhance the visual perception of the driver during an overtaking manoeuvre [85]. It was designed to assist the driver when trying to overtake a vision-obstructing vehicle, which is a very dangerous manoeuvre, with a significant death per crash ratio. By using V2V communication, STS is able to deliver real-time video streaming of the windshield camera's perspective from the preceding vehicle to the driver that wants to pursue the overtaking manoeuvre.

Originally [85], STS' driver interface was presented as a simple and non-manipulated video stream displayed on the centre console, as illustrated in Fig. 5.9.

In [53], a refined version of STS was introduced, extending the original representation of the video streamed from the camera in the vehicle in front using an AR-based interface that tries to transform vision-obstructing vehicles into transparent tubular objects. The main idea is to present a 2D representation of a 3D tubular and transparent object that reflects the length of the preceding vehicle in order to give depth perspective. Figure 5.10 depicts this tubular object and its 3D-looking frame representation. The frame generation is explained in detail in [53]. This frame is designed to highlight the blind-spot that the length of the preceding vehicle and the distance-of-view of its camera originates. It is important for the driver to have this perception, in order to better decide the best moment to pursue the overtaking manoeuvre.

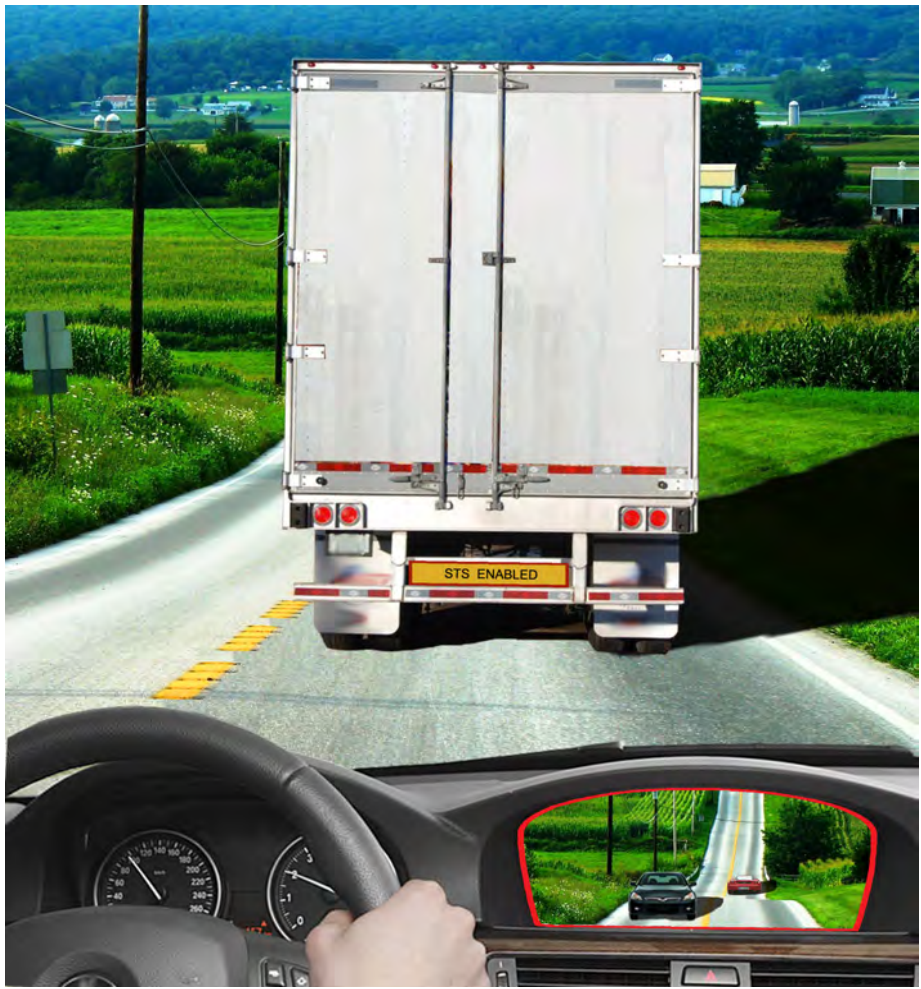

Figure 5.9: STS video stream displayed on the centre console [85].

The vehicle that intends to pursue the overtaking manoeuvre is represented by box A, and the preceding vehicle by the dashed box B. All the variables used in the frame generation and their description can be found in Table 5.1. Using computer vision STS detects the rear of the preceding vehicle, which is represented by the outside frame in Fig 5.10. With the detected width $w_0$ and height $h_0$, and the previously known width $w$ and height $h$ and distance between vehicles $d$, STS is able to calculate the ratio. This ratio will be used to calculate the width $w_i$ and height $h_i$ that will define the inside frame that will embed the video-streaming transmitted by the preceding vehicle with its front perspective. To be more accurate with the frame's depth perception, STS adds the distance from the camera to the ground capturing point $e$ to the length of the preceding vehicle, $l$ in these equations. The image conveyed to the driver needs not only give a real depth perception of the real distance of the objects in the video-streaming, but also to exhibit possible limitations that can arise from this system, such as the blind-spot. This limitation emerges as a consequence of the fact that the video-streaming only displays the view beyond the distance at which the camera starts to capture the road.

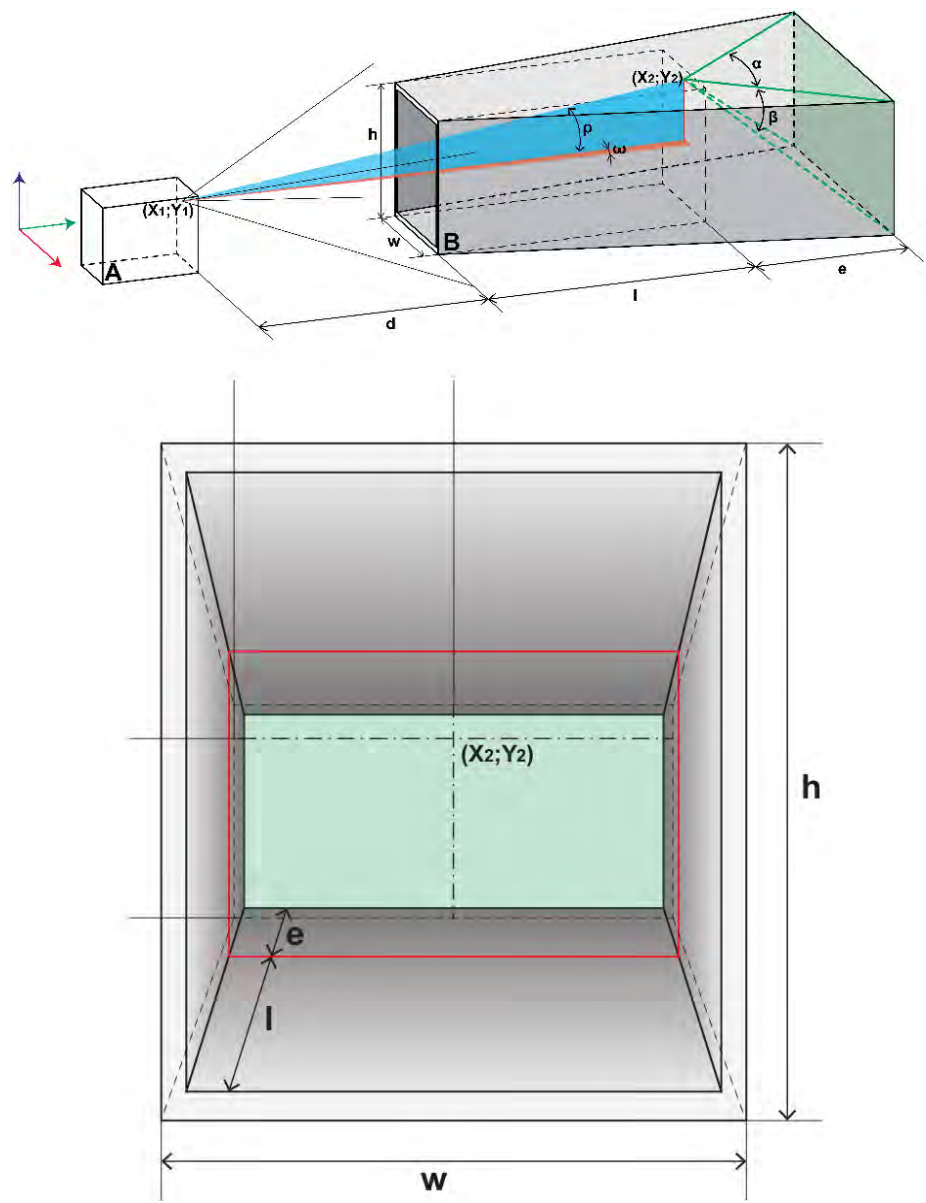

Figure 5.10: Schematics of the computed image representing the 3D-looking frame on which the video-streaming is superimposed. This is explained in detail in [53].

We have designed a simple scenario in OSG which comprises a straight road with one lane in each direction, populated by vehicles which includes cars as well as long trucks. DIVERT simulates the mobility of these vehicles, with trucks travelling at slower speeds than cars. A car driven through the driving simulator based on the Smart ForTwo space frame is also included in the simulation. VNS implements the network simulation, which for this particular driver assistance system has to convey an appropriate delay in the video streaming packets. Note that this delay is of critical importance, as an excessive delay will cause the representation of vehicles in outdated positions, which could mislead the assessment of the safety for the initiation of an over-taking manoeuvre. The network simulation layer, based on IEEE 802.11p, realistically simulated this delay, which would be excessively high using alternative mobile communication paradigms (cellular, for instance), but remains feasible using 802.11p (around 100 ms).

Table 5.1: Description of the variables used for framing the video

| Variable | Description |
| --- | --- |
| $(x_1, y_1)$ | position of the driver's eye point |
| $(x_2, y_2)$ | position of the camera |
| d | distance between vehicles |
| e | distance from camera to the ground capturing point |
| h, l, w | height, length, width of the vehicle B |
| $\alpha$ | horizontal view angle of the camera of the vehicle B |
| $\beta$ | vertical view angle of the camera of the vehicle B |
| $\omega$ | horizontal angle between the cameras positions |
| $\rho$ | vertical angle between the cameras positions |

When a driver in the Smart ForTwo simulator activates STS, we replace the 3D object of the preceding truck with the 3D tubular object represented in Fig. 5.10. This object includes a virtual billboard where the video stream coming from the truck's camera is represented. Note that this video stream already includes the frame delays introduced by the network simulator (NS-3). Figure 5.11 presents a view of the simulation of STS, with and without the system activated.

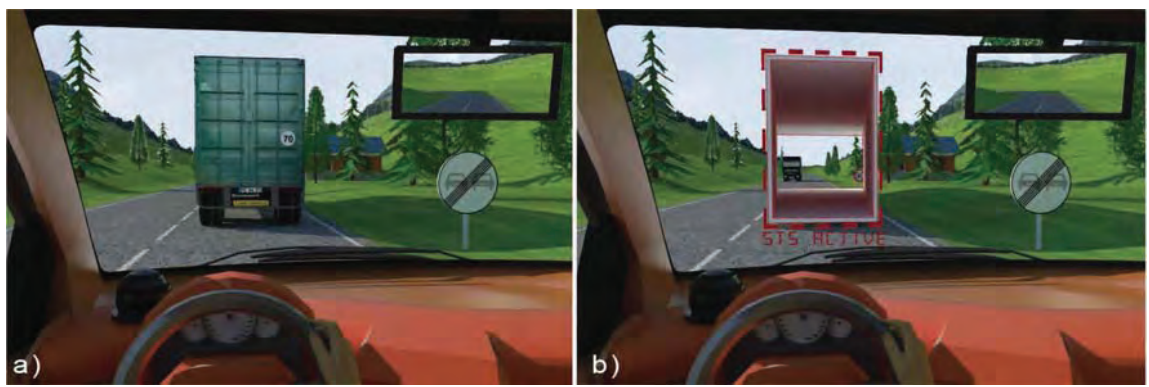

Figure 5.11: In-vehicle perspective of the functional overtaking assistant. In frame a) we depict the driver's view of the road when driving behind a vision-obstructing truck. In frame b) the driver activates the See-Through-System (STS) to assist with the overtaking manoeuvre [53].

### 5.3.2   Virtual Surround Sound

Driving is a multi-sensory experience and it is reasonable to state that it mainly relies on vision for regular tasks and audio for alerts. As mentioned in Chapter 3 sirens have been in use since the 1920's and horns date back to the first horseless carriages that were introduced long before automobiles. From a human reaction point-of-view, audible stimuli [115] produces a much faster reaction time than visual stimuli. Furthermore, most people are able to perceive the directionality of the sound. As mentioned before, some ADAS for parking assistance are only based on audio cues and employ the vehicle surround sound to convey the directionality of the nearby obstacles. This allows drivers to increase their awareness, especially for obstacles outside their field-of-view.

We propose the virtual surround sound (VSS) as an audio AR for alerting drivers of safety related situations. A possible system that we have prototyped based on VSS is a virtual siren, which can be used by ambulances, as well as normal vehicles in emergency situations. The system relies on the constant beaconing provided by the DSRC-enabled vehicles, which provide an information-rich situational awareness. When a CAM message with *LightBarInUse* or *SireneInUse* flags is received, the relative position of the emergency vehicle is calculated and a synthetic siren sound is reproduced through the vehicle's surround sound. The relative speed between the vehicles is used to shift the pitch of the sound in order to reflect the Doppler effect, while the distance is used to determine the volume of the sound. The osgAudio library [90] for the OSG framework provides features such as positionable listeners and sound sources, attenuation and Doppler shifting. In our prototype implementation in the context of the Smart ForTwo driving simulator, we configure a listener representing the human driver seated in the driver's seat of the Smart ForTwo space frame. The relative position of the remaining vehicles simulated in OSG with respect to this listener affect the sound sources, attenuation and Doppler shifting, using a surround sound set of speakers installed in the Smart ForTwo space frame to produce the sound effects that create acoustic AR content. We next present an example of the code using osgAudio library primitives to implement the virtual siren warning.

```
#include <stdio.h>
#include <stdlib.h>
#include "AL/al.h"
#include "AL/alc.h"
#include "AL/alut.h"
#include <memory.h>
#include <unistd.h>
#include <math.h>
#include <ncurses.h>
#include <sys/types.h>
#define NUM_BUFFERS 1

ALuint g_Buffers[NUM_BUFFERS];
ALvoid DisplayALError(ALbyte *szText, ALint errorcode){
 printf("%s%s", szText, alGetString(errorcode));
}

int main(int argc, char* argv[]){
 time_t t1,t2;
```

```
ALint error;
ALCcontext *Context;
ALCdevice *Device;
ALfloat listenerPos[]={0.0,0.0,0.0};
ALfloat listenerVel[]={0.0,0.0,0.0};
ALfloat listenerOri[]={0.0,0.0,-1.0, 0.0,1.0,0.0}; // Listener virado para o ecra

Device = alcOpenDevice(NULL);
Context=alcCreateContext(Device,NULL);
alcMakeContextCurrent(Context);

//atributos do listener
alListenerfv(AL_POSITION,listenerPos);  // Posicao
alListenerfv(AL_VELOCITY,listenerVel);  // Velocidade
alListenerfv(AL_ORIENTATION,listenerOri);  // Orientacao
alGenBuffers(NUM_BUFFERS, g_Buffers);  // Gerar Buffers

ALbyte* WaveNames[] = {
 "ambulance.wav"
};

ALsizei size, freq;
ALenum format;
ALvoid *data;
ALboolean loop;

alutLoadWAVFile(WaveNames[0],&format,&data,&size,&freq,&loop);

if ((error = alGetError()) != AL_NO_ERROR){
 DisplayALError("alutLoadWAVFile : ", error);
 alDeleteBuffers(NUM_BUFFERS, g_Buffers);
 exit(-1);
}

alBufferData(g_Buffers[0],format,data,size,freq);
alutUnloadWAV(format,data,size,freq);

//----------- Source -------------
```

```
ALuint source[2];
ALfloat sourcePos[]={ 10.0, 0.0, -10.0}; //Frente e direita do listener
ALfloat sourceVel[]={ 10.5, 0.0,  10.5};
alGenSources(2,source);

if ((error = alGetError()) != AL_NO_ERROR){
 DisplayALError("alGenSources 2 : ", error);
 return;
}

alSourcefv(source[0],AL_POSITION,sourcePos);
alSourcefv(source[0], AL_VELOCITY, sourceVel);
alSourcei(source[0],AL_BUFFER, g_Buffers[0]);
alSourcei(source[0],AL_LOOPING,AL_TRUE);

//alDopplerFactor(4.0);

alSourcePlay(source[0]);
(void) time(&t1);

while(1){
  (void) time(&t2);
  if((t2-t1)>=1){
    (void) time(&t1);
    sourcePos[0] = sourcePos[0] + sourceVel[0];
    sourcePos[1] = sourcePos[1] + sourceVel[1];
    sourcePos[2] = sourcePos[2] + sourceVel[2];
    alSourcefv(source[0], AL_POSITION, sourcePos);
  }
}
/*char ch;
do{
 ch = getchar();
}while(ch != 'q');*/

alSourceStop(source[0]);
alSourceStopv(2, source);
alDeleteSources(2, source);

return;
```

}

Note that this system allows emergency vehicles to avoid using sirens in restricted zones such as nearby hospitals. Furthermore, a regular vehicle in an emergency situation could also alert neighbouring vehicles without requiring an actual siren to be installed in the vehicle.

The virtual surround system allows many more applications than just the emergency vehicle scenario, using acoustic AR to reproduce sounds tailored for each specific situation. For instance, vehicles equipped with emergency brake assist usually employ hazard warning lights with rapid blinking pattern, which are activated for several seconds after the emergency brake assist is used. However, these rear lights are effective for the car travelling immediately behind the braking car, being much less effective for the cars that have other cars in-between. Using an appropriate CAM message that signals the emergency brake event through DSRC, we make the VSS reproduce a virtual tyre skidding sound in all vehicles in the region of interest (ROI), parametrized by the distance to the braking vehicle. We have performed a preliminary evaluation of the reaction time of drivers driving two cars behind the braking car, using the simulation framework described previously. Response time decreases substantially, being able to avoid the rear-end collision in all our experiments. A complete evaluation of such a system in terms of driver reaction time is out of the scope of this thesis.

### 5.3.3 Virtual Traffic Lights

As mentioned previously, the fundamental idea of VTLs is to replace physical traffic light devices with virtual ones that are displayed inside the vehicle and supported through V2V communication. It has been shown that traffic control through an adequate placement of traffic lights can reduce delay for road traffic and thus moderate the occurrence of collisions [59]. This novel approach to regulate traffic intends to improve traffic flow through the ubiquitous optimized management of individual intersections enabling thus the creation of a traffic lights when and where needed. Figure 5.12 exemplifies the functioning of the VTL system. In frame a) a conflict-free intersection is shown. Vehicle A uses periodic beaconing to advertise its position and heading as it approaches the intersection. No conflicts are detected and it is not necessary to create a VTL. Frame b) shows how the periodic beaconing of concurrent vehicles results in the detection of a crossing conflict and in the need to create a VTL. One of the conflicting vehicles is elected as the intersection leader, and will create and control the VTL. This leader stops at the intersection and temporarily replaces a road-based traffic light in the control of the intersection. While stopped at the intersection, as shown in frame c), the leader optimizes the functioning of the VTL based on the number of vehicles in each approach and periodically broadcasts VTL messages with the colour of each approach/lane. When the cycle ends and the green light is assigned to the leader's lane, the current leader selects a new leader from the vehicles stopped under red lights, as seen in frame d). This new leader continues the cycles and if there are no vehicles under red lights, then the VTL ceases to exist.

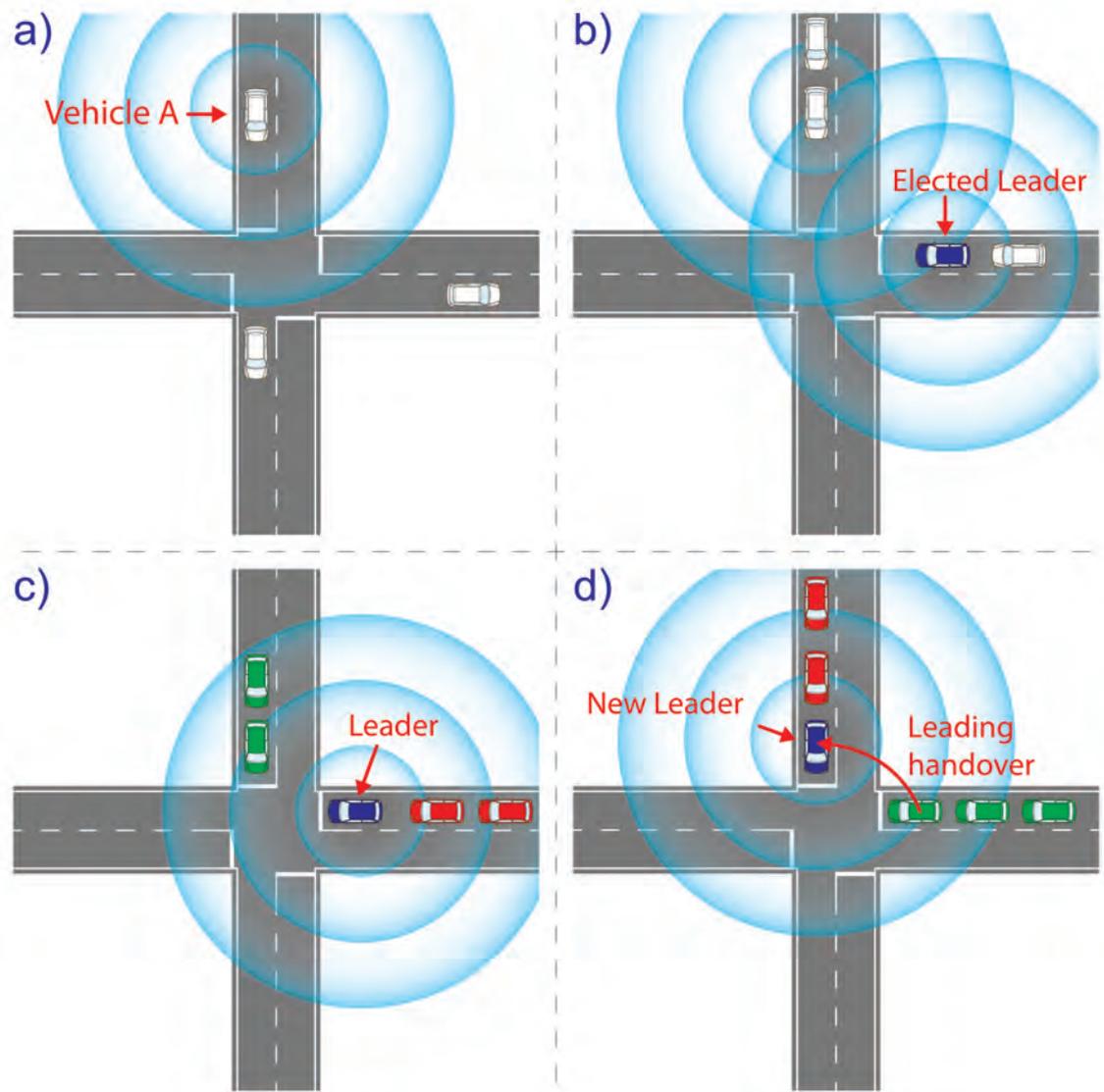

Figure 5.12: Functioning of the Virtual Traffic Lights.

In [86] we presented a Human Machine Interface (HMI) to display the Virtual Traffic Lights and we conducted some evaluation tests to explore the driver's reactions when confronted with this new interface. We evaluated the system to identify possible negative effects of the VTLs in respect to the primary driving task.

The main challenge of virtually representing traffic lights is to accurately reflect the characteristics of conventional traffic lights, as well as making the transition to a new visualization process as smooth as possible. A good interface design complies with specific requirements. In a driving environment, special attention needs to be given to safety and usability. As a consequence, the interface has to be simple and easy to use without interfering with the primary driving task. This means that the time it takes to recognize the displayed information has to be as short as possible. Additionally the information conveyed by the Human Machine Interface has to reflect the traffic conditions found in the real world: good visibility of the displayed information needs to be ensured through good luminance, brightness and contrast. In addition, weather conditions also need to be accounted for, as well as the best possible location for the user interface. Some of the main characteristics of the virtual representation of traffic lights are:

- **Design:** We used a Head Up Display (HUD) to project the virtual objects onto the vehicle's windshield. The images used in the projections correspond to elements found in a real road environment, such as the traffic lights ahead warning signal, arrows and physical traffic lights. Since we also used unfamiliar symbols a text label showing the distance to the intersection completes the information provided to the driver.

- **Placement and Operation:** Following the specifications in [109] we projected the Virtual Traffic Lights information 2.5 to 4 meters away from the driver's eyes, in his lateral field of view. This avoids a road vision obstruction that can result from the projection in the central field of view.

- **Maintenance and Uniformity:** Due to the electronic nature of the system's implementation to display the VTLs, the maintenance is similar to other electronic components found in the vehicle. In addition, the installation of the sensors and the V2V communications enables a similar functioning of all the traffic lights virtually displayed.

- **Colour Code:** Luminance contrast requirements were followed to ensure that there was no visual interference with the road traffic environment.

- **Symbols used:** The VTLs should be functional in situations where an adequate stopping sight distance at the intersection is not available. This occurs when physical infrastructures that are visible to the driver as a reference point are non-existent. Therefore, the VTL approach displays a traffic light ahead warning sign, as seen in Fig. 5.13 (A1 and B1) so that the driver has information about an approaching intersection. Through the windshield projection of the traffic lights, the driver is able to see the traffic light's state at all times. This characteristic makes our approach unique, since it prevents situations where the view of traffic lights mounted on the road can be obstructed by objects. In Portugal the traffic light ahead vertical sign should be placed between 300 and 150 meters from the place where the physical traffic lights are located [98]. As such, the virtual traffic light ahead sign was displayed on the windshield 200 meters before the intersection.

- **Signal Timing:** Our VTL system assures an effective response to changes in traffic conditions through a robust detection system. The traffic light phase awareness additionally allows for the driver to be warned if a traffic violation occurs. Each vehicle maintains an internal database with information about intersections where a virtual traffic light can be created. When approaching such an intersections, if a VTL message is detected, the current state of the VTLs is presented to the driver through the In-Vehicle display. Our detection system is based on beaconing and location table features of VANET's geographical routing protocols, such as Geocast. When vehicles are approaching intersections and do not detect the messages from the VTL, they consult their location tables and the road map topology to infer crossing conflicts that will give rise to the collaborative creation of a VTL. We assume lane-level accuracy on the location tables and a common digital road map that also has lane-level information of topology.

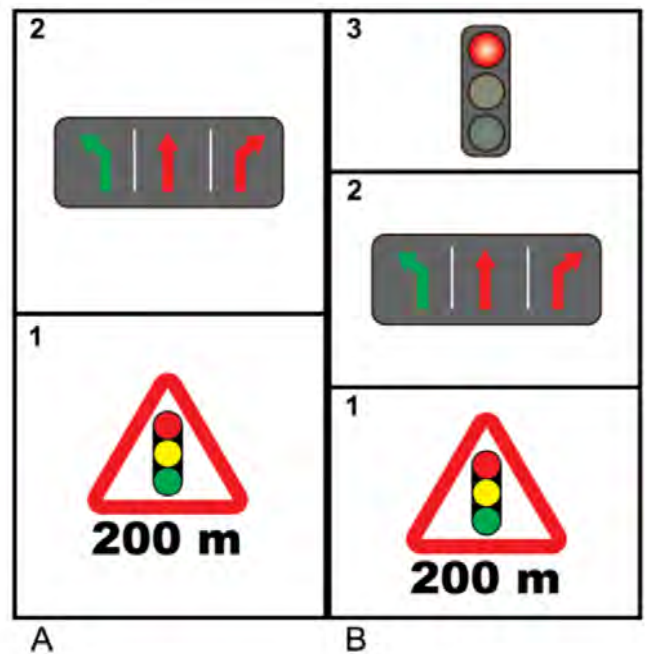

Figure 5.13: Images projected on the driver's windshield corresponding to two different user interface design solutions. A) The image represents the interface through the traffic light ahead sign and driving priority through green or red coloured arrows. B) The image indicates the traffic light ahead, driving priority and driving permission through a traffic light image. [86].

### 5.3.3.1   Methodology

To evaluate the driving performance we first defined the events that cause a variation in the speed, applying them in the same way for each participant.

The experiment consisted on driving through the eight scenario mentioned previously, without secondary tasks in a medium to high traffic density without any critical events. We logged the indicators that caused variation in the scenario such as a red traffic light, and the speed and distance to the intersection. We calculated the brake activity through the deceleration change rate by measuring the speed and calculating the change in deceleration. Additionally we collected driving performance data and subjective ratings through a post task questionnaire. The collected data length was the same for all the participants and it allowed comparing every point overlapping similar speed data sections and determining differences in the speed variation.

In an early stage test phase we performed a formative evaluation to compare different user interface design approaches. We then re-designed the first virtual prototype based on the evaluation results of subjective rates. Figure 5.13 shows the two interface approaches. Approach A consisted of two signs: A1 shows the traffic lights ahead warning sign with its correspondent distance to the intersection label (200 meters); A2 shows the driving priority of the driver and of the other vehicles in the vicinity. The green or red coloured arrows reflect the same behaviour than a conventional traffic light. This sign was displayed when the vehicle was at a distance of 150 meters from the intersection and its size varies depending on the distance to from the intersection. Approach B also presented the driver with the 1 and 2 signs, having the same meaning as A1 and A2. Although approach B presented the B2 sign at a distance between 150 to 50 meters from the intersection, also varying in size according to the distance from the intersection. Sign B3 represent a conventional traffic sign and was displayed when the vehicle was at a distance of 50 to 0 meters from the intersection. While in the formative evaluation, sign A1/B1 was also compared against another sign representing intersection ahead, shown in Fig. 5.14.

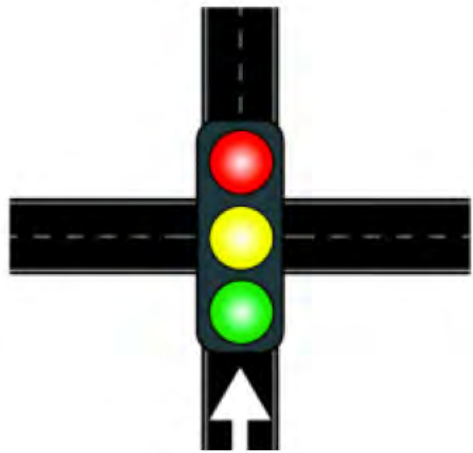

Figure 5.14: Image representing an intersection ahead. [86].

A summative evaluation was also conducted to compare the VTL HMI with that of regular physical traffic lights. To measure the driving performance, we logged the speed data points to get the speed variation and brake performance. Additionally we recorded the traffic light state. Figure 5.15 shows the scenario with the virtual traffic lights, represented through colour arrows on the left and the physical traffic lights on the right.

### 5.3.3.2    Preliminary Evaluation Results

The data resulting from the early stage tests helped to improve the human-centred design of the VTLs user interface, since 100% of the participants agreed that the sign indicating intersection ahead shown in Fig. 5.14 was not intuitive enough. Based on this feedback we redesigned the interface and prompted the participants with the A1/B1 image from Fig. 5.13. Further tests with both approaches indicated that B reflected the idea of a traffic light in a more intuitive way. This was confirmed by 83% of the participants that considered the design in B to be simpler and easier to understand than the design in A. Nonetheless, neither one of the designs were considered dangerous or unsafe by anyone. 90% of the participants classified the information presented by the user interface for Virtual Traffic Lights as clear and intuitive and they did not find the system distracting or unsafe. 55% of the participants considered the Virtual Traffic Lights at least as safe as the conventional set of traffic lights.

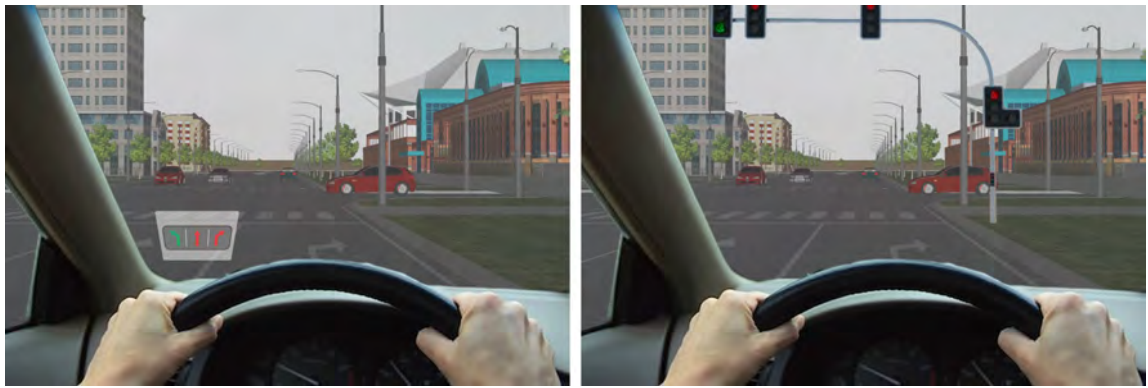

Figure 5.15: Virtual traffic lights (left) and physical traffic lights (right) [86].

As for the brake activity the deceleration change rates differed slightly when using the Virtual Traffic Lights (0.0914) and when using the physical traffic lights (0.0626). However this difference in regard to the brake activity was not significant (t(9)=1.615, p=0.141). Regarding the speed variation metrics, no differences could be determined in the performance with one or the other system. As expected a high data variation regarding the speed could be observed depending on the participant. For example, Fig. 5.16 shows the decrease in the speed in a red traffic light situation for two participants with the virtual traffic light (VTL) system and the conventional traffic light (PTL) representation at a distance less than 25 meters from the intersection.

## 5.4 In-Vehicle Prototyping

In addition to the simulator-based prototyping described in the previous section, it is also important to test the proposed system using real-vehicles in real driving scenarios. In the context of the limited resources involved in an MSc. thesis, this can constitute an overwhelming task, almost impossible to implement. Nevertheless, and given the research projects that support the research developed within this thesis, we were able to implement and test-drive some simple prototypes of driver assistance systems that leverage on AR and VANET communication. In this section we briefly describe such prototyping.

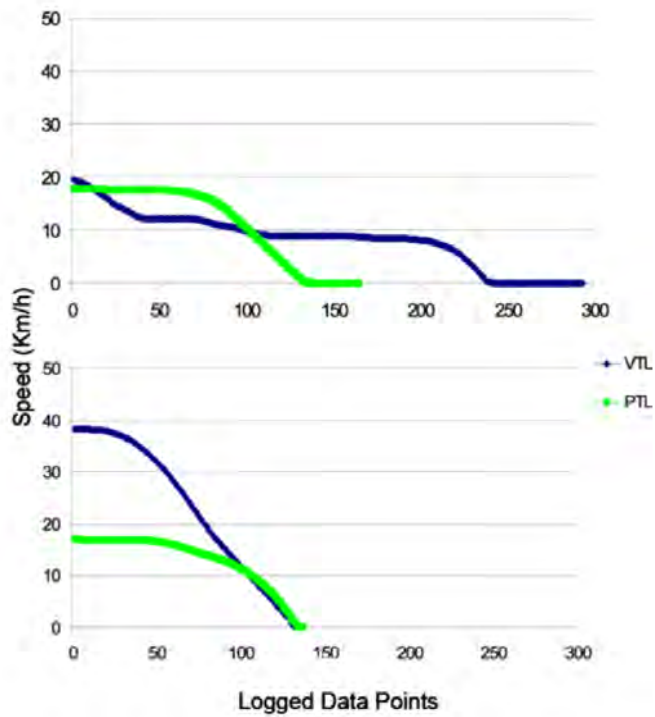

Figure 5.16: Speed variation depending on the traffic control device. The graphic illustrates the data corresponding to two drivers. [86].

### 5.4.1    An In-Vehicle Prototype of the See-Through System

Two critical components are necessary for the in-vehicle implementation of this prototype: the DSRC radio, which supports inter-vehicle video-streaming; and the AR equipment, which presents the vision-obstructing vehicles as transparent tubular objects.

#### 5.4.1.1    The DSRC radio prototype and video streaming

To implement the inter-vehicle communication through 802.11p, we have used the radios developed within the DRIVE-IN project [36]. A detailed description of the development of these radios is available in [4]. These radios are based on a single-board computer, featuring a IEEE 802.11a/b/g wireless card and an IEEE 802.11p wireless card. They also have a native GPS chip. The operating system is based on Debian Linux Squeeze, and the wireless driver for 802.11p is based on a modification of the ath5k driver. This driver supports channel coordination, channel routing, multi-channel synchronisation, and alternate and continuous channel access. In Fig. 5.17 we illustrate one of these radios, where the in-vehicle antennas are visible. Mounting in-vehicle antennas is much simpler when no dedicated vehicles are available for our experiments.

To implement the video streaming, a GoPro camera is installed on the windshield of the truck. The Smoke video codec is used to minimise delays introduced by video-frame compression [55].

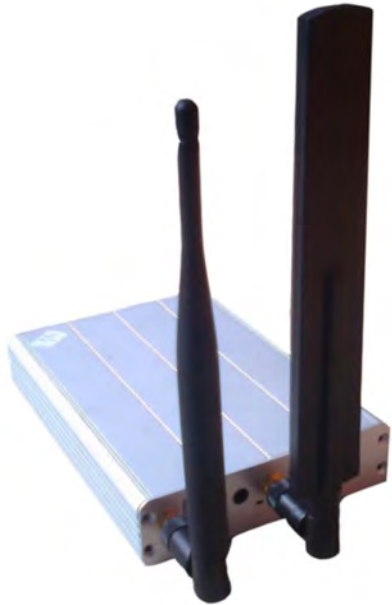

Figure 5.17: DSRC radio developed within the DRIVE-IN project.

### 5.4.1.2 AR Generation

The implementation of the AR content on the windshield of the vehicle is the more complex task. In [55], Gomes et al. used a transparent LCD installed on the windshield to implement an optical see-through system from the displaying of AR content. A test-drive of the prototype implementation of STS done using a transparent LCD can be seen in the video available at [37]. Using this technology, some problems arise from the level of optical transparency that is provided by the LCD. It causes a blurred vision of the non-digital content, resulting in a less comfortable visual experience with such an implementation of a virtual windshield. In addition, the fixed position of the LCD in relation to the driver's eye point also results in misalignment in some situations in the super-imposition of the digital content and the real-worlds object. Eye tracking systems could mitigate this problem.

To overcome these problems, we implemented the STS' human-computer interface using AR smart-glasses. We believe that in a near future AR-enabled cars could contain factory-installed integrated smart glasses, which would connect to the car's other sensors, and be used by drivers as naturally as sun glasses or seat belts.

For the implementation of STS using smart glasses we have used the Vuzix®STAR 1200 XL [1] equipment. These glasses feature a 1080p camera, adjustable eye-separation and 35-degree diagonal field-of-view. The Vuzix®glasses' transparent widescreen video displays provide a see-through perspective even when we overlay 2D or 3D content, which is perfect for the STS. We used laptops connected to all the devices in each vehicle to run the experiment. The distance between vehicles used in the frame generation is provided by the GPS receivers placed in both vehicles. The use of Vuzix glasses overcomes the issue with the driver's perspective display calibration exhibited in the implementation used a transparent LCD. The blurred vision of non-digital content is also totally non-existent using the VUZIX equipment. Figure 5.18 show the usage of the Vuzix®glasses in our test-drive experiment.

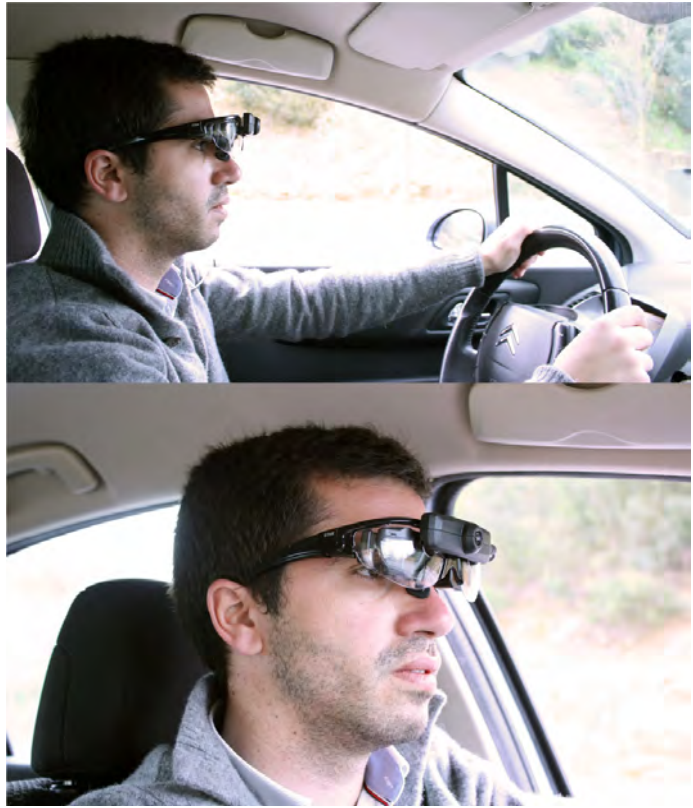

Figure 5.18: Vuzix®STAR 1200XL AR glasses used in the vehicular environment [46].

These glasses have an embedded camera on the bridge of the nose that records the exact perspective of the driver's eyes. Furthermore, the glasses' displays are placed right in front of the eyes, thus producing a more accurate AR immersion. Using the embedded camera and computer vision we are able to detect the rear of the preceding vehicle in such a way that we can overlap the 3D-looking frame with the embedded video-streaming. The STS implementation using the Vuzix®glasses can be seen in Fig. 5.19 . In this figure we are able to observe the seamless blend of digital objects with the real-world content, where the STS's 3D-looking frame is almost undistinguishable from the real objects. Another important issue of the implementation of STS with smart glasses compared to the transparent LCD implementation, is that the complex setup of the transparent LCD solution, together with the legal issues involving windshield-installed systems in public driving scenarios, forced for a test-drive in a closed area with these solution, while using the smart glasses we were able to make the driving experiments in common and public driving scenarios.

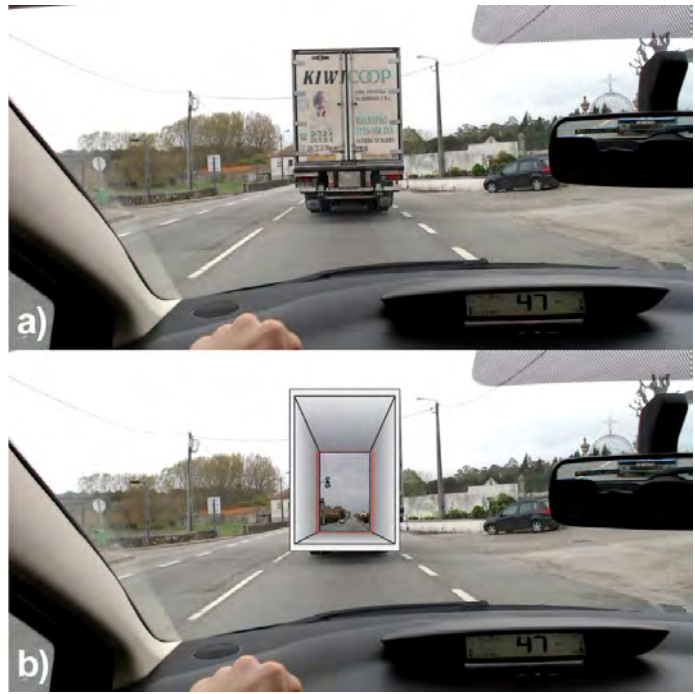

Figure 5.19: See-Through System as an overtaking assistant, implemented using Vuzix®STAR 1200XL AR glasses. In frame a) we show the driver's perspective with STS disabled. In frame b) we show the AR result with STS enabled. Note that the snapshot shown in frame b) results from an offline photomontage that merges a video recorded using the glasses (without AR) with the AR content (3D boundary frame and video) that is actually sent to the glasses display. It is impossible to obtain an actual snapshot of the driver's vision while using the glasses. [46].

### 5.4.2 Displaying Virtual Traffic Signs

Using VUZIX®smart glasses, the creation of virtual traffic signs in our in-vehicle experiments becomes simple. The VUZIX®glasses include a 3 degree of freedom Wrap Head Tracker, with compass (pitch, yaw, roll), which provide the parameters that characterise the orientation of the drivers' vision. Essentially, this glasses work as an external monitor connected to a laptop PC running an instance of OSG. A viewer in OSG is configured to have the same perspective as the VUZIX®glasses, using parameters coming from the Wrap Head Tracker and compass. A GPS is also connected to the laptop PC, providing accurate positioning of the driver that is using VUZIX®glasses. Traffic signs represented as 3D objects, with different levels of detail as a function of distance, are placed in the OSG virtual space, using coordinates in the same coordinate system of the GPS receiver. As the vehicle approaches the location of the virtual traffic sign, it will be transparently created by OSG and displayed through VUZIX®glasses. Care has to be taken in relation to the real obstructions that could hide this virtual sign and that are not represented in the OSG scenario. In particular, the case of other vehicles obstructing such virtual objects can generate a non-realistic perspective of the virtual traffic sign.

As with the implementation of STS using VUZIX®glasses, the recreation of the visual notion of the driver using smart glasses is impossible to convey through images presented in this thesis. This experience requires the actual use of the VUZIX®equipment.

Note that the representation of more sophisticated roadside traffic objects, such as traffic lights, is also possible using the same framework, with animations performed by OSG. Instead of basing the creation of such virtual traffic signage objects in static information read by OSG, a just-in-time creation can be started based on V2V communication (a virtual emergency triangle, for instance), or V2I communication (temporary road signs, for instance).

## 5.5 Video See-Through Implementation

As is clear from the presentation in the previous sections, the creation of complete and detailed 3D scenarios to perform a simulation-based evaluation of our proposed driver assistance systems is an overwhelming task, requiring a level of resources that usually exceeds academic projects. Identically, performing road-based experiments is also very difficult within the constraints of academic research. Not only the access to vehicles and budget to support on-road experiments, but mainly the dangers involved in the evaluation of novel prototypes in real driving conditions and interactions with other non-participating vehicles.

We thus designed an intermediate approach between simulation-based evaluation and test-drive evaluation. We collected video recordings from the point of view of the driver, annotated with the GPS coordinates of each frame that was recorded. We could thus setup a scenario is OSG where the viewer is configured with these GPS coordinates, with the remaining parameters (field of view degree, etc) also set to the same parameters of the video recording. In OSG we set a scenario which includes a billboard object, which always faces the viewer, where the video recording is displayed. The inclusion of additional 3D objects, such as traffic signs or traffic lights is done as in the previously describe manner for 3D modelled OSG scenarios. This allows for a very realistic and simple to execute evaluation of the displaying of virtual information merged with the previously collected video. In Fig. 5.20 we show the result of displaying a virtual traffic light using this Video See-Through Implementation.

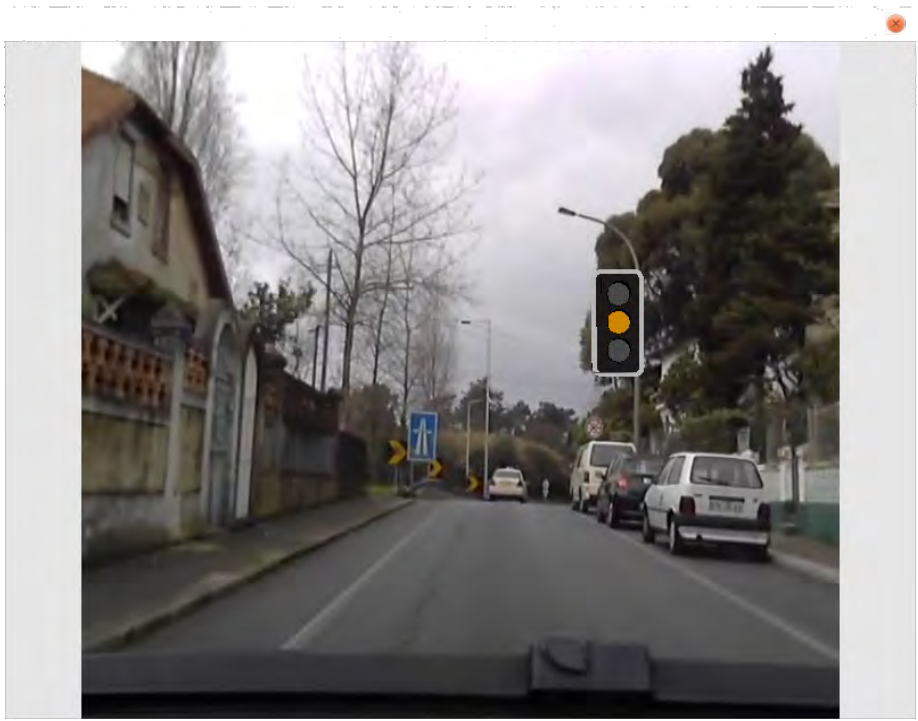

Figure 5.20: Virtual traffic light implemented with Video See-Through.

# Chapter 6

# Conclusions and Future Work

## 6.1 Conclusions

With this thesis we aimed at designing and developing novel driver infotainment systems based on the creation of augmented reality superimposed over the windshield. Our goal was to identify ways in which the overall driving experience can be improved in terms of safety, navigation and traffic control efficiency.

An analyses of augmented reality technology has provided us with the knowledge necessary to understand different display and tracking methods, as theses highly influence the level of immersion users feel. We found that the enclosed aspect of a car is an ideal tool to provide drivers with an enhanced and immersive driving experience, as the vehicles windshield can operate as a transparent canvas over which digital content can be superimposed.

As such, we explored driving as a task in which drivers' senses are triggered. We conducted an extensive overview of devices and systems that aim at providing more and better awareness to drivers. We started with an analysis of traditional and more common devices found not only within vehicle, but also in road scenarios and we note that the necessity to create augmented scenarios has been explored even before the development of motorized vehicles. With the development of technology it is visible that augmented reality is already being explored e a vehicular context. We listed several examples that show how vehicles already implement many visual, audible and some tactile augmented reality systems for human computer interaction with advanced driving assistance systems. However, these are limited due to week sensing capabilities and a lack of data available in the vehicles.

We proposed leveraging the richness of information and situational awareness provided by vehicular networks in order to create cooperative ADAS systems designed to improve road safety. We implemented a smart glass version of the see-through system, a visual AR system that uses real-time video streaming through VANET to virtually transform the preceding vehicle into a partially transparent object, in order to improve the safety of overtaking manoeuvres. We implemented the Virtual Surround Sound, an audio AR system that transforms the information provided by the VANET of an approaching emergency vehicle into an audible representation. A first proposal of a graphical user interface for the Virtual Traffic Lights was implemented and we found that participants could easily identify a new visualization scheme.

Overall, we show that VANETs can provide important safety and awareness capabilities and that augmented reality allows for intuitive driving interfaces.

We also believe our results contribute to the formulation of a vision where the vehicle is perceived as an extension of the body which permeates the human senses to the world outside the vessel, where the car is used as a better, multi-sensory immersive version of a mobile phone that integrates touch, vision and sound enhancements, leveraging unique properties of vehicular ad hoc networking.

## 6.2   Research papers published as a result of the work in this

In this section, we list the work which has been published in the scope of this thesis.

- [86] C. Olaverri-Monreal, P. Gomes, M. K. Silveria, M. Ferreira. In-Vehicle Virtual Traffic Lights: a Graphical User Interface, Seventh Iberian Conference on Information Systems and Technologies, Madrid, Spain, 2012.

- [46] Michel Ferreira, Pedro Gomes, Michelle Krüger Silvéria, Fausto Vieira. Augmented Reality Driving Supported by Vehicular Ad Hoc Networking. Mixed and Augmented Reality (ISMAR), 2013 IEEE International Symposium on. 2013. (To appear.)

## 6.3   Future Work

Our research will continue with the exploration of augmented reality in a vehicular context with further testing of our current systems.

- We intend to better our driving simulator to provide a greater sense of realism while testing systems.

- We will continue to explore augmented reality in a driving environment with the smart glasses.

- Our Virtual Surround Sound will continue under development.

- We will explore augmented reality through digital advertising super-imposed on physical billboards.